\documentclass[prl,showpacs,twocolumn,amsmath,amssymb,floatfix,superscriptaddress]{revtex4-2}
\usepackage{amsmath}
\usepackage{amssymb}
\usepackage{graphicx}
\usepackage{bm}
\usepackage{color}
\usepackage{hyperref}
\usepackage{graphicx}
\usepackage{upgreek}
\usepackage{scalerel}
\usepackage{multirow}
\usepackage{footnote}

\usepackage{pgffor}
\usepackage{pdfpages}
\usepackage{mathdots}
\usepackage{float}
\usepackage{silence}
\usepackage{physics} 
\usepackage{lscape}   
\usepackage{color, colortbl}
\usepackage[table]{xcolor}
\usepackage{comment}  
\makeatletter
\AtBeginDocument{\let\LS@rot\@undefined}
\makeatother

\begin{document}
\title{Crossed surface flat bands in three-dimensional superconducting altermagnets}

\author{Yuri Fukaya}
\affiliation{Faculty of Environmental Life, Natural Science and Technology, Okayama University, 700-8530 Okayama, Japan}

\author{Bo Lu}
\affiliation{Center for Joint Quantum Studies, Tianjin Key Laboratory of Low Dimensional Materials Physics and Preparing Technology, Department of Physics, Tianjin University, Tianjin 300354, China}

 \author{Keiji Yada}
 \affiliation{Department of Applied Physics, Nagoya University, 464-8603 Nagoya, Japan}
 
\author{Yukio Tanaka}
\affiliation{Department of Applied Physics, Nagoya University, 464-8603 Nagoya, Japan}

\author{Jorge Cayao}
\affiliation{Department of Physics and Astronomy, Uppsala University, Box 516, S-751 20 Uppsala, Sweden}
 
\date{\today} 
\begin{abstract}
Superconducting altermagnets have proven to be a promising ground for emergent phenomena, but their study has involved two-dimensional systems. 
In this work, we investigate three-dimensional $d$- and $g$-wave altermagnets with spin-singlet chiral $d$-wave superconductivity and show the formation of crossed surface flat bands due to the interplay between superconducting and altermagnetic symmetries.
We find that these crossed flat bands are topologically protected, appear at zero energy in the surface along $z$ due to the superconducting nodal lines in the $xy$-plane, and their number of corners is determined by the crystal symmetry of altermagnets.  We also show that the superconducting nodal lines give rise to Bogoliubov-Fermi surfaces, which then affect the appearance of zero-energy arcs in the surface along $x$. 
Moreover, we demonstrate that the crossed   flat bands or surface arcs, and Bogoliubov-Fermi surfaces give rise to the coexistence of three  distinct dependences of the charge conductance on the normal transparency, hence offering a solid way for their detection and paving the way for realizing higher-dimensional topological phases using altermagnets. 
\end{abstract}
\maketitle

%%%%%%%%%%%%%%%%%%%%%%%%%%%%%%%
% SECTION 1:                  Introduction.                                  %
%%%%%%%%%%%%%%%%%%%%%%%%%%%%%%%
The discovery of altermagnetism has recently triggered great interest  due to its potential for realizing emergent phenomena and spintronic applications \cite{Bailing,MazinPRX22,Song2025,FukayaJPCM2025}. Altermagnets (AMs) are characterized by  a zero net magnetization as a result of   their momentum dependent  spin splitting of energy bands  and anisotropic spin-polarized Fermi surfaces  \cite{noda2016momentum,NakaNatCommun2019,Hayami19,Ahn2019,Yuanprb20,LiborSAv,NakaPRB2020,Yuanprm21,LiborPRX22,MazinPRX22,landscape22}. While these   properties have proven crucial in the normal state \cite{Bailing}, it has been   shown that combining AMs with superconductivity offers unprecedented opportunities  \cite{FukayaJPCM2025,liu2025review}. 
For instance, altermagnetism has been shown to induce key bulk and surface phenomena  \cite{FukayaJPCM2025,liu2025review}, including  anomalous Josephson transport \cite{Sun23,Papaj23,Ouassou23,Beenakker23,Bo2024,sun2024,Cheng24,fukaya2024}, topological superconductivity~\cite{cano23,Zhongbo23,CCLiu1,CCLiu2,mondal2024,hadjipaschalis2025Mzm,hodge2025platform,mcardle2026topologicalsuper,sharma2026doublepeakmajorana,wan2026chiral}, exotic superconducting pairs \cite{zhang2024,Maeda2025, chakraborty2024,khodas2025strain,PhysRevB.111.054520,parshukov2025,mazin2025notes,fu2025light,fu2025floquet,Yokoyama25floquet,mukasa2025finite,heinsdorf2025,monkman2025perscurrent,ChangPRB2025},  nonreciprocal transport \cite{Banerjeediode24,chengdiode24,Chakraborty25,sharma2025diode,sharmadiode2026},  nontrivial light-matter coupling \cite{fu2025light,fu2025floquet}, and distinct types of stable subgap states \cite{lu2025subgap,liu2025FFLOBFS}.  
% For instance, altermagnetism has been shown to induce key bulk and surface phenomena  \cite{FukayaJPCM2025,liu2025review}, including anomalous Josephson transport \cite{zhang2024,Ouassou23,Beenakker23,Bo2024,sun2024,Cheng24,fukaya2024}, {\color{blue}topological superconductivity \cite{cano23,Zhongbo23,CCLiu1,CCLiu2,mondal2024,hadjipaschalis2025Mzm,hodge2025platform,mcardle2026topologicalsuper}, distinct types of stable subgap states \cite{lu2025subgap,liu2025FFLOBFS}, etc~\cite{Maeda2025, chakraborty2024,khodas2025strain,PhysRevB.111.054520,parshukov2025,mazin2025notes,fu2025floquet,Yokoyama25floquet,mukasa2025finite,liu2026altermagnetism,heinsdorf2025,monkman2025perscurrent,ChangPRB2025,Banerjeediode24,chengdiode24,Chakraborty25,sharma2025diode,fu2025light,fu2025floquet}.}  

The appearance of stable subgap states  is of particular relevance because they  signal unusual superconductivity   protected by symmetry and topology \cite{KT2000,tanaka2012symmetry,sato2016,sato2017,Cayao2020odd,tanaka2024theory}. In this regard, altermagnetism has been shown to induce Bogoliubov-Fermi surfaces (BFSs), flat bands, and Andreev bound states (ABSs), all with direct ties to the   altermagnetic and superconducting symmetries \cite{lu2025subgap,liu2025FFLOBFS}. While these ideas   show the potential of AMs for   topological superconducting phases, the   existing studies   only addressed two-dimensional (2D) systems,   leaving  unexplored the large family of 3D AMs. 
One of the most interesting 3D systems is Sr$_2$RuO$_4$, which has very recently been predicted to host altermagnetic order \cite{Autieri25,Ramires2025} as a phase that coexists with a very likely 3D chiral $d$-wave superconductivity \cite{RamiresPRB2019,SuhPRR2020,maeno2024NatPhys,maeno2024JPSJ} probed by nuclear magnetic resonance \cite{pustogow2019Nature,ishida2020JPSJ}. 
It is thus natural to wonder about the design and detection of higher-dimensional topological superconducting phases by exploiting the symmetry and topology of 3D superconducting AMs.

\begin{figure}[t!]
    \centering
    \includegraphics[width=8.5cm]{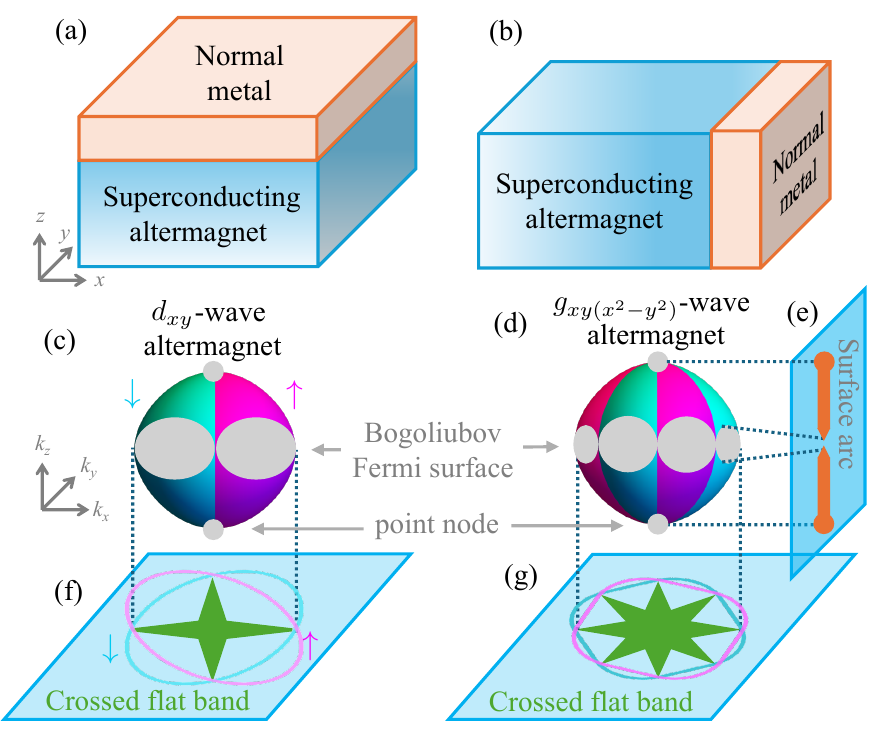}
    \caption{(a,b) Sketches  of   the studied  junctions  along  $z$- (a) and $x$-directions (b), formed by a 3D superconducting altermagnet (light blue) and a normal metal (light red). (c-f)   Bulk and surface properties of   $d_{xy}$- and  $g_{xy(x^2-y^2)}$-wave altermagnets with  chiral $d$-wave superconductivity. (c,d)   Bulk Fermi volumes for $d_{xy}$- (c) and $g_{xy(x^2-y^2)}$-wave (d) altermagnets, where the cyan and magenta colors indicate  up and down spins. The line nodes of the Fermi volumes give rise to Bogoliubov-Fermi surfaces (gray) under  chiral $d$-wave  superconductivity. (e) A projection of the    Fermi volumes on the  [100] surface along $x$ leads to surface arc states   (orange), whose ends are marked by the point nodes.   (f,g) When projecting the   Fermi volumes on the [001] surface along $z$, crossed flat bands emerge (green) with their corners defined by the nodes of the  2D spin-polarized altermagnetic Fermi surfaces, indicated by  cyan and magenta ellipses for down and up spins. The crossed flat bands appear at zero energy, which is ensured by the nodal lines of chiral  $d$-wave   of superconductivity. }
    \label{Fig1}
\end{figure}

% {\color{blue}Although the study of superconducting altermagnets has started in 2D systems, most materials of altermagnets are 3D.} 
% In this work, we consider 3D  superconducting AMs [Fig.\,\ref{Fig1}] and demonstrate the emergence of crossed flat bands as a generic topological surface phenomenon due to the interplay between intrinsic superconducting and altermagnetic symmetries. 
In this work, we consider 3D superconducting AMs [Fig.\,\ref{Fig1}] and demonstrate the emergence of crossed surface flat bands as a generic topological phenomenon due to the interplay between superconducting and altermagnetic symmetries. 
In particular, we focus on $d$- and $g$-wave AMs with chiral $d$-wave superconductivity, which are very likely to appear in Sr$_2$RuO$_4$, see \cite{Autieri25,Ramires2025} and
\cite{RamiresPRB2019,SuhPRR2020,maeno2024NatPhys,maeno2024JPSJ,csireYF2026directional}. 
We discover that the $xy$-plane nodal lines of chiral $d$-wave superconductivity ensure that the crossed flat bands appear at zero energy on the [001] surface, while their corners are determined by the altermagnetic nodes. 
The crossed surface flat bands are topologically protected and represent the higher-dimensional analog of zero-energy ABSs \cite{Hu94,KT2000} in 3D superconducting AMs, thus broadening their initial conception in Weyl superconductors \cite{LuPRL2015} and the idea that flat bands inherit the shape of the projected Fermi surface \cite{SatoPRB2011,BrydonPRB2011}.
Moreover, we find that the superconducting nodal lines induce BFSs, which then modify the formation of arcs on the [100] surface.
% We further show that the crossed flat bands, arc states, and BFSs produce very distinctive conductance signatures, which can be used for their detection. 
We further show that the crossed flat bands or arc states, and BFSs coexist and produce very distinctive transparency dependent conductance signals, which can be used for their detection.  Our findings establish 3D  superconducting AMs as a fertile ground for designing   symmetry-protected topological phenomena.

%%%%%%%%%%%%%%%%%%%%%%%%%%%%%%%
% SECTION 2:                            Model:                                %
%%%%%%%%%%%%%%%%%%%%%%%%%%%%%%%
\textit{3D superconducting AMs}.---
We consider a 3D superconducting AM modelled by the following Bogoliubov-de Gennes (BdG) Hamiltonian,
 \begin{equation}
 \label{HBdG}
    \hat{H}^{\alpha}_\mathrm{BdG}(\bm{k})=
     \varepsilon(\bm{k})\hat{\sigma}_{0}\hat{\tau}_{3}+\hat{H}^{\alpha}_\mathrm{AM}(\bm{k})+\hat{H}_\mathrm{SC}(\bm{k})\,,
     \end{equation}%
where $\varepsilon(\bm{k})=-\mu-2t\cos{k_x}-2t\cos{k_y}-2t\cos{k_z}$, while $\hat{H}^{\alpha}_\mathrm{AM}(\bm{k})$ and $\hat{H}_\mathrm{SC}(\bm{k})$ describe the altermagnetic and superconducting orders, with $\alpha$ denoting the type of altermagnetism (see below). Here,  $\bm{k}=(k_x,k_y,k_z)$, $\mu$ is the chemical potential,   $t$ is the hopping integral, while $\hat{\sigma}_{j}$ and $\hat{\tau}_{j}$ are the $j$-th Pauli matrices in spin and particle-hole subspaces, respectively. Altermagnetism is modelled by
\begin{equation}
    \hat{H}^{\alpha}_\mathrm{AM}(\bm{k})=M_{\bm{k}}^{\alpha}\hat{\sigma}_{3}\hat{\tau}_{3}\,,
\end{equation}
where $M_{\bm{k}}^{\alpha}$ characterizes $\alpha$-wave altermagnetic order with $\alpha=d,g$.
We focus on $d_{xy}$- and $d_{x^{2}-y^{2}}$-wave AMs described by $M_{\bm{k}}^{d1}=2t_{d1}\sin{k_x}\sin{k_y}$, $M_{\bm{k}}^{d2}=t_{d2}(\cos{k_x}-\cos{k_y})$. We also consider $g$-wave AMs with $M_{\bm{k}}^{g}=4t_{g}\sin{k_x}\sin{k_y}(\cos{k_x}-\cos{k_y})$.   Thus, $t_{d1}$, $t_{d2}$, and $t_{g}$ represent the strength of $d_{xy}$-, $d_{x^{2}-y^{2}}$-, and $g_{xy(x^2-y^2)}$-wave AMs. 
%{\color{blue}Thus, $t_{\alpha}$ with $\alpha=d1,d2,g$ represents the strength of $d_{xy}$-, $d_{x^{2}-y^{2}}$-, and $g_{xy(x^2-y^2)}$-wave AMs.}
While $\hat{H}_\mathrm{AM}(\bm{k})$ is described by an effective 2D model, our findings remain in 3D AMs \cite{EzawaPRB}, see S3 in the Supplementary Material (SM)~\cite{SM}, which includes Refs.~\cite{UedaSigristReview,tsuchiuraJPSJ1995,MicnasReview,Ketterson_Song_super,annett2004superconductivity}. 
Moreover, we assume spin-singlet 3D chiral superconductivity   modelled by  $\hat{H}_\mathrm{SC}(\bm{k})=-\Delta\sin{k_z}[\sin{k_x}\hat\tau_2+\sin{k_y}\hat\tau_1]\hat{\sigma}_{2}$, with pair amplitude $\Delta$;  $t_{\alpha}$ is varied within $\Delta$ to ensure the coexistence of altermagnetism and superconductivity~\footnote{The chosen values of the superconducting and altermagnetic amplitudes are obtained by ensuring the coexistence of altermagnetism and 3D chiral $d$-wave superconductivity in the self-consistent gap equation~\cite{SM}. 
This allowed us to choose altermagnetic and superconducting amplitudes of the same order}, as shown in S1.2 of \cite{SM}.  In fact,  in S1.2 of~\cite{SM}, we show that the chosen spin-singlet 3D chiral $d$-wave pairing   coexists with altermagnetism by self-consistently solving the gap equation with a second-nearest neighbor attractive interaction  \cite{ChakrabortyPRB2025zerofield,ChourasiaPRB2025}. Thus, this justifies our choice of  3D chiral $d$-wave pairing in 3D superconducting AMs and can apply the interorbital pairing of Sr$_2$RuO$_4$~\cite{RamiresPRB2019,SuhPRR2020,maeno2024NatPhys,maeno2024JPSJ} because the relating symmetry and gap structure are similar to interlayer pairings~\cite{SuhPRR2020}. Also, Refs.\,\cite{Autieri25,Ramires2025} predict altermagnetic order in Sr$_2$RuO$_4$, and a very likely 3D chiral $d$-wave superconductivity in Refs.\,\cite{RamiresPRB2019,SuhPRR2020,maeno2024NatPhys,maeno2024JPSJ}.  
% Does this material also be a candidate material? and UPt$_3$ in a recent experiment revealing with the 3D chiral $d$-wave pairing~\cite{bisset2025UPt3}. 
Although we focus on a particular 3D nodal superconductor, our   analysis can be straightforwardly generalized to other  superconducting AMs with unconventional superconductivity.
 
 %%%%%%%%%%%%%%%%%%%%%%%%%%%%%%%
% SECTION 3:          Crossed  flat bands:                              %
%%%%%%%%%%%%%%%%%%%%%%%%%%%%%%%

\textit{Emergence of crossed surface flat bands}.---We are interested in unveiling the impact of altermagnetism on superconductivity in a 3D superconducting AM modelled by Eq.\,(\ref{HBdG}). 
For this purpose, it is useful to first inspect the energy bands, which are obtained as 
\begin{equation}
\label{EkBdG}
E_{\bm{k},\pm}^{s,\alpha}=s M_{\bm{k}}^{\alpha}\pm\sqrt{\varepsilon^{2}(\bm{k})+|\bar{\Delta}(\bm{k})|^{2}}\,,
\end{equation}
where $s=\pm$ and $\bar{\Delta}(\bm{k})=\Delta\sin{k_z}[\sin{k_x}+i\sin{k_y}]$. 
These energy bands help understand the appearance of bulk phases. 
In the absence of AMs,    the 3D chiral superconductor has line nodes in the $xy$-plane, while point nodes in the  $z$-plane [Fig.\,\ref{Fig1}(c,d)]. 
These line and point nodes give rise, respectively, to zero-energy surface flat bands on the [001] surface and to surface arc states on the [100] surface~\cite{Hu94,KT2000,SatoPRB2011,KobayashiPRB2015}; the shape of the flat bands is determined by that of the projected Fermi surface, with a disk shape for a spherical Fermi surface~\cite{KobayashiPRB2015}. 
Here, we obtain $ E_{\bm{k},+}^{-,\alpha}=0$ for $ [M_{\bm{k}}^{\alpha}]^{2}=\varepsilon^{2}(\bm{k})+|\bar{\Delta}(\bm{k})|^{2}$.
This defines the formation of a BFS, and shows that the line nodes of the chiral $d$-wave superconductor become BFSs due to altermagnetism, except for the nodal lines of AMs.
Interestingly, altermagnetism strongly deforms the surface states of the chiral superconductor, giving rise to \emph{crossed flat bands} and modified surface arcs in the [001] and [100] surfaces [Figs.\,\ref{Fig1}(e,f,g)].  
While the surface arcs are inherited from the chiral superconductor, the crossed flat bands are entirely due to the interplay between altermagnetism and chiral $d$-wave superconductivity: by projecting the altermagnetic Fermi surfaces, we identify that the altermagnetic nodes (spin-degenerate lines in 3D) define the number of corners of the crossed flat band. 
Moreover, due to the nodes of the 3D chiral $d$-wave superconductivity, the crossed flat bands appear at zero energy.  This analysis also applies to spin-triplet $p_z$-wave   superconductivity \cite{tanaka2012symmetry}.  
Thus, the emergence of crossed flat bands is a generic phenomenon resulting from the interplay between altermagnetism and nodal superconductivity and hence tied to the inherent crystal symmetries.

%%%%%%%%%%%%%%%%%%%%%%%%%%%%%%%%%%%%%%%%%%
% SECTION 3:            Topological stability of the Crossed  flat bands:                      %
%%%%%%%%%%%%%%%%%%%%%%%%%%%%%%%%%%%%%%%%%%

\textit{Topological origin of crossed surface flat bands}.---The symmetries required for the emergence of crossed flat bands also imply that they have a topological origin~\cite{YadaPRB2011,SatoPRB2011,BrydonPRB2011,tanaka2012symmetry,tanaka2024theory,PhysRevB.102.075430}, which we show here.
In 3D superconductors with nodal lines but without altermagnetism, a nonzero  1D winding number can be obtained with the chiral operator by combining the pseudo-time-reversal (pTRS) and particle-hole (PH) symmetries~\cite{KobayashiPRB2015}, where pTRS is the product of conventional TRS and local gauge transformation. This local gauge transformation is introduced to cancel out the inner phase of the pair potential. In the presence of altermagnetism, the chiral symmetry is broken because of the broken TRS by AMs \cite{Bailing,FukayaJPCM2025}. Still, the Hamiltonian keeps the pseudo-magnetic mirror symmetry (pMMS), which is given by the product of pTRS and mirror symmetry.  For $d_{xy}$- and $d_{x^{2}-y^{2}}$-wave AMs, the  pMMSs are given, respectively, by
 \begin{align}
 \hat{\Theta}^{'}_{\bm{k}} \hat{H}^{d1}_\mathrm{BdG}(k_x, k_y, k_z)\hat{\Theta}_{\bm {k}}^{'\dagger}&=
[\hat{H}^{d1}_\mathrm{BdG}(-\nu k_x, \nu k_y, k_z)]^{*}\label{eq:mms1}\\
 \hat{\Theta}^{'}_{\bm{k}} \hat{H}^{d2}_\mathrm{BdG}(k_x, k_y, k_z)\hat{\Theta}_{\bm {k}}^{'\dagger}&=
[\hat{H}^{d2}_\mathrm{BdG}(-\nu k_y, -\nu k_x, k_z)]^{*}\label{eq:mms2}
 \end{align}
where $\hat{\Theta}^{'}_{\bm k}=\hat{U}_{\phi_{\bm k}}^\dagger\hat{\Theta} \hat{U}_{\phi_{\bm k}}$  denotes the pMMS, $\hat{\Theta}=i\hat{\sigma}_2\hat{\tau}_0$ is the conventional TRS, $\hat{U}_{\phi_{\bm k}}=e^{-i\phi_{\bm k}\hat{\sigma}_0\hat{\tau}_{3}/2}$ is the local gauge transformation, and  $\phi_{\bm k}=\tan^{-1}(\sin k_y/\sin k_x)$ is the phase of the pair potential. Here, $\nu=+1$ and $\nu=-1$ in Eq.\,(\ref{eq:mms1}) indicate pMMSs for $d_{xy}$-wave AMs in the $xz$- and $yz$-planes, respectively. 
Also, $\nu=+1$ and $\nu=-1$ in Eq.\,(\ref{eq:mms2}) indicate pMMSs for $d_{x^2-y^2}$-wave AMs in the diagonal planes on $x=y$ and $x=-y$, respectively.
For $g_{xy(x^2-y^2)}$-wave AMs, the system has both pMMSs given by  Eq.\,(\ref{eq:mms1}) and Eq.\,(\ref{eq:mms2}). 
In addition to the pMMSs, $\hat{H}^{\alpha}_\mathrm{BdG}({\bm k})$ has particle-hole symmetry given by
\begin{align}
\hat{C} \hat{H}^{\alpha}_\mathrm{BdG}(k_x, k_y, k_z)\hat{C}^\dagger=-[\hat{H}_\mathrm{BdG}^{\alpha}(-k_x, -k_y, -k_z)]^{*}\label{eq:phs}
\end{align}
with $\hat{C}=\hat{\sigma}_{0}\hat{\tau}_1$.
Then, $\hat{H}^{\alpha}_\mathrm{BdG}(\bm{k})$ anticommutes with the chiral operator $\hat{\Gamma}_{\bm{k}}=-i\hat{C}\hat{\Theta}^{'}_{\bm{k}}$ on the magnetic mirror plane.
 Concrete forms of $\hat\Gamma_{\bm k}$ on each mirror plane are given in S1.4 of  \cite{SM}. 
Importantly, these magnetic mirror and chiral symmetries are still available on the smooth surfaces, which are perpendicular to the magnetic mirror planes, allowing for the symmetry-protection of surface states as a unique effect due to altermagnetism and nodal superconductivity. 
For example, a vertical pMMS is given in Eq.\,(\ref{eq:mms1}) and the corresponding chiral symmetry is accessible on the $k_x$- and $k_y$-axes in the surface Brillouin zone. 
Hence, a winding number can be obtained on the symmetric lines,  whose nonzero values signal the appearance of   topological phases with topologically protected surface states \cite{SM}.  Thus, a nonzero winding number along   the $k_x, k_y$-axis and/or the diagonal direction  implies that the crossed flat bands along the [001] direction discussed in the previous section are topologically protected  [Fig.\,\ref{Fig1}(f,g)]. Also, because of the pMMS on the $xz$-plane, topologically protected zero-energy surface states appear on the [100] surface, which we  identified   as surface arcs [Fig.\,\ref{Fig1}(e)]. As a result, both  surface arcs and crossed flat bands are topologically  protected  and can be realized in 3D superconducting AMs~\footnote{This topological protection does not happen in the case when BFS exists because the winding number is ill-defined for the gapless dispersion. Thus, the surface arc state at $k_z\sim 0$ is not expected to produce resonant states at zero energy.}.

\begin{figure}[t!]
    \centering
    \includegraphics[width=8.5cm]{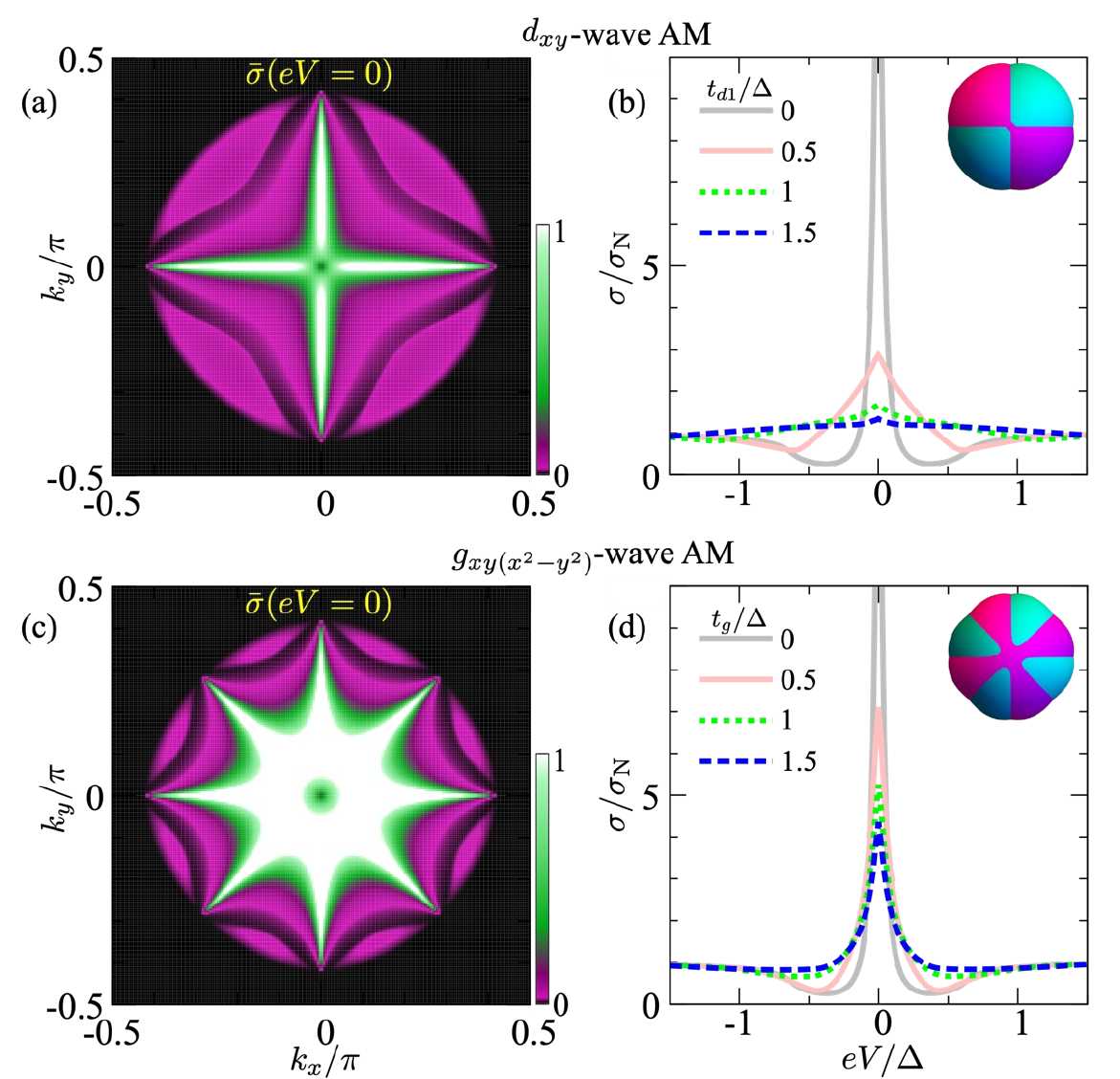}
    \caption{(a,c) Zero-bias conductance along the $z$-direction [001] as a function of momenta $k_{x,y}$ for a superconducting AM with $d_{xy}$-wave (a) and  $g_{xy(x^2-y^2)}$-wave (c) altermagnetism.  (b,d) Normalized total conductance  along the $z$-direction as a function of $eV$ for distinct values of the altermagnetic strength $t_{\alpha}$. $\sigma_\mathrm{N}$ indicates the conductance in the normal state at $eV=0$. The insets in (b,d) indicate the normal state Fermi surfaces projected onto the [001] surface for up (magenta) and down spins (cyan). Parameters:  $t_{\alpha}=\Delta$ (a,c), $\mu=-4.5t$, $\Delta=0.01t$,   $U_\mathrm{b}=5t$, and $\delta=0.01\Delta$.}
    \label{Fig2}
\end{figure}
 
%%%%%%%%%%%%%%%%%%%%%%%%%%%%%%%
% SECTION 4:              Conductance signatures:                   %
%%%%%%%%%%%%%%%%%%%%%%%%%%%%%%%
\textit{Conductance signatures of crossed  flat bands}.---Having shown the emergence of crossed flat bands and surface arcs, as well as their  topological origin, here we explore how they can be detected by means of conductance.  For this purpose, we couple the 3D superconducting AM to a normal metal, such that we can assess transport along $z$ and $x$ directions [Fig.\,\ref{Fig1}(a,b)]. We obtain the conductance following the common tunneling Hamiltonian approach  based on the Lee-Fisher formula~\cite{Lee_Fisher,OhashiPRB2024,fukaya2025PUM,lu2025subgap}:
\begin{equation}
\label{conductance1}
    \sigma(eV)=\frac{1}{2\pi}\int^{\pi}_{-\pi}\int^{\pi}_{-\pi}d\tilde{\bm{k}}\,\bar\sigma(\tilde{\bm{k}},eV)\,,
\end{equation}%
with   $\tilde{\bm{k}}=(k_u,k_v)$ for $u=x,y$ and $v=y,z$, while $\bar\sigma(\tilde{\bm{k}},eV)$ is the momentum-dependent conductance obtained as
\begin{equation}
\label{conductance2}
    \bar\sigma(\tilde{\bm{k}},eV)=\mathrm{Tr'}
    \sum_{i,j={0,1}}[
    \hat{G}_{ii}\hat{V}_{ji}^{\dagger}\hat{G}_{jj}\hat{V}_{ji}-\hat{G}_{ij}\hat{V}_{ji}\hat{G}_{ij}\hat{V}_{ji}
    ]\,,
\end{equation}
where we have dropped the momenta for simplicity, and $\mathrm{Tr'}$ means that the trace is taken in electron space.  
Here, $\hat{V}_{10}=-\hat{t}_{0}\otimes\hat{\tau}_{0}$, with $\hat{t}_{0}$ characterising the hopping matrix in metallic lead [Fig.\,\ref{Fig1}(a,b)]. 
Moreover, $\hat{G}_{nm}$ are Green's functions numerically obtained by using the recursive Green's function method~\cite{Umerski97,san2013multiple,TakagiPRB2020,OhashiPRB2024,fukaya2025PUM,lu2025subgap}, and the details of this formulation are presented in   S2 of the SM~\cite{SM}. 

We first assess the conductance signatures along $z$ direction [001], see Fig.\,\ref{Fig2}~\footnote{The characteristic energy scale that determines the conductance profile originates from the relative amplitude of the altermagnetic strengths $t_{\alpha}$ and the pair potential $\Delta$.}. 
We begin by exploring the momentum-dependent conductance $\bar\sigma(\tilde{\bm{k}},eV=0)$ at zero bias $eV=0$, which is presented in Fig.\,\ref{Fig2}(a,c) for superconducting AMs with $d_{xy}$- and $g_{xy(x^{2}-y^{2})}$-wave altermagnetism. 
The most relevant signature of the momentum-dependent conductance at zero bias is that the strongest intensities come from the contribution of the crossed surface flat bands [Fig.\,\ref{Fig2}(a,c)], which emerge at the [001] surface as discussed in the previous section. 
Depending on the type of underlying altermagnetism, the zero-bias conductance  $\bar\sigma(\tilde{\bm{k}},eV=0)$ unveils crossed flat bands with different number of corners [Fig.\,\ref{Fig2}(a,c)], as expected since the corners of the emergent crossed flat bands are determined by the nodes of the AM, where chiral symmetry can be defined. 
For $d_{xy}$-wave AM, $\bar\sigma(\tilde{\bm{k}},eV=0)$ reveals four corners [Fig.\,\ref{Fig2}(a)], while eight corners for  $g_{xy(x^{2}-y^{2})}$-wave AM [Fig.\,\ref{Fig2}(c)].  
For $d_{x^{2}-y^{2}}$-wave AMs, $\bar\sigma(\tilde{\bm{k}},eV=0)$ also exhibits four corners but with a crossed flat band rotated by $\pi/4$, see Sec.\,E1 and Fig.\,\ref{Fig1_EM}(a) in the End Matter. 
Moreover, $\bar\sigma(\tilde{\bm{k}},eV=0)$ in Fig.\,\ref{Fig2}(a,c) acquires lower but nonzero intensities around the crossed flat bands, which arise due to   the BFS. 
This is confirmed by the zero-energy bulk density of states integrated in $k_z$, see Sec.\,E2 and Fig.\,\ref{Fig2_EM} in the End Matter~\footnote{The BFS in our study is similar to that in the original multiorbital model of BFS studies~\cite{Agterbergprl17,Timmprb17,Brydonreview} from the point of view of time-reversal symmetry breaking. However, in our case, we adopt a single-orbital model, and the time-reversal symmetry breaking is entirely due to altermagnetism, which produces a zero net magnetization, hence being different from conventional magnetism needed in initial studies of BFSs \cite{Agterbergprl17,Timmprb17,Brydonreview}.}.
There also appears a darkish area between the BFS and crossed flat band in Fig.\,\ref{Fig2}(a,c)  originated from a gapped dispersion where quasiparticles attain vanishing velocity along $z$, see S1.3 in SM \cite{SM}.    
Notably, we find that, for the crossed flat bands, BFSs and gapped regions, respectively,
we obtain $\bar\sigma(\tilde{\bm{k}},eV=0)\sim a\bar{\sigma}_{\rm N}^{0}+b\bar{\sigma}_{\rm N}^{1}+c\bar{\sigma}_{\rm N}^{2}$, with the normal state conductance $\bar{\sigma}_{\rm N}$, unveiling three distinct types of fundamental power laws that coexist in 3D superconducting AMs
\footnote{We note that $\bar\sigma(\tilde{\bm{k}},eV=0)$ given by $2\bar\sigma_{\rm N}^{0}$ and $\bar\sigma_{\rm N}^{2}$ originates due to perfect resonance of Andreev reflections \cite{TK95,KT2000,tanaka2024theory} and Andreev reflection without resonance \cite{Beenakker,tanaka2024theory}.}; see S2 in  \cite{SM}. 
The total conductances $\sigma(eV)$ are then shown in Fig.\,\ref{Fig2}(b,d) as a function of $eV$ for distinct altermagnetic strengths $t_{d1,g}$:
$\sigma(eV)$ develops a zero-bias peak whose height decreases as the altermagnetic strength increases; see also Fig.\,\ref{Fig1_EM}(b). 
The presence of a large zero-bias conductance peak (ZBCP) without any altermagnetism is due to the zero-energy ABSs of chiral $d$-wave superconductors~\cite{TK95,KT2000,KobayashiPRB2015,tanaka2024theory}. 
Altermagnetism, however, reduces this ZBCP:  Its reduced value remains even at strong values of altermagnetic strengths [Fig.\,\ref{Fig2}(b,d)]. 
While the same reduction occurs for $d_{xy}$-, $d_{x^{2}-y^{2}}$-, and $g_{xy(x^{2}-y^{2})}$-wave AMs, the latter exhibits a stronger and well-formed ZBCP; see Sec.\,E5 and Fig.\,\ref{Fig5_EM} of the End Matter. The signature of the BFS (altermagnetic nodes) appears in the width (height) of the ZBCP  of $\sigma(eV)$~\cite{SM}. 
In all cases, the ZBCP constitutes direct evidence of the crossed surface flat bands formed at the [001] surface of 3D superconducting AMs.

\begin{figure}[t!]
    \centering
    \includegraphics[width=8.5cm]{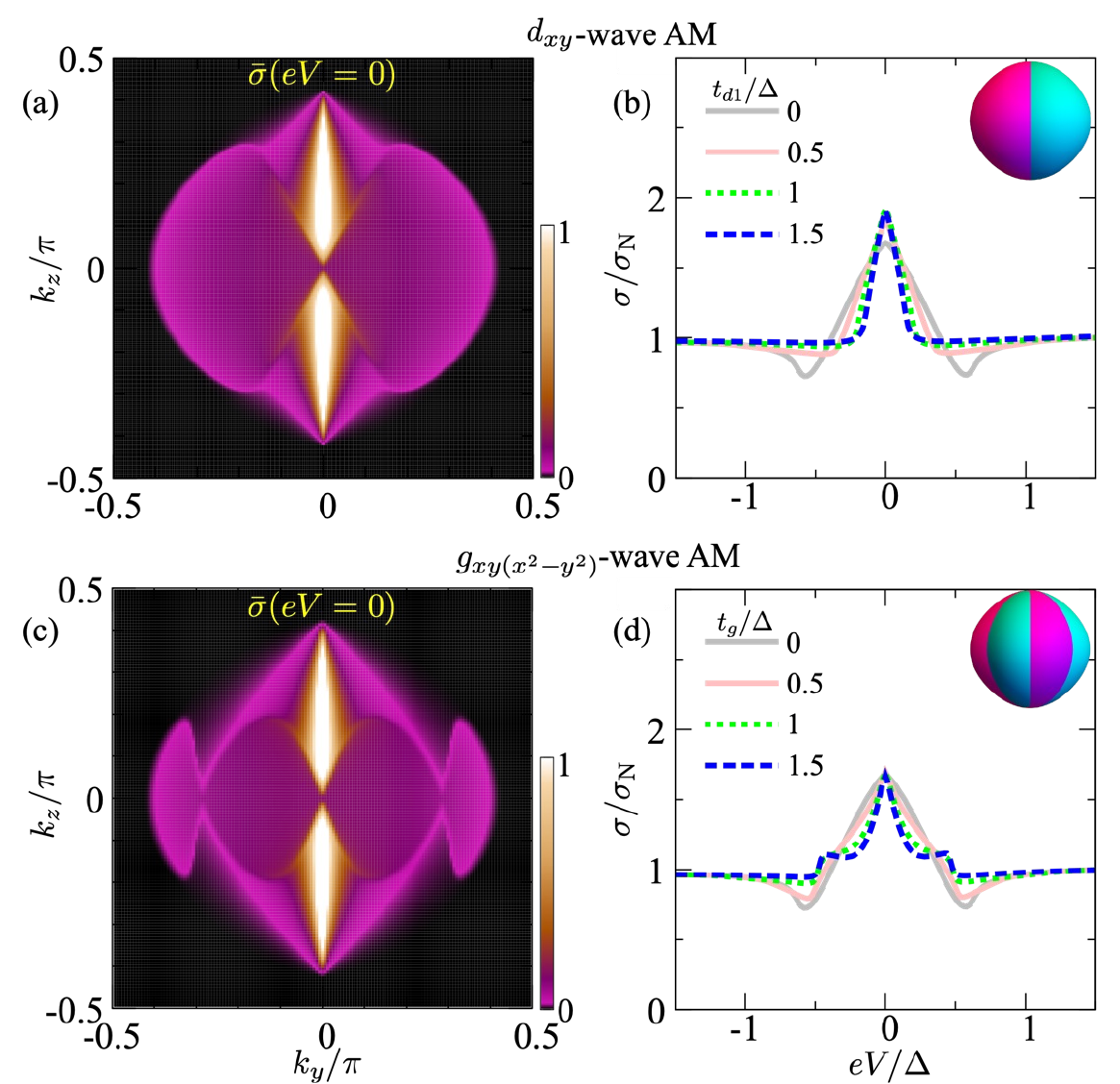}
    \caption{(a,c) Zero-bias conductance along the $x$-direction [100] as a function of  $k_{y,z}$  for a superconducting AM with $d_{xy}$-wave (a) and  $g_{xy(x^2-y^2)}$-wave (c) altermagnetism.  
    (b,d) Normalized total conductance  along the $x$-direction [100] as a function of $eV$ for distinct values of the altermagnetic strength $t_{\alpha}$. $\sigma_\mathrm{N}$ indicates the conductance in the normal state at $eV=0$. 
    The insets in (b,d) indicate the normal state Fermi surfaces projected onto the [100] surface for up (magenta) and down spins (cyan). Parameters:  $t_{\alpha}=\Delta$ (a,c), $\mu=-4.5t$, $\Delta=0.01t$,  $U_\mathrm{b}=5t$, $\delta=0.01\Delta$.}
    \label{Fig3}
\end{figure}%

%%%%%%%%%%%%%%%%%%%%%%%%%%%%%%%%%%%%%%%
% SECTION 5:    Directional dependent transport and signatures of BFS :      %
%%%%%%%%%%%%%%%%%%%%%%%%%%%%%%%%%%%%%%%

\textit{Conductance along $x$ direction and signatures of surface arcs}.---
Conductance along $x$ reveals distinct signatures of the 3D superconducting AM, giving rise to direction-dependent transport in these systems. 
In Fig.\,\ref{Fig3}(a,c), we present the zero bias momentum-dependent conductance $\bar\sigma(\tilde{\bm{k}},eV=0)$   for superconducting AMs with $d_{xy}$- and $g_{xy(x^{2}-y^{2})}$-wave altermagnetism;   Sec.\,E3 and Fig.\,\ref{Fig3_EM}(a) of the End Matter  shows $\bar\sigma(\tilde{\bm{k}},eV=0)$ for $d_{x^{2}-y^{2}}$-wave AMs. 
We see that
$\bar\sigma(\tilde{\bm{k}},eV=0)$ develops high intensities around $k_{y}=0$ due to the contribution of zero-energy surface arc states formed at the [100] surface [Fig.\,\ref{Fig3}(a,c)].   
As discussed before, these zero-energy surface arcs originate from the pMMS on the $xz$-plane. 
Interestingly, the surface arc conductance signal is interrupted around $k_{z}=0$ due to a reduced contribution coming from the BFSs, see Fig.\,\ref{Fig1}(d); the role of the BFS can indeed be confirmed by the bulk density of states integrated in $k_x$, see Fig.\,\ref{Fig4_EM}.
For $d_{x^{2}-y^{2}}$-wave AMs in Fig.\,\ref{Fig3_EM}(a),  $\bar\sigma(\tilde{\bm{k}},eV=0)$ reveals the formation of spin-split surface arc states, which disperse with $k_{y,z}$ and is smeared by the BFSs. %exhibit a vanishing effect due to BFSs. 
As for transport along $z$, here we also find that $\bar\sigma(\tilde{\bm{k}},eV=0)$  is given by $2\bar\sigma_{\rm N}^{0}$, $\bar\sigma_{\rm N}^{1}$, and $\bar\sigma_{\rm N}^{2}$ in the presence of zero-energy surface arcs, BFS, and gapped region; see S2.3 in \cite{SM}. In the case of the total conductance $\sigma(eV)$,  Fig.\,\ref{Fig3}(b,d) shows it as a function of $eV$ for distinct strengths of $d_{xy}$ and $g_{xy(x^2-y^2)}$-altermagnetism.   
It exhibits a broad peaked profile around zero bias without altermagnetism [Fig.\,\ref{Fig3}(b,d)], resulting from   the chiral edge modes \cite{KobayashiPRB2015}. This broad conductance profile sharpens at finite altermagnetism for $d_{xy}$ and $g_{xy(x^2-y^2)}$-wave AMs due to the change in the dispersion of the chiral edge modes, producing a clear ZBCP [Fig.\,\ref{Fig3}(b,d)] that stems from the surface arc states~\cite{SM}. Interestingly, for $d_{x^{2}-y^{2}}$-wave altermagnetism, the resonance in $\sigma(eV=0)$ splits into two, and the resonance shift depends on the altermagnetic strength [Fig.\,\ref{Fig3_EM}(b)]; this spin-split total conductance originates from the spin-split surface arcs observed in  $\bar\sigma(\tilde{\bm{k}},eV=0)$ [Fig.\,\ref{Fig3_EM}(a)].
This behavior is similar to the reported transport in Sr$_2$RuO$_4$~\cite{KashiwayaPRL2011}, supporting the idea that conductance along $x$ is determined by the surface arc states and the BFS.

%%%%%%%%%%%%%%%%%%%%%%%%%%%%%%%
% SECTION 5:                       CONCLUSIONS                       %
%%%%%%%%%%%%%%%%%%%%%%%%%%%%%%%
In conclusion, we have demonstrated that crossed surface flat bands emerge in three-dimensional superconducting altermagnets as a generic topological phenomenon protected by intrinsic crystal symmetries.
Also, we have shown that the crossed flat bands can be detected by means of conductance, which depends on the transport direction and can also assess the formation of topological surface arcs and Bogoliubov-Fermi surfaces. We have found that the zero-bias momentum-resolved conductance exhibits three distinct types of fundamental power laws, offering a solid way for detecting the emergent topological phases, e.\,g., by Doppler shift measurements as those predicted in Refs.\,\cite{TanakaJPSJ2002,TanumaPRB2002_Sep,TanumaPRB2002_Nov,Tanaka2003JPSJ}, ARPES \cite{RevModPhys.75.473,RevModPhys.93.025006,zhang2022angle,RevModPhys.96.015003}, and quasiparticle interference experiments~\cite{allan2012anisotropic,allan2013imaging,marques2021tomographic,wang2025NatPhys}, which would require   low transparencies and low temperatures for observing the crossed flat bands; see Sec.\,E5 in End Matter.  
Our results apply to the large family of predicted 3D AMs, such as CrSb~\cite{long2026CrSb} and La$_2$CuO$_4$~\cite{LiborPRX22}, and superconducting UPt$_3$~\cite{bisset2025UPt3} and half-Heusler compounds~\cite{Brydonreview}. Notably, our work holds direct relevance in  Sr$_2$RuO$_4$, where altermagnetism and chiral $d$-wave superconductivity are expected to coexist \cite{pustogow2019Nature,ishida2020JPSJ,RamiresPRB2019,SuhPRR2020,maeno2024NatPhys,maeno2024JPSJ,Autieri25,Ramires2025,csireYF2026directional}, being  thus a promising material to test our predictions.
Our work, therefore, establishes three-dimensional superconducting altermagnets as an outstanding ground for realizing higher-dimensional topological phases, which can also pave the way for future tunneling spectroscopy experiments in three-dimensional superconductors. 
Based on our topological analysis, we argue that found crossed flat bands are robust against symmetry-preserving perturbations, such as multi-orbital, spin-orbit coupling, and scalar disorder, while symmetry-breaking perturbations are very likely detrimental and require a thorough investigation.

%%%%%%%%%%%%%%%%%%%%%%%%%%%%%%%
%                        ACKNOWLEDGMENTS                               %
%%%%%%%%%%%%%%%%%%%%%%%%%%%%%%%

Y.\,F.\,acknowledges financial support from the Sumitomo Foundation and JSPS with Grants-in-Aid for Scientific Research (KAKENHI Grant No.\ 26K17096), and the calculation support from Okayama University and the computer resource offered under the category of general project by Research Institute for Information Technology, Kyushu University.   
B.\,L.\,acknowledges financial support from the National Natural Science Foundation of China (project 12474049) and Beijing National Laboratory for Condensed Matter Physics (2025BNLCMPKF011). 
K.\,Y., Y.\,F., and Y.\,T.\,acknowledge financial support from JSPS with Grants-in-Aid for Scientific Research (KAKENHI Grants No.\,25K07203).  
Y.\,T.\,also acknowledges financial support from JSPS with Grants-in-Aid for Scientific Research (KAKENHI Grants No.\,23K17668 and 24K00583).  
J.\,C.\,acknowledges financial support from the Swedish Research Council (Vetenskapsr\aa det Grant No.\,2021-04121).  
 
 \bibliography{biblio}
\onecolumngrid
\newpage
\begin{center}
\textbf{\textsc{END MATTER}}
\end{center}
\twocolumngrid
\vspace{1em}

\textit{\textbf{E1.} Conductance along $z$-direction  with $d_{x^{2}-y^{2}}$-wave AMs}.---The conductance along $z$-direction for a superconducting AM junction with $d_{x^{2}-y^{2}}$-wave altermagnetism is presented in  Fig.\,\ref{Fig1_EM}. In the case of the momentum dependent conductance at zero-bias voltage $\bar\sigma(\tilde{\bm{k}},eV=0)$ in Fig.\,\ref{Fig1_EM}(a), it develops high intensity values along the diagonals $k_{x}=\pm k_{y}$ in contrast to the case of $d_{xy}$-wave altermagnetism having large values at $k_{x,y}=0$, see Fig.\,\ref{Fig2}(a). These large zero-bias conductance values reveal the emergence of crossed flat bands in the [001] surface of superconducting AMs with $d_{x^{2}-y^{2}}$-wave altermagnetism. As noted in the main text, $\bar\sigma(\tilde{\bm{k}},eV=0)$ also reflects the BFS, seen as an area that surrounds the crossed flat band in Fig.\,\ref{Fig1_EM}(a). Between the BFS and the area of the crossed flat band appears a darkish region, which is gapped and quasiparticles acquire vanishing velocity and hence vanishing conductance. Interestingly, $\bar\sigma(\tilde{\bm{k}},eV=0)$  acquires a unique dependence on its zero-bias normal state counterpart:  $\bar\sigma(\tilde{\bm{k}},eV=0)$ is roughly given by $2\bar{\sigma}_{\rm N}^{0}$,  $\bar{\sigma}_{\rm N}^{1}$, and $\bar{\sigma}_{\rm N}^{2}$ for the regime with crossed flat bands, BFSs, and gapped region, respectively; see S2 in   \cite{SM}. Also, the total conductance $\sigma(eV)$ varies from a very peaked profile at zero bias in the absence of $d_{x^{2}-y^{2}}$-wave altermagnetism to a reduced but visible peak at finite as the altermagnetic strength increases and surpasses the magnitude of the pair potential [Fig.\,\ref{Fig1_EM}(b)]. Thus, the reduced peaked profile manifests the formation of crossed flat bands on the [001] surface.

% \textit{\textbf{E2.} Projected density of states on the [001] surface}.---
\textit{\textbf{E2.} Bulk density of states along the [001] direction}.---
We have seen that the gapless structure of the energy dispersion in Eq.\,(\ref{EkBdG}) originates from BFSs. 
It is thus expected that the density of states (DOS) enhances due to BFSs. 
To show this, we calculate the bulk DOS as
\begin{equation}
\label{DOSEq}
    % D(E)=-\frac{1}{\pi}\int^{\pi}_{-\pi}d\bm{k}\mathrm{Tr'}[\mathrm{Im}\tilde{G}^\mathrm{R}(\bm{k},E)],
    \tilde{D}(\bm{k},E)=-\frac{1}{\pi}\mathrm{Tr'}[\mathrm{Im}\tilde{G}^\mathrm{R}(\bm{k},E)],
\end{equation}%
where $\mathrm{Tr'}$ indicates the trace taken in electron space, $\tilde{G}^\mathrm{R}(\bm{k},E)$ is the retarded Green's function   $\tilde{G}^\mathrm{R}(\bm{k},E)=[E+i\delta-\hat{H}_\mathrm{BdG}(\bm{k})]^{-1}$, with $\hat{H}_\mathrm{BdG}(\bm{k})$ given by Eq.\,(\ref{HBdG}), $E$ and $\delta$ are the energy and an infinitesimal positive number. 
Using Eq.\,(\ref{DOSEq}), we obtain the bulk DOS on the [001] direction by integrating for $k_z$.
% along the transport direction. 
To unveil the BFS, in Fig.\,\ref{Fig2_EM} we plot the bulk DOS on the [001] direction at zero energy for $d_{xy}$-, $d_{x^{2}-y^{2}}$-, and $g_{xy(x^{2}-y^{2})}$-wave AMs. The most important feature is that the BFS produces the highest intensity in the projected zero-energy DOS, see Fig.\,\ref{Fig2_EM}. 
Interestingly, the BFSs surround regions with tiny values, which possess corners depending on the type of AM. 
These inner regions correspond to the formation of crossed flat bands, in Figs.\,\ref{Fig2}(a,c) and \ref{Fig1_EM}(a). 
The bright point at $k_{x,y}=0$ is due to the nodal point where the energy gap closes. The  BFS signals in the bulk DOS remain in more elaborated 3D AMs, see Sec.\,S3 in SM~\cite{SM}.

\begin{figure}[t!]
    \centering
    \includegraphics[width=8.4cm]{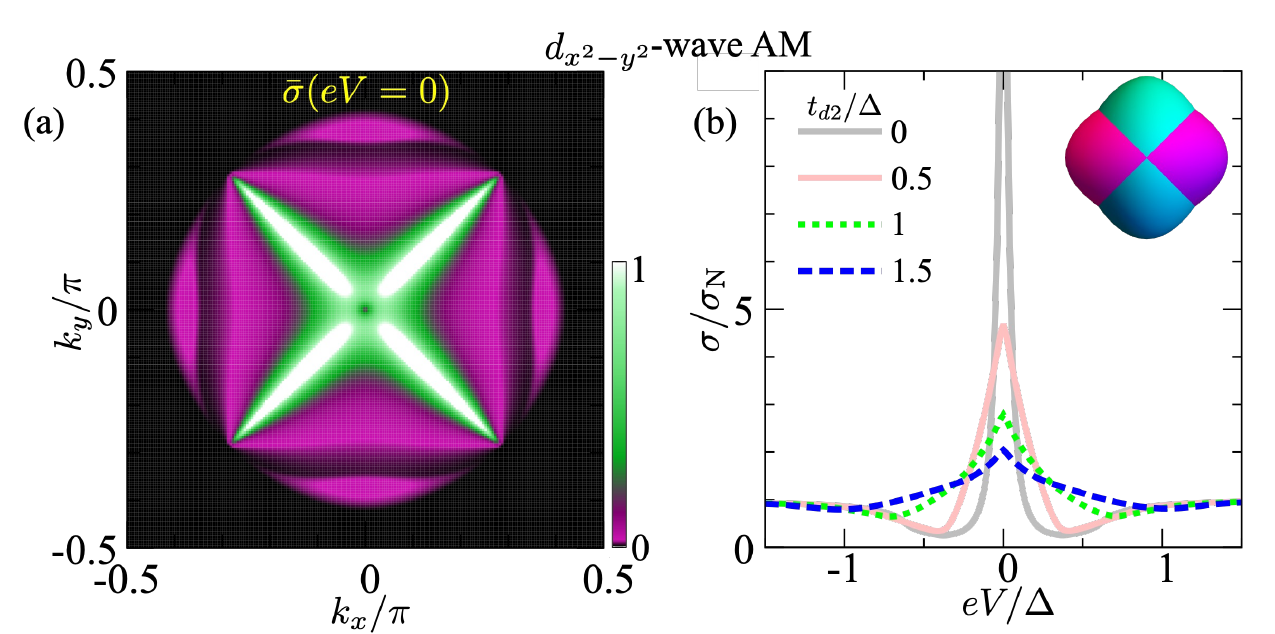}
    \caption{(a) Zero-bias conductance along the $z$-direction [001] for a superconducting AM with $d_{x^2-y^2}$-wave   altermagnetism as a function of   $k_{x,y}$. (b) Normalized total conductance  along the $z$-direction as a function of $eV$ for distinct values of the altermagnetic strength $t_{d2}$; $\sigma_\mathrm{N}$ indicates the conductance in the normal state at $eV=0$. The inset in (b) shows the normal state Fermi surfaces projected onto the [001] surface for up (magenta) and down spins (cyan).  Parameters:  $t_{d2}=\Delta$ (a), $\mu=-4.5t$, $\Delta=0.01t$,   $U_\mathrm{b}=5t$, $\delta=0.01\Delta$.}
    \label{Fig1_EM}
\end{figure}

\begin{figure}[t!]
    \centering
    \includegraphics[width=8.5cm]{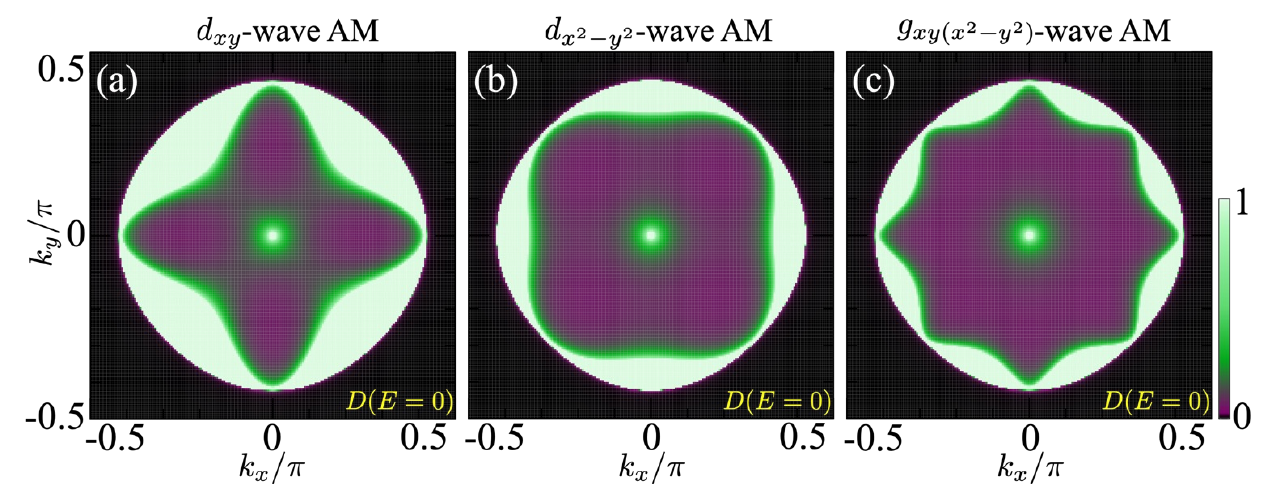}
    \caption{
    (a-c) Zero-energy bulk DOS integrated in $k_z$ as a function of $k_{x,y}$ for a superconducting AM  with $d_{xy}$-  (a),  $d_{x^2-y^2}$-  (b), and $g_{xy(x^2-y^2)}$-wave altermagnetism.
    % (a-c) Projected zero-energy DOS on the   [001] surface as a function of   $k_{x,y}$ for a superconducting AM  with $d_{xy}$-  (a),  $d_{x^2-y^2}$-  (b), and $g_{xy(x^2-y^2)}$-wave altermagnetism. 
    Parameters:  $t_{\alpha}=\Delta$, $\mu=-4.5t$, $\Delta=0.01t$,   $U_\mathrm{b}=5t$, $\delta=0.01\Delta$.}
    \label{Fig2_EM}
\end{figure}

\textit{\textbf{E3.} Conductance along $x$-direction  with $d_{x^{2}-y^{2}}$-wave AMs}.---Conductance along $x$ in a superconducting AM with $d_{x^{2}-y^{2}}$-wave altermagnetism is shown in   Fig.\,\ref{Fig3_EM}(a) and  we observe the formation of   arcs on the [100] surface. Here, the momentum dependent  zero-bias conductance $\bar\sigma(\tilde{\bm{k}},eV=0)$ shows that the arcs are spin-split and disperse with momentum, unlike those due to $d_{xy}$-wave altermagnetism in Fig.\,\ref{Fig3}(a). The region with tiny but nonzero intensities corresponds to BFSs, while even weaker values of $\bar\sigma(\tilde{\bm{k}},eV=0)$ occur  in a gapped region, see S1 in  SM \cite{SM}. Here, we have also verified that $\bar\sigma(\tilde{\bm{k}},eV=0)$ is given by $2\bar{\sigma}_{\rm N}^{0}$,  $\bar{\sigma}_{\rm N}^{1}$, and $\bar{\sigma}_{\rm N}^{2}$ for the surface arcs, BFS, and gapped region; see S2 in   \cite{SM}. 
Moreover, the presence of altermagnetism splits the zero-bias total conductance peak of $\sigma(eV)$ in Fig.\,\ref{Fig3_EM}(b), producing two broad resonances around zero bias as a signature of surface arcs.

\begin{figure}[h!]
    \centering
    \includegraphics[width=8.4cm]{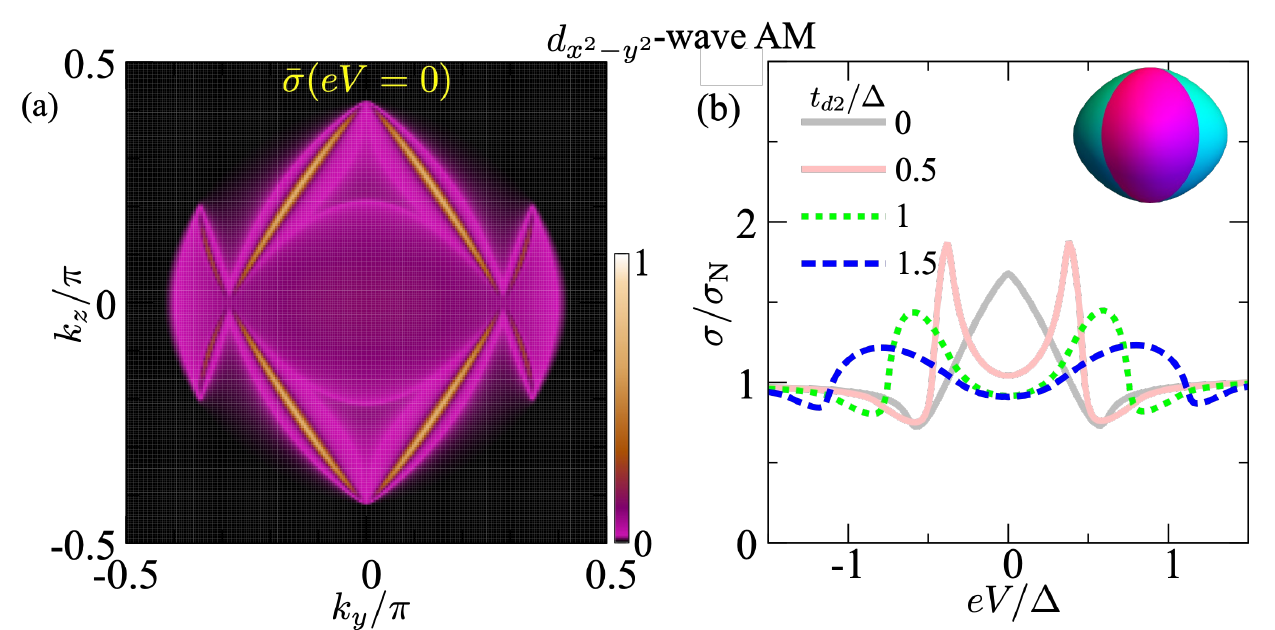}
    \caption{(a) Zero-bias conductance along the $x$-direction [100] for a superconducting AM with $d_{x^2-y^2}$-wave  altermagnetism as a function of   $k_{x,y}$.  (b) Normalized total conductance along the $x$-direction [100] as a function of $eV$ for distinct values of the altermagnetic strength $t_{d2}$. $\sigma_\mathrm{N}$ is the conductance in the normal state at $eV=0$. The inset in (b) shows the normal state  Fermi surfaces projected onto the $x$-direction for up (magenta) and down spins (cyan).  Parameters:  $t_{d2}=\Delta$ (a), $\mu=-4.5t$, $\Delta=0.01t$,   $U_\mathrm{b}=5t$, $\delta=0.01\Delta$.}
    \label{Fig3_EM}
\end{figure}

\begin{figure}[b!]
    \centering
    \includegraphics[width=8.5cm]{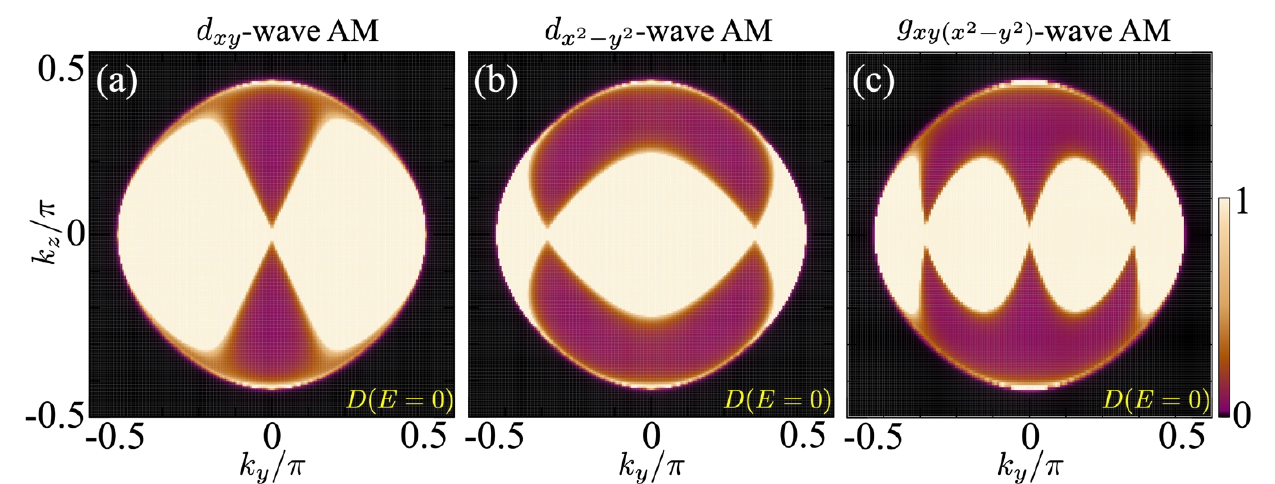}
    \caption{(a-c) Zero-energy bulk DOS integrated in $k_x$ as a function of  $k_{y,z}$ for a superconducting AM  with $d_{xy}$-  (a),  $d_{x^2-y^2}$-  (b), and $g_{xy(x^2-y^2)}$-wave altermagnetism.
    % Projected zero-energy DOS on the  [100] surface as a function of  $k_{y,z}$ for a superconducting AM  with $d_{xy}$-  (a),  $d_{x^2-y^2}$-  (b), and $g_{xy(x^2-y^2)}$-wave altermagnetism. 
    Parameters:  $t_{\alpha}=\Delta$, $\mu=-4.5t$, $\Delta=0.01t$,   $U_\mathrm{b}=5t$, $\delta=0.01\Delta$.}
    \label{Fig4_EM}
\end{figure}

% \textit{\textbf{E4.} Projected density of states on the [100] surface}.---
\textit{\textbf{E4.} Bulk density of states along the [100] direction}.---
For the bulk DOS along the [100] direction using Eq.\,(\ref{DOSEq}), we find that BFSs appear around $k_z=0$ for all the cases of altermagnetism [Fig.\,\ref{Fig4_EM}]. 
Depending on the type of AM, the bulk DOS shows the BFSs coexisting with regions having tiny but nonzero intensities that correspond to the zero-energy surface arcs on the [100] direction [Fig.\,\ref{Fig3}(a,c) and Fig.\,\ref{Fig4_EM}(a)]. We also confirm the BFS signatures in proper 3D AMs, see Sec.\, S2 in the SM~\cite{SM}.

\begin{figure}[t!]
    \centering
    \includegraphics[width=8.3cm]{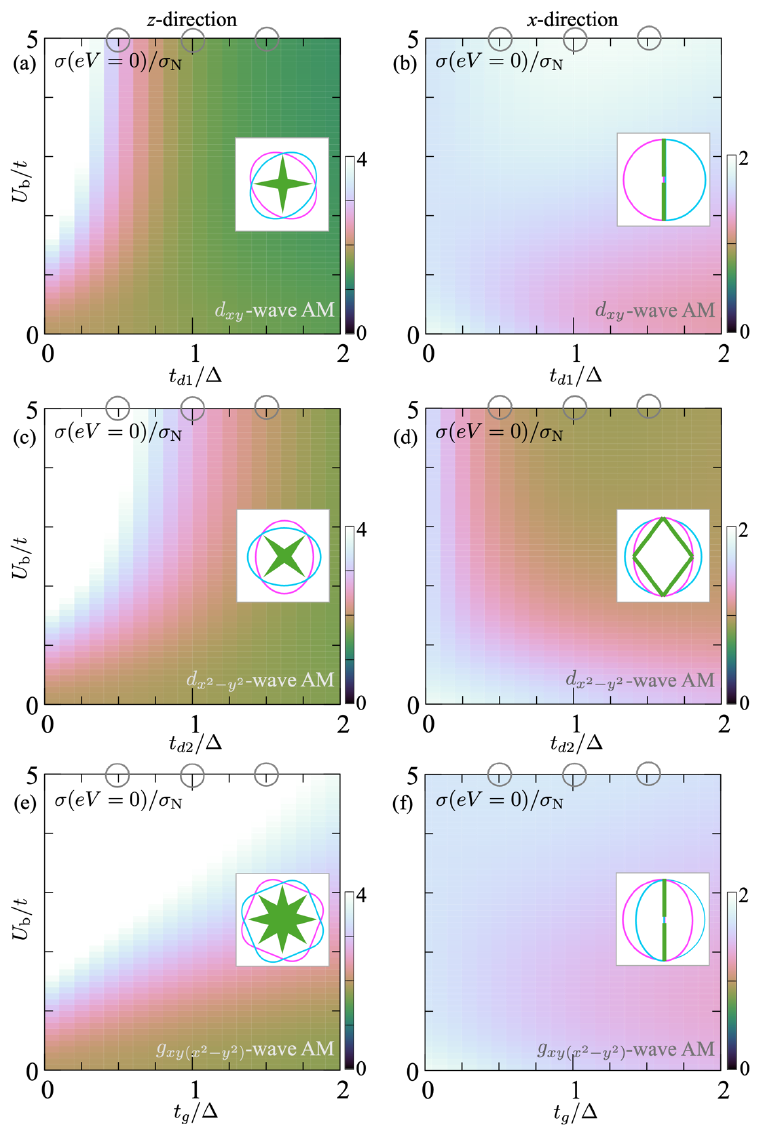}
    \caption{Normalized zero-bias conductance along $z$ (a,c,e) and $x$ (b,d,f) directions as a function of   $U_\mathrm{b}$ and $t_{\alpha}$ for  superconducting AMs with $d_{xy}$-, $d_{x^2-y^2}$-, and $g_{xy(x^2-y^2)}$-wave altermagnetism. Here, $\sigma_\mathrm{N}$ denotes the conductance in the normal state at $eV=0$, and we calculate it for each $t_{\alpha}$ and $U_\mathrm{b}$.  Insets: crossed flat bands and   normal state Fermi surfaces for up (magenta) and down (cyan) spins.  The gray circles mark the values of $U_{b}$ and $t_{\alpha}$ chosen for Fig.~\ref{Fig2} and Fig.~\ref{Fig3}.   Parameters: $\mu=-4.5t$, $\Delta=0.01t$, $\delta=0.01\Delta$. 
    }
    \label{Fig5_EM}
\end{figure}%

\textit{\textbf{E5.} Stability of the ZBCP}.---Having shown the formation of crossed flat bands, surface arcs and BFSs, we now explore    the ZBCP $\sigma(eV=0)$ under variations of the  altermagnetic strength $t_{\alpha}$ and barrier potential $U_\mathrm{b}$. This is shown in Fig.~\ref{Fig5_EM} for transport along $z$ and $x$, in both cases for  $d_{xy}$-, $d_{x^{2}-y^{2}}$-, and $g_{xy(x^{2}-y^{2})}$-wave altermagnetism. For transport along $z$, the largest zero-bias conductance  $\sigma(eV=0)$ is concentrated below $t_{\alpha}<\Delta$ in the case of $d_{xy}$- and $d_{x^{2}-y^{2}}$-wave AMs, see Figs.~\ref{Fig5_EM}(a,c). For $g_{xy(x^{2}-y^{2})}$-wave, the ZBCP remains robust even for larger values of $t_{\alpha}$, see Figs.~\ref{Fig5_EM}(e). These behaviors originate because the crossed flat bands in the respective cases become smaller with the increase of altermagnetic strength $t_{\alpha}$. Despite the reduced  ZBCPs, their values are still peaked at zero bias for $d_{xy}$- and $g_{xy(x^{2}-y^{2})}$-wave AMs,   signalling that the crossed flat bands remain  for $t_{\alpha}>\Delta$, as we indeed show in Fig.\,\ref{Fig2}(b,d). We also note that the crossed flat bands exist even at $U_\mathrm{b}=0$, since the corners of the crossed flat bands are already formed without any barrier. However, a finite $U_\mathrm{b}$ sharpens the conductance, and much larger zero-bias conductance values occur along momenta of the altermagnetic nodal lines. Hence, the visibility of the crossed flat band is enhanced by large $U_\mathrm{b}$, which corresponds to   low transparencies. Along the $x$-direction shown in Figs.~\ref{Fig5_EM}(b,d,f), the value of $\sigma(eV=0)$ takes values ranging from 1 to 2, and it does not change significantly for variations of $U_\mathrm{b}$ and $t_{\alpha}$, even though  the surface arc signatures   partly disappear under BFSs. Also, $\sigma(eV=0)$ becomes smaller as $t_{d2}$ increases because the surface arc states are split by $d_{x^2-y^2}$-wave altermagnetism  [Fig.~\ref{Fig3_EM}]. The stability of the ZBCP discussed here also persists for 3D AMs, see  S3 in  \cite{SM}.

\end{document}

\title{Supplementary material for ``Crossed surface flat bands in three-dimensional superconducting altermagnets''}
	
\author{Yuri Fukaya}
\affiliation{Faculty of Environmental Life, Natural Science and Technology, Okayama University, 700-8530 Okayama, Japan}

\author{Bo Lu}
\affiliation{Department of Physics, Tianjin University, 300354 Tianjin, China}

 \author{Keiji Yada}
 \affiliation{Department of Applied Physics, Nagoya University, 464-8603 Nagoya, Japan}

\author{Yukio Tanaka}
\affiliation{Department of Applied Physics, Nagoya University, 464-8603 Nagoya, Japan}

\author{Jorge Cayao}
\affiliation{Department of Physics and Astronomy, Uppsala University, Box 516, S-751 20 Uppsala, Sweden}

	\maketitle
	%\widetext
	
	%%%%%%%%%% Merge with supplemental materials %%%%%%%%%%
	%\widetext
	%%%%%%%%%% Merge with supplemental materials %%%%%%%%%%
	%%%%%%%%%% Prefix a "S" to all equations, figures, tables and reset the counter %%%%%%%%%%
	\setcounter{equation}{0} \setcounter{figure}{0} \setcounter{table}{0} %
	\setcounter{page}{1} \makeatletter
	\renewcommand{\theequation}{S\arabic{equation}} \renewcommand{\thefigure}{S%
		\arabic{figure}} \renewcommand{\bibnumfmt}[1]{[S#1]} \renewcommand{%
		\citenumfont}[1]{S#1}
	%%%%%%%%%% Prefix a "S" to all equations, figures, tables and reset the counter %%%%%%%%%%

In this supplementary material, we provide additional details of the bulk system studied in the main text, discuss how we calculate conductance, and demonstrate that our main findings remain for a system involving three-dimensional (3D) altermagnetism with 3D chiral $d$-wave superconductivity.

\section{S1: Bulk properties of three-dimensional superconducting altermagnets} \label{Sec_S1}

First of all, we present the details of the model Hamiltonian of 3D superconductors (SCs) with two-dimensional (2D) altermagnets (AMs). Then we will show their corresponding bulk properties.

\subsection{S1.1: Hamiltonian model with a 3D superconductor and 2D altermagnet} \label{Sec_S1_1}

In this subsection, we discuss further details of the model Hamiltonian. 
We start by considering the normal part of the Hamiltonian, which is given by
\begin{align}
    \hat{H}_{\alpha}(\bm{k})=\hat{H}_{0}(\bm{k})+M^{\alpha}_{\bm{k}}\hat{\sigma}_{3}, 
\end{align}
where $\hat{H}_{0}(\bm{k})$ is the kinetic term and $M^{\alpha}_{\bm{k}}$ characterizes altermagnetism.
More precisely, the kinetic part is given by
\begin{align}
    \hat{H}_{0}(\bm{k})&=\varepsilon(\bm{k})\hat{\sigma}_{0},\\
    \varepsilon(\bm{k})&=-\mu-2t\cos{k_x}-2t\cos{k_y}-2t\cos{k_z},
\end{align}%
where $\bm{k}=(k_x,k_y,k_z)$ is the momentum, $\hat{\sigma}_{0,1,2,3}$ is the Pauli matrices in spin space, $\mu=-4.5t$ is the chemical potential, $t$ is the hopping integral along the $(x,y)$ and $z$-directions. 
We consider the $d$ and $g$-wave altermagnetism modelled by
\begin{align}
    M^{d1}_{\bm{k}}&=2t_{d1}\sin{k_x}\sin{k_y},\label{2D_AM1}\\
    M^{d2}_{\bm{k}}&=t_{d2}(\cos{k_x}-\cos{k_y}),\label{2D_AM2}\\
    M^{g}_{\bm{k}}&=4t_{g}\sin{k_x}\sin{k_y}(\cos{k_x}-\cos{k_y})\label{2D_AM3},
\end{align}%
where $t_{d1}$, $t_{d2}$, and $t_{g}$ are the strength of $d_{xy}$, $d_{x^2-y^2}$ and $g_{xy(x^2-y^2)}$-wave AMs, respectively.

With superconductivity, the system is described by the Bogoliubov--de Gennes (BdG) Hamiltonian:
\begin{align}
    \hat{H}^{\alpha}_\mathrm{BdG}(\bm{k})&=\varepsilon(\bm{k})\hat{\sigma}_{0}\hat{\tau}_{3}+\hat{H}^{\alpha}_\mathrm{AM}(\bm{k})+\hat{H}_\mathrm{SC}(\bm{k}),\\
    \hat{H}^{\alpha}_\mathrm{AM}(\bm{k})&=M^{\alpha}_{\bm{k}}\hat{\sigma}_{3}\hat{\tau}_{3}, \\
    \hat{H}_\mathrm{SC}(\bm{k})&=-\Delta\sin{k_z}[\sin{k_x}\hat\tau_2+\sin{k_y}\hat\tau_1]\hat{\sigma}_{2},
\end{align}%
with $\hat{\tau}_{0,1,2,3}$ the Pauli matrices in Nambu space and $\Delta=0.01t$ the pair amplitude of the chiral $d$-wave state.
% Here, $\hat{\Delta}(\bm{k})$ is the spin-singlet pair potential given by
% \begin{align}
%     \hat{\Delta}(\bm{k})=\psi(\bm{k})i\hat{\sigma}_{2},
% \end{align}%
% with the spin-singlet order parameter $\psi(\bm{k})$.
% In this study, we focus on the three-dimensional (3D) chiral $d$-wave pairing:
% \begin{align}
%     \psi(\bm{k})=\Delta\sin{k_z}[\sin{k_x}+i\sin{k_y}],
% \end{align}%
% with the energy gap amplitude $\Delta=0.01t$.
The eigenvalues of $\hat{H}^{\alpha}_\mathrm{BdG}(\bm{k})$ are obtained as: 
\begin{align}
    E^{s,\alpha}_{\bm{k},\pm}(\bm{k})&=sM^{\alpha}_{\bm{k}}\pm\sqrt{\varepsilon^2(\bm{k})+|\bar\Delta(\bm{k})|^2},\label{Ebdg}
\end{align}%
with $s=\pm$ and $\bar\Delta(\bm{k})=\Delta\sin{k_z}[\sin{k_x}+i\sin{k_y}]$.
% \begin{align}
%     E_{1}(\bm{k})&=M^{\alpha}_{\bm{k}}+\sqrt{\varepsilon^2(\bm{k})+|\psi(\bm{k})|^2},\\
%     E_{2}(\bm{k})&=M^{\alpha}_{\bm{k}}-\sqrt{\varepsilon^2(\bm{k})+|\psi(\bm{k})|^2},\\
%     E_{3}(\bm{k})&=-M^{\alpha}_{\bm{k}}+\sqrt{\varepsilon^2(\bm{k})+|\psi(\bm{k})|^2},\\
%     E_{4}(\bm{k})&=-M^{\alpha}_{\bm{k}}-\sqrt{\varepsilon^2(\bm{k})+|\psi(\bm{k})|^2}.
% \end{align}%
Eq.\ (\ref{Ebdg}) is presented in the main text as Eq.\ (3) and helps us identify the formation of Bogoliubov-Fermi surfaces (BFSs).

\subsection{S1.2: Energy gap equation for the three-dimensional chiral $d$-wave pairing in the presence of the altermagnetism}

In this subsection, we discuss the generation of 3D chiral $d$-wave pairing and comparison with spin-triplet $f$-wave pairings~\cite{UedaSigristReview} in the presence of altermagnetism in a simplified model.
In order to express spin-singlet 3D chiral $d$-wave and spin-triplet $f$-wave pairings in the lattice model, we assume the effective attractive interaction in the 3D extended Hubbard model. 
In the following, we obtain the coexisting phase of altermagnetism and 3D chiral $d$-wave pairing.   
We choose the extended Hubbard model with an attractive interaction for the second nearest-neighbor, as in Ref.\ \cite{tsuchiuraJPSJ1995}, where the coexistence of spin-singlet $d_{x^2-y^2}$-wave and $s$-wave pairings was studied.   
Within mean-field approximation, the gap equation in 3D systems is given by~\cite{UedaSigristReview}:
\begin{align}
    \bar\Delta_\mathrm{S}(\bm{k})={k_\mathrm{B}T}\int^{\pi}_{-\pi}\int^{\pi}_{-\pi}\int^{\pi}_{-\pi}{dq_{x}dq_{y}dq_{z}}W(\bm{k}-\bm{q})\sum_{\omega_n}\tilde{F}_\mathrm{S}(\bm{q},i\omega_{n}), \label{gapeq_singlet}
\end{align}%
for spin-singlet 3D chiral $d$-wave,
\begin{align}
    \bar\Delta_\mathrm{T}(\bm{k})={k_\mathrm{B}T}\int^{\pi}_{-\pi}\int^{\pi}_{-\pi}\int^{\pi}_{-\pi}{dq_{x}dq_{y}dq_{z}}W(\bm{k}-\bm{q})\sum_{\omega_n}\tilde{F}_\mathrm{T}(\bm{q},i\omega_{n}),
\end{align}%
for spin-triplet $f_{z(x^2-y^2)}$-wave ($S_z=0$), and
\begin{align}
    \bar\Delta_{\uparrow\uparrow}(\bm{k})&=2{k_\mathrm{B}T}\int^{\pi}_{-\pi}\int^{\pi}_{-\pi}\int^{\pi}_{-\pi}{dq_{x}dq_{y}dq_{z}}W(\bm{k}-\bm{q})\sum_{\omega_n}\tilde{F}^{\uparrow\uparrow}(\bm{q},i\omega_{n}),\label{Gapeq_triplet}\\
    \bar\Delta_{\downarrow\downarrow}(\bm{k})&=2{k_\mathrm{B}T}\int^{\pi}_{-\pi}\int^{\pi}_{-\pi}\int^{\pi}_{-\pi}{dq_{x}dq_{y}dq_{z}}W(\bm{k}-\bm{q})\sum_{\omega_n}\tilde{F}^{\downarrow\downarrow}(\bm{q},i\omega_{n}),
\end{align}%
for spin-triplet $f_{x(z^2-y^2)}$ and $f_{y(x^2-z^2)}$-wave ($S_z=\pm1$) pairings, with $\bm{k}=(k_x,k_y,k_z)$ and $\bm{q}=(q_x,q_y,q_z)$. Here, $W(\bm{k}-\bm{q})$ describes the attractive interaction, while $\tilde{F}_\mathrm{S,T}(\bm{q},i\omega_{n})$ and $\tilde{F}^{\sigma\sigma'}(\bm{q},i\omega_{n})$ are the anomalous electron-hole Green's function, which depends on the type of superconductor. Moreover,  
$\omega_{n}=(2n+1)\pi k_\mathrm{B}T$ for the integer $n$ denotes the fermionic Matsubara frequency.

In the spin-singlet (S) 3D chiral $d$-wave pairing model, the anomalous  Green's functions $\tilde{F}_\mathrm{S}(\bm{q},i\omega_n)$ are given by
\begin{align}
    \tilde{F}_\mathrm{S}(\bm{q},i\omega_n)&=\sum_{\omega_{n}}[\bar{F}_\mathrm{S}^{\uparrow\downarrow}(\bm{q},i\omega_{n})-\bar{F}_\mathrm{S}^{\downarrow\uparrow}(\bm{q},i\omega_{n})],\\
    \bar{F}_\mathrm{S}^{\uparrow\downarrow}(\bm{k},i\omega_{n})&=\frac{\bar\Delta_\mathrm{S}(\bm{k})}{[\omega_{n}+iM^{\alpha}_{\bm{k}}]^{2}+\xi^{2}(\bm{k})+|\bar\Delta_\mathrm{S}(\bm{k})|^{2}},\\
    \bar{F}_\mathrm{S}^{\downarrow\uparrow}(\bm{k},i\omega_{n})&=-\frac{\bar\Delta_\mathrm{S}(\bm{k})}{[\omega_{n}-iM^{\alpha}_{\bm{k}}]^{2}+\xi^{2}(\bm{k})+|\bar\Delta_\mathrm{S}(\bm{k})|^{2}},
\end{align}%
where
\begin{align}
    \xi(\bm{k})&=-\mu-2t\cos{k_x}-2t\cos{k_y}-2t\cos{k_z},\\
    \bar\Delta_\mathrm{S}(\bm{k})&=\Delta\sin{k_z}[\sin{k_x}+i\sin{k_y}],\\
        M^{d1}_{\bm{k}}&=2t_{d1}\sin{k_x}\sin{k_y},\\
    M^{d2}_{\bm{k}}&=t_{d2}(\cos{k_x}-\cos{k_y}),\\
    M^{g}_{\bm{k}}&=4t_{g}\sin{k_x}\sin{k_y}(\cos{k_x}-\cos{k_y}).
\end{align}%
Here,  $\mu$, $t$, $\Delta$, and $t_{\alpha}$ for $\alpha=s,d1,d2,g$ denote the Matsubara frequency, chemical potential, hopping integral, the magnitude of the gap function, and the $d_{xy}$, $d_{x^2-y^2}$, and $g$-wave altermagnetic order, respectively. 

Moreover, the attractive interaction $W(\bm{k})$ is  modelled as
\begin{align}
    W(\bm{k})&=2W[\cos({k_x+k_z})+\cos({-k_x+k_z})+\cos({k_y+k_z})+\cos({-k_y+k_z})\notag\\
    &+\cos({k_x+k_y})+\cos({-k_x+k_y})], \label{Wk}
\end{align}%
where $W>0$ is the strength of the second-nearest neighbor attractive interaction.
This type of interaction is motivated by the second-nearest neighbor attractive channel for spin-singlet 3D chiral $d$-wave pairing.
In this second-nearest neighbor attractive channel, a higher-angular momentum pairing is more favorable, and for spin-triplet pairing, the $p$-wave can not be adopted in Eq.\ (\ref{Wk}) because $p$-wave pairing requires the nearest neighbor attractive channel~\cite{UedaSigristReview}. 
For this reason, we compare spin-singlet 3D chiral $d$-wave with the spin-triplet $f$-wave pairings: $f_{z(x^2-y^2)}$, $f_{x(z^2-y^2)}$, and $f_{y(x^2-z^2)}$-wave pairings~\cite{UedaSigristReview}.
As we show below, there exists a competition between spin-singlet chiral $d$-wave pairing and spin-triplet $f$-wave pairings for the second-nearest neighbor attractive interaction.
For spin-triplet $f_{z(x^2-y^2)}$-wave superconducting pairing, the anomalous Green's functions are given by
\begin{align}
    \tilde{F}_\mathrm{T}(\bm{q},i\omega_n)&=\sum_{\omega_{n}}[\bar{F}_\mathrm{T}^{\uparrow\downarrow}(\bm{q},i\omega_{n})+\bar{F}_\mathrm{T}^{\downarrow\uparrow}(\bm{q},i\omega_{n})],\\
    \bar{F}_\mathrm{T}^{\uparrow\downarrow}(\bm{k},i\omega_{n})&=\frac{\bar\Delta_\mathrm{T}(\bm{k})}{[\omega_{n}+iM^{\alpha}_{\bm{k}}]^{2}+\xi^{2}(\bm{k})+|\bar\Delta_\mathrm{T}(\bm{k})|^{2}},\\
    \bar{F}_\mathrm{T}^{\downarrow\uparrow}(\bm{k},i\omega_{n})&=\frac{\bar\Delta_\mathrm{T}(\bm{k})}{[\omega_{n}-iM^{\alpha}_{\bm{k}}]^{2}+\xi^{2}(\bm{k})+|\bar\Delta_\mathrm{T}(\bm{k})|^{2}},\\
    \bar\Delta_\mathrm{T}(\bm{k})&=\Delta\sin{k_z}(\cos{k_x}-\cos{k_y}),
\end{align}%
while for   spin-triplet $f_{x(z^2-y^2)}$ and $f_{y(x^2-z^2)}$-wave pairings by
\begin{align}
    \tilde{F}^{\uparrow\uparrow}(\bm{q},i\omega_n)&=\frac{\bar\Delta^{\uparrow\uparrow}(\bm{k})}{\omega^2_{n}+[\xi(\bm{k})+M^{\alpha}_{\bm{k}}]^{2}+|\bar\Delta^{\uparrow\uparrow}(\bm{k})|^{2}},\\
    \tilde{F}^{\downarrow\downarrow}(\bm{k},i\omega_{n})&=\frac{\bar\Delta^{\downarrow\downarrow}(\bm{k})}{\omega^2_{n}+[\xi(\bm{k})-M^{\alpha}_{\bm{k}}]^{2}+|\bar\Delta^{\downarrow\downarrow}(\bm{k})|^{2}},
\end{align}%
\begin{align}
    \bar\Delta_\mathrm{\uparrow\uparrow}(\bm{k})&=-d_{x}(\bm{k})+id_{y}(\bm{k}),\\
    \bar\Delta_\mathrm{\downarrow\downarrow}(\bm{k})&=d_{x}(\bm{k})+id_{y}(\bm{k}),\\
    d_{x}(\bm{k})&=\Delta\sin{k_x}(\cos{k_z}-\cos{k_y}),\\
    d_{y}(\bm{k})&=\Delta\sin{k_y}(\cos{k_x}-\cos{k_z}).
\end{align}%

At this stage, we point out that we would like to assess what type of superconductivity is possible under altermagnetism, with the properties discussed above. 
The idea is to find $\Delta$, and hopefully the critical temperature, for each type of superconductor as a function of the altermagnetic strength. For this purpose, we first put in an appropriate form the expressions for $\bar{\Delta}_{\rm S}$, $\bar{\Delta}_{\rm T}$, and $\bar{\Delta}_{\uparrow\uparrow}$, taking into account the above equations. 
Thus, to determine $\Delta$~\cite{MicnasReview} for spin-singlet 3D chiral $d$-wave pairing, we multiply $\sin{k_x}\sin{k_z}$ on both sides of Eq.\ (\ref{gapeq_singlet}) and we obtain:
\begin{align}
    \bar\Delta_\mathrm{S}(\bm{k}) \sin{k_x}\sin{k_z}={k_\mathrm{B}T}\int^{\pi}_{-\pi}\int^{\pi}_{-\pi}\int^{\pi}_{-\pi}{dq_{x}dq_{y}dq_{z}}W(\bm{k}-\bm{q})\sum_{\omega_n}\tilde{F}_\mathrm{S}(\bm{q},i\omega_{n}) \sin{k_x}\sin{k_z}.
\end{align}%
Hence, from this manipulation and integral over $\bm{k}$, we obtain
\begin{align}
    \Delta={4k_\mathrm{B}T W}\int^{\pi}_{-\pi}\int^{\pi}_{-\pi}\int^{\pi}_{-\pi}{dq_{x}dq_{y}dq_{z}}\tilde{F}_\mathrm{S}(\bm{q},i\omega_{n})\sin{q_x}\sin{q_z}.
\end{align}%
Because $d_{yz}$-wave pairing is degenerate with $d_{zx}$-wave, we can only focus on the $d_{zx}$-wave sector, and we can solve the gap equation:
\begin{align}
    1&={4k_\mathrm{B}TW}\int^{\pi}_{-\pi}\int^{\pi}_{-\pi}\int^{\pi}_{-\pi}{dq_{x}dq_{y}dq_{z}}\\
    &\times\sum_{\omega_{n}}\left[\frac{\sin{q_x}\sin{q_z}}{[\omega_n+iM^{\alpha}_{\bm{q}}]^2+\xi^{2}(\bm{q})+|\bar\Delta_\mathrm{S}(\bm{q})|^2}+\frac{\sin{q_x}\sin{q_z}}{[\omega_n-iM^{\alpha}_{\bm{q}}]^2+\xi^{2}(\bm{q})+|\bar\Delta_\mathrm{S}(\bm{q})|^2}\right]\sin{q_x}\sin{q_z}.\notag
\end{align}%
This gap equation allows us to estimate the critical temperature as a function of the altermagnetic order.

In a similar manner as for the spin-singlet component above, for spin-triplet $f_{x(z^2-y^2)}$ and $f_{y(x^2-z^2)}$-wave pairings, we multiply $\sin{k_x}(\cos{k_z}-\cos{k_y})$ on the both sides of Eq.\ (\ref{Gapeq_triplet}):
\begin{align}
    &\bar\Delta_{\uparrow\uparrow}(\bm{k}) \sin{k_x}(\cos{k_z}-\cos{k_y})\\
    &=2{k_\mathrm{B}T}\int^{\pi}_{-\pi}\int^{\pi}_{-\pi}\int^{\pi}_{-\pi}{dq_{x}dq_{y}dq_{z}}W(\bm{k}-\bm{q})\sum_{\omega_n}\tilde{F}^{\uparrow\uparrow}(\bm{q},i\omega_{n}) \sin{k_x}(\cos{k_z}-\cos{k_y}),\notag
\end{align}%
and, performing an integral over $\bm{k}$, obtain
\begin{align}
    1&={4k_\mathrm{B}TW}\int^{\pi}_{-\pi}\int^{\pi}_{-\pi}\int^{\pi}_{-\pi}{dq_{x}dq_{y}dq_{z}}\times\sum_{\omega_{n}}\left[\frac{\sin{q_x}(\cos{q_z}-\cos{k_y})}{\omega^2_{n}+[\xi(\bm{q})+M^{\alpha}_{\bm{q}}]^{2}+|\bar\Delta^{\uparrow\uparrow}(\bm{q})|^{2}}\right]\sin{q_x}(\cos{q_z}-\cos{q_y}). \label{gap_eq_triplet_x}
\end{align}%
It is evident that, in Eq.\ (\ref{gap_eq_triplet_x}), the critical temperature $T_\mathrm{c}$ for $f_{y(x^2-z^2)}$-wave pairing is degenerate with that for $f_{x(z^2-y^2)}$-wave pairing.
\begin{figure}[t!]
    \centering
    \includegraphics[width=13cm]{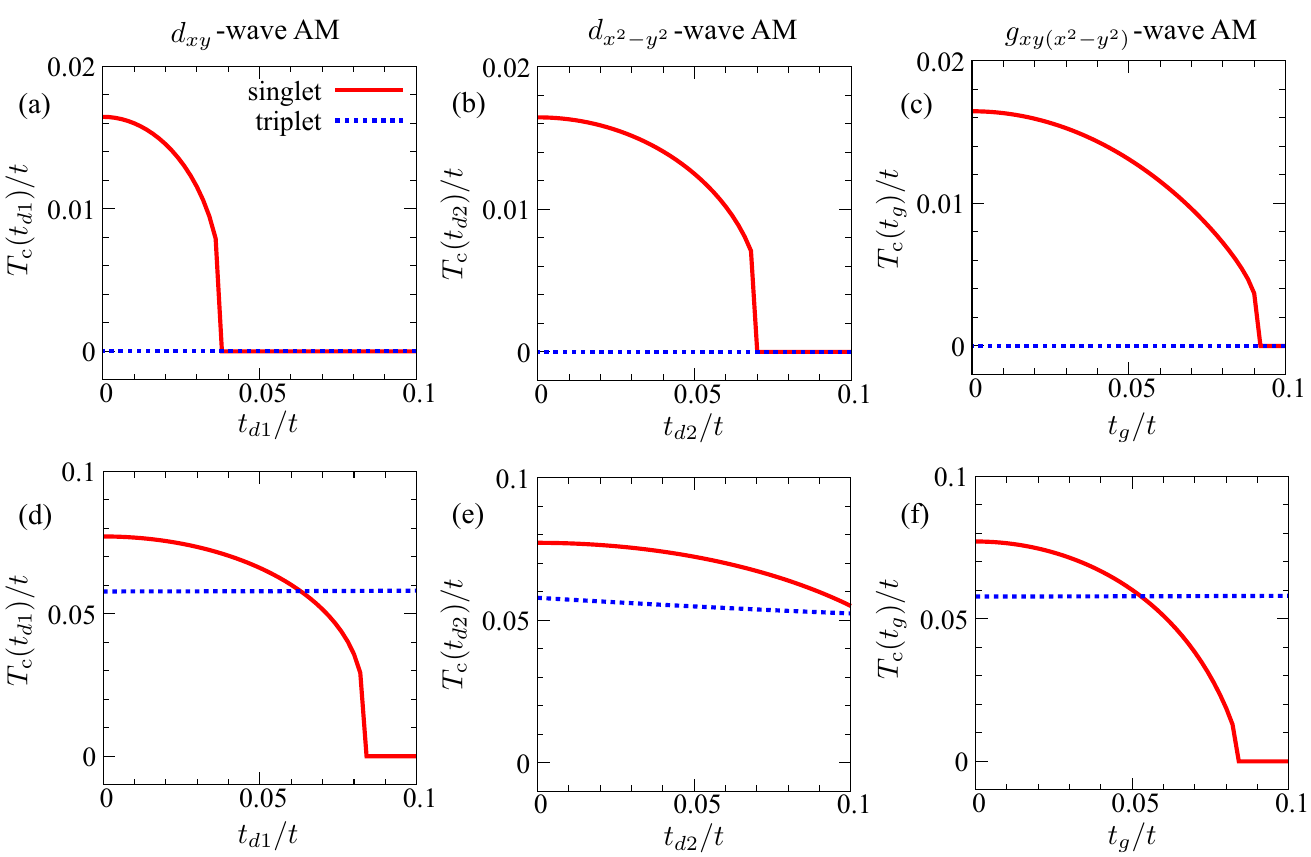}
    \caption{Critical temperature obtained from the gap equations as a function of the distinct altermagnetic strengths $t_{\alpha}$ ($\alpha=d1,d2,g$) for the spin-triplet $f_{x(z^2-y^2)}$-wave (blue-dotted curves) and spin-singlet 3D chiral $d$-wave pairings (red-solid curves). 
    Top row corresponds to chemical potential $\mu=-4.5t$ and interaction strength $W=2.5t$, while bottom row to $\mu=-2.5t$ and  $W=t$.}
    \label{gap_eq}
\end{figure}

We can now determine the critical temperatures $T=T_\mathrm{c}$ for each $t_{\alpha}$ by $\Delta\rightarrow0$, and compare those for the spin-singlet 3D chiral $d$-wave and the spin-triplet $f_{x(z^2-y^2)}$-wave ($S_z=\pm 1$) pairings.
In Fig.~\ref{gap_eq}, we show the critical temperatures $T_\mathrm{c}$ as a function of the altermagnetic strength $t_{\alpha}$ for each altermagnet at $\mu=-4.5t$ and $W=2.5t$ in Figs.~\ref{gap_eq}(a-c), while at $\mu=-2t$ and $W=t$ in Figs.~\ref{gap_eq}(d-f). 
Figs.~\ref{gap_eq}(a-c) correspond to $d_{xy}$-, $d_{x^2-y^2}$-, and $g_{xy(x^2-y^2)}$-wave altermagnetism; the same for Figs.~\ref{gap_eq}(d-f).
For $\mu=-4.5t$ and $W=2.5t$ shown in Figs.~\ref{gap_eq}(a-c), without altermagnetism $t_{\alpha}=0$, we find the solution of the gap equation for spin-singlet chiral $d$-wave pairing, and then spin-singlet chiral $d$-wave pairing is realized. Here, the critical temperature for spin-singlet chiral $d$-wave pairing is finite and decreases as the altermagnetic strengths are increased, see red curves in Figs.~\ref{gap_eq}(a-c); this behavior occurs for all studied altermagnets, but the values of the altermagnetic field at which the critical temperature vanishes are different in each case. Interestingly, for the spin-triplet $f_{x(z^2-y^2)}$-wave pairing, we do not obtain the solution of the gap equation, giving a vanishing critical temperature for  $f_{x(z^2-y^2)}$-wave pairing, see blue curves-dashed in Figs.~\ref{gap_eq} (a-c). This effect seems to be an expected property of low fillings, as shown in Ref.~\cite{MicnasReview}, see more below.
Thus, finite values of critical temperature are obtained for the spin-singlet 3D chiral $d$-wave pairing for the altermagnetic strengths with the following conditions: $t_{d1}<0.038t$, $t_{d2}<0.068t$, and $t_{g}<0.092t$; within these ranges,  the spin-singlet 3D chiral $d$-wave pairing is favorable because spin-triplet $f_{x(z^2-y^2)}$-wave pairing can not overcome spin-singlet chiral $d$-wave pairing without conventional magnetism.

To further understand the favourable superconducting pairing, critical temperatures in Fig.1(d-f)  reveal that lower fillings enable their finite values in the case of spin-triplet $f_{x(z^2-y^2)}$-wave pairing. 
Nevertheless, the spin-singlet 3D chiral $d$-wave pairing still shows a dominant behavior below a certain value of altermagnetism. 
At this point, we remind that it is known that the pairing with higher critical temperatures $T_\mathrm{c}$ is favorable in actual systems because it allows for larger order parameters and smaller free energies~\cite{MicnasReview,UedaSigristReview,Ketterson_Song_super,annett2004superconductivity}. 
In consequence, the spin-singlet chiral $d$-wave pairing considered in our work can be still seen as a dominant superconductivity even at low fillings.  
Hence, although $T_\mathrm{c}$ of spin-triplet pairing strongly depends on the chemical potential $\mu$, spin-singlet 3D chiral $d$-wave superconductivity can generally be favorable and overcome the spin-triplet pairing in 3D superconducting AMs in the presence of the second-nearest neighbor attractive interaction.

In conclusion, by selecting a chemical potential of $\mu = -4.5t$ and an altermagnetic strength $t_{\alpha}$ that self-consistently satisfy the gap equation, our model specifically describes a regime where spin-singlet chiral $d$-wave pairing remains dominant despite the altermagnetically induced spin-splitting of the Fermi surfaces. These parameters are explicitly used in the main text, implying that our main findings are physically possible.

\subsection{S1.3: Structure of quasiparticle energy dispersions in the bulk} \label{Sec_S1_2}

\begin{figure}[t!]
    \centering
    \includegraphics[width=17.5cm]{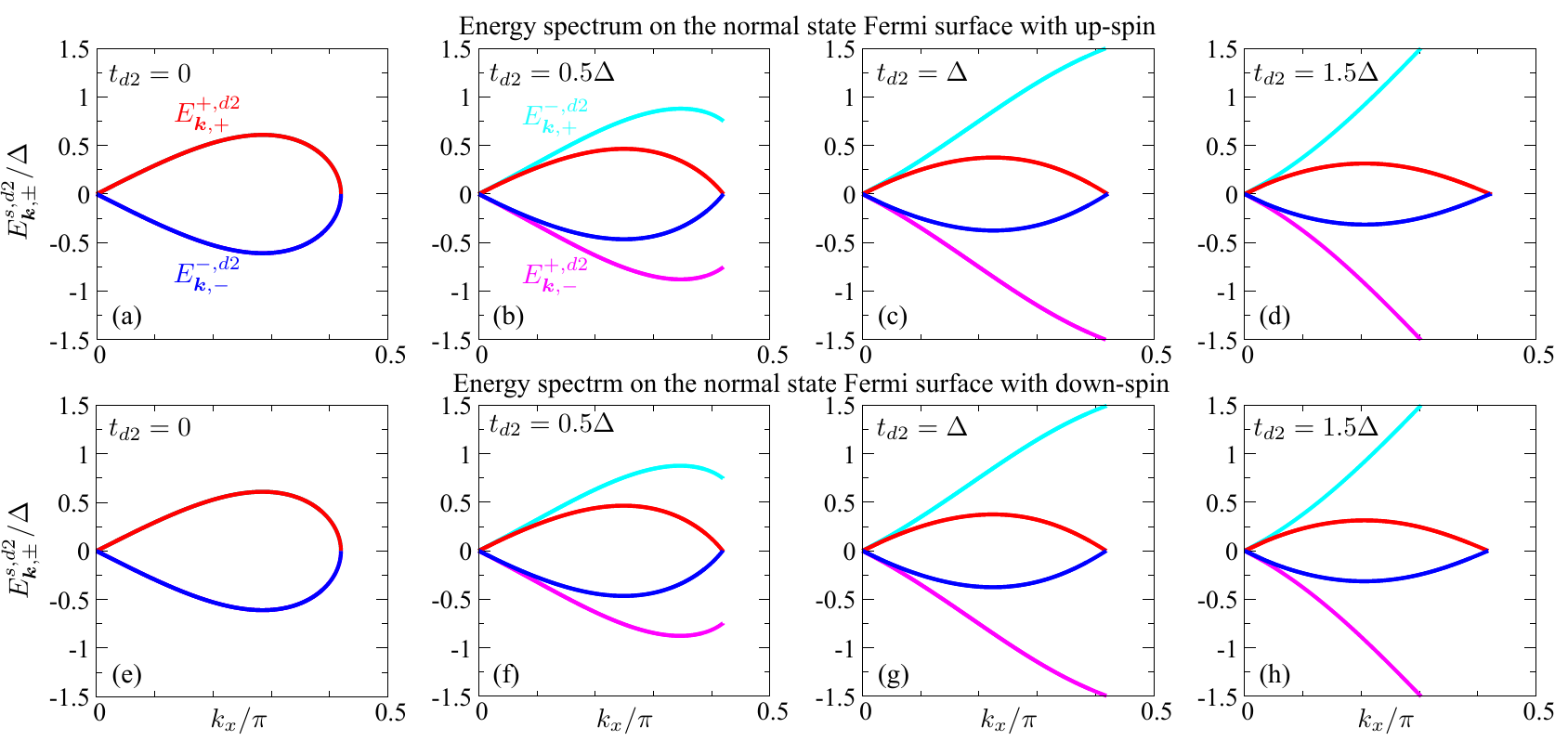}
    \caption{$E^{s,d2}_{\bm{k},\pm}$ for the $d_{x^2-y^2}$-wave altermagnet (AM) as a function of $k_x$ at $k_y=0$ scanning on the Fermi surface with (a-d) up and (e-g) down spins.
    We choose the altermagnetic order as (a,d) $t_{d2}=0$, (b,e) $t_{d2}=0.5\Delta$, (c,f) $t_{d2}=\Delta$, and (d,g) $t_{d2}=1.5\Delta$.
    In the normal state, at $k_y=0$, $k_z$ is determined by the normal state energy dispersion $\varepsilon(\bm{k})\pm M^{\alpha}_{\bm{k}}=0$ for each $k_x$.
    For each $(k_x,k_z)$, we plot $E^{s,d2}_{\bm{k},\pm}$ as a function of $k_x$ at $k_y=0$. 
    In (a,e), we obtain $E^{+,d2}_{\bm{k},+}=E^{-,d2}_{\bm{k},+}$ and $E^{+,d2}_{\bm{k},-}=E^{-,d2}_{\bm{k},-}$ for each $k_x$, and then the shape of zero-energy flat bands becomes a disk.
    In the presence of the altermagnetism, if $E^{+,d2}_{\bm{k},+}\sim E^{-,d2}_{\bm{k},+}$ and $E^{+,d2}_{\bm{k},-}\sim E^{-,d2}_{\bm{k},-}$, zero-energy flat bands remain, and the resulting zero-energy flat bands becomes crossed.
    We select the parameter as: $\mu=-4.5t$ and $\Delta=0.01t$. }
    \label{fig_sm:Ek_vs_kx_FS}
\end{figure}%
\begin{figure}[t!]
    \centering
    \includegraphics[width=17.5cm]{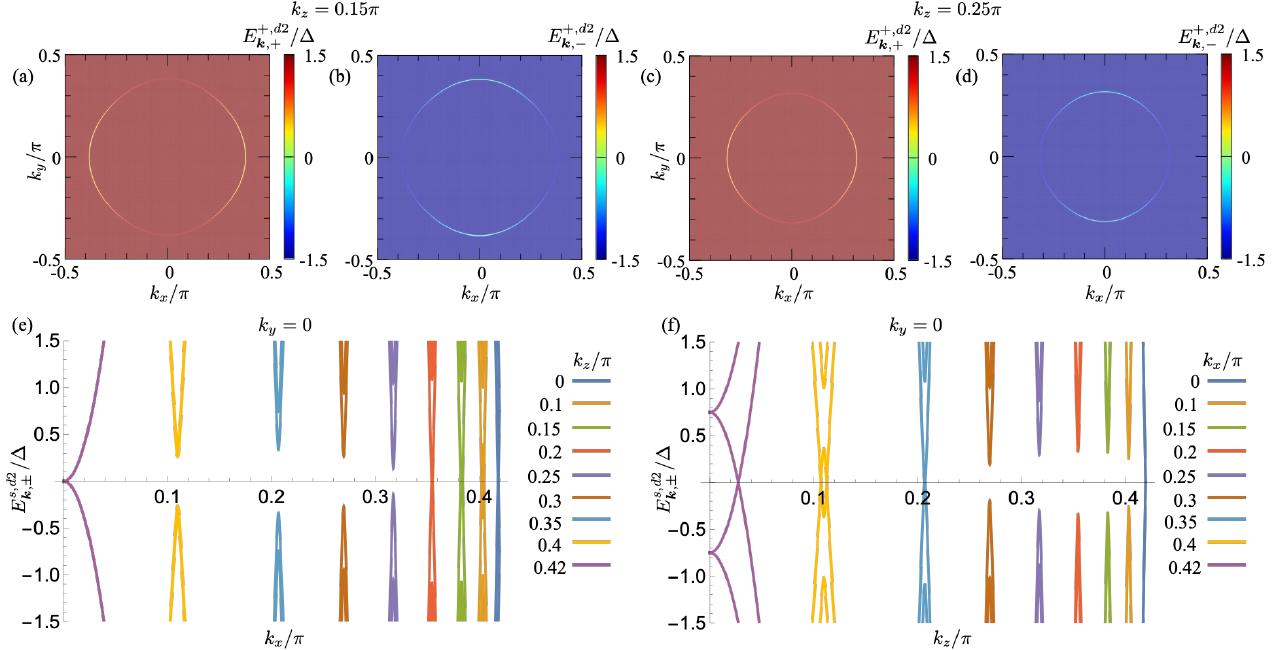}
    \caption{(a,c) $E^{+,d2}_{\bm{k},+}=-E^{-,d2}_{\bm{k},-}$ and (b,d) $E^{+,d2}_{\bm{k},-}=-E^{-,d2}_{\bm{k},+}$ for $(k_x,k_y)$ space for the $d_{x^2-y^2}$-wave AM at (a,b) $k_z=0.15\pi$ and (c,d) $0.25\pi$.
    (e) $E^{s,d2}_{\bm{k},\pm}$ for the $d_{x^2-y^2}$-wave AM as a function of $k_x$ at $k_y=0$ by changing $k_z$.
    (f) $E^{s,d2}_{\bm{k},\pm}$ for the $d_{x^2-y^2}$-wave AM as a function of $k_z$ at $k_y=0$ by changing $k_x$.
    We select the parameters as $\mu=-4.5t$, $\Delta=0.01t$, and $t_{d2}=\Delta$.}
    \label{fig_sm:Ek_vs_k}
\end{figure}%
In the following, we show how to determine the width of crossed flat bands.
We start with the quasiparticle energy dispersion $E^{s,d2}_{\bm{k},\pm}$ as a function of $k_x$ at $k_y=0$ by scanning on the Fermi surface with up [Fig.~\ref{fig_sm:Ek_vs_kx_FS} (a-d)] and down spins [Fig.~\ref{fig_sm:Ek_vs_kx_FS} (e-g)], for the $d_{x^2-y^2}$-wave AM.
Then $k_z$ is determined by the normal state energy dispersion $\varepsilon(\bm{k})\pm M^{d2}_{\bm{k}}=0$ for each $k_x$ at $k_y=0$.
Because the spin splitting of the Fermi surface is small, the results for up and down spins are almost the same.
Without altermagnetism ($M^{\alpha}_{\bm{k}}=0$), we obtain $E^{+,d2}_{\bm{k},+}=E^{-,d2}_{\bm{k},+}$ and $E^{+,d2}_{\bm{k},-}=E^{-,d2}_{\bm{k},-}$, that is, $E^{+,d2}_{\bm{k},+}-E^{-,d2}_{\bm{k},+}=E^{+,d2}_{\bm{k},-}-E^{-,d2}_{\bm{k},-}= 2M^{\alpha}_{\bm{k}}= 0$ for each momentum, and then we also obtain a disk of zero-energy flat bands.
In the presence of altermagnetism ($M^{\alpha}_{\bm{k}}\ne0$), we generally get $E^{+,d2}_{\bm{k},+}\ne E^{-,d2}_{\bm{k},+}$ and $E^{+,d2}_{\bm{k},-}\ne E^{-,d2}_{\bm{k},-}$.
Nevertheless, the relations of $E^{+,d2}_{\bm{k},+}\sim E^{-,d2}_{\bm{k},+}$ and $E^{+,d2}_{\bm{k},-}\sim E^{-,d2}_{\bm{k},-}$, that is, $E^{+,d2}_{\bm{k},+}-E^{-,d2}_{\bm{k},+}=E^{+,d2}_{\bm{k},-}-E^{-,d2}_{\bm{k},-}= 2M^{\alpha}_{\bm{k}}\sim0$ can be satisfied, and then the zero-energy surface Andreev bound states remain.
Along the $k_z$-direction, although the dispersions of $E^{+,d2}_{\bm{k},+}=E^{-,d2}_{\bm{k},+}$ and $E^{+,d2}_{\bm{k},-}=E^{-,d2}_{\bm{k},-}$ for each $(k_x,k_y)$ are not exactly satisfied, we can obtain their approximate expression at small value of $M^{\alpha}_{\bm{k}}$.
In the case of the $d_{x^2-y^2}$-wave AM, $E^{+,d2}_{\bm{k},+}$  and $E^{-,d2}_{\bm{k},-}$ are partly split, and $E^{+,d2}_{\bm{k},+}\sim E^{-,d2}_{\bm{k},+}$ and $E^{+,d2}_{\bm{k},-}\sim E^{-,d2}_{\bm{k},-}$ are obtained [Figs.~\ref{fig_sm:Ek_vs_kx_FS} (b,c,d,f,g,h)].
Then the momentum of this overlapping region is $k_{s}\sim0.08\pi$ at $t_{d2}=0.5\Delta$ [Figs.~\ref{fig_sm:Ek_vs_kx_FS} (b,e)], $k_{s}\sim0.06\pi$ at $t_{d2}=\Delta$ [Figs.~\ref{fig_sm:Ek_vs_kx_FS} (c,f)], and $k_{s}\sim0.04\pi$ at $t_{d2}=1.5\Delta$ [Figs.~\ref{fig_sm:Ek_vs_kx_FS} (d,g)].
If $E^{+,d2}_{\bm{k},+}\sim E^{-,d2}_{\bm{k},+}$ and $E^{+,d2}_{\bm{k},-}\sim E^{-,d2}_{\bm{k},-}$, the zero-energy flat bands remain, and as a result, this momentum $k_{s}$ corresponds to the size of the crossed flat band as discussed in Fig.\,4(a) of the main text.
Indeed, the energy dispersion of the quasiparticle $E^{s,d2}_{\bm{k},\pm}$ remains the degeneracy within the momentum $k_{s}$ even with the altermagnetism, and it contributes to the size of the crossed flat band. 
Along the $k_x$-direction, for the $d_{xy}$-wave AM, when $k_y=0$ ($k_x=0$), we obtain $M^{\alpha}_{\bm{k}}=0$ for each $k_z$.
It means that surface arc states, which appear along the $k_z$-axis, are not affected by the $d_{xy}$-wave AM; however, they are only modified by the BFS. 
While, for the $d_{x^2-y^2}$-wave AM, $M^{\alpha}_{\bm{k}}=0$ satisfies at $|k_x|=|k_y|$ for each $k_z$, and arc surface states appears at $(k_y,k_z)$ with $|k_y|=|k_x|$. 
For the $g_{xy(x^2-y^2)}$-wave AM, if $k_y=0$ ($k_x=0$) or $|k_x|=|k_y|$, we obtain $M^{\alpha}_{\bm{k}}=0$.
As a result, arc surface states are partly determined as shown in Fig.\ 3(c) of the main text.
Thus, crossed flat bands can have the width, and the surface arc states can remain shown in Figs.\,2(a,c), 3(a,c), 4(a), and 6(a) of the main text.

Next, we discuss the velocity of the quasiparticle along the $k_z$-direction.
In Fig.~\ref{fig_sm:Ek_vs_k}, we plot $E^{s,d2}_{\bm{k},\pm}$ for the $d_{x^2-y^2}$-wave AM at $t_{d2}=\Delta$ with $\Delta=0.01t$.
$E^{+,d2}_{\bm{k},+}=-E^{-,d2}_{\bm{k},-}$ and $E^{+,d2}_{\bm{k},-}=-E^{-,d2}_{\bm{k},+}$ for $(k_x,k_y)$ space at $k_z=0.15\pi,0.25\pi$ are shown in Figs.~\ref{fig_sm:Ek_vs_k} (a,b,c,d).
The amplitude of $E^{+,d2}_{\bm{k},\pm}$ becomes the smallest at the Fermi level.
To see more details, we plot $E^{s,d2}_{\bm{k},\pm}$ as a function of $k_x$ at $k_y=0$ for each $k_z$.
The energy gap is opening for $0.2\pi\le k_z\le0.42\pi$.
The BFS appears for $k_z\ge0.2\pi$, and a point node for $k_z=0.42\pi$.
We also show $E^{s,d2}_{\bm{k},\pm}$ as a function of $k_z$ at $k_y=0$ by changing $k_x$ in Fig.~\ref{fig_sm:Ek_vs_k} (f).
The energy gap is also opening for $0\le k_z\le0.35\pi$.
The BFS also exhibits for $k_x\ge0.35\pi$, and a point node for $k_x=0$.
These behaviors indicate that the energy gap closes around $k_{x0}=\pm0.35\pi$ and the BFS is realized for $k_{x}\ge |k_{x0}|$.
$E^{s,d2}_{\bm{k},\pm}$ at the minimal energy of the upper band can have zero velocity at $k_x=k_{x0}$.
When the minimal energy is zero, the velocity along the $k_z$-direction of the quasiparticle also becomes zero. 
Thus, the total conductance vanishes at $k_x=k_{x0}$.
Hence, this idea explains the vanishing conductance in the gapped regions of Figs.\,2(a,c) and 4(a), as well as Figs.\,3(a,c) and 6(a) of the main text.

\subsection{S1.4: Details on the topological stability of zero-energy surface states} \label{Sec_S1_3}

Here, we discuss the details of the winding number, which protects the crossed flat bands and arc states.
To define the winding number, a chiral operator that anticommutes with the Hamiltonian is required~\cite{YadaPRB2011,SatoPRB2011,BrydonPRB2011,tanaka2012symmetry,tanaka2024theory}.
The most common definition of the chiral operator is the one constructed from time-reversal (TRS) and particle–hole symmetries (PHS)~\cite{YadaPRB2011,SatoPRB2011,BrydonPRB2011}.
However, in the present system, both AM and pair potential break the TRS,
and therefore, this chiral symmetry is broken.
Nevertheless, to consider chiral symmetry in the present system, it is necessary to take into account the remaining symmetries of the system.
For example, the breaking of time-reversal symmetry due to the pair potential can be recovered by the following local gauge transformation~\cite{KobayashiPRB2015}:
\begin{align}
    \hat H'_{\mathrm{SC}}({\bm k})&=\hat U_{\phi_{\bm k}}\hat H_\mathrm{SC}({\bm k})\hat U^\dagger_{\phi_{\bm k}}\nonumber\\
    &=-\Delta \sqrt{\sin^2 k_x+\sin^2k_y}\sin k_z\hat{\sigma}_{2}\hat{\tau}_{2},
\end{align}%
with $\hat U_{\phi_{\bm k}}=e^{-i\phi_{\bm k}\hat{\sigma}_{0}\hat{\tau}_{3}/2}=\cos(\phi_{\bm k}/2)\hat{\sigma}_{0}\hat{\tau}_{0}-i\sin(\phi_{\bm k}/2)\hat{\sigma}_{0}\hat{\tau}_{3}$ and $\phi_{\bm k}=\tan^{-1}(\sin k_y/\sin k_x)$, i. e., 
\begin{equation}
\begin{split}
\cos \phi_{\bm k}&=\sin k_x/\sqrt{\sin^2 k_x+\sin^2k_y}\,,\\
\sin \phi_{\bm k}&=\sin k_y/\sqrt{\sin^2 k_x+\sin^2k_y}\,.
\end{split}
\end{equation}
%$\cos \phi_{\bm k}=\sin k_x/\sqrt{\sin^2 k_x+\sin^2k_y}$ and $\sin \phi_{\bm k}=\sin k_y/\sqrt{\sin^2 k_x+\sin^2k_y}$. 
This local gauge transformation does not affect the kinetic and altermagnetic parts: $\hat U_{\phi_{\bm k}}\varepsilon({\bm k})\hat{\sigma}_{0}\hat{\tau}_{3}\hat U^\dagger_{\phi_{\bm k}}=\varepsilon({\bm k})\hat{\sigma}_{0}\hat{\tau}_{3}$ and $\hat U_{\phi_{\bm k}}\hat H_{\mathrm{AM}}^\alpha({\bm k})\hat U^\dagger_{\phi_{\bm k}}=\hat H_{\mathrm{AM}}^\alpha({\bm k})$. 
By this local gauge transformation, the normal part and pair potential term of the Hamiltonian have the following pseudo-time-reversal (pTRS) and particle-hole symmetries:
\begin{align}
\hat{\Theta}\hat H'_{\mathrm{NM+SC}}({\bm k})\hat{\Theta}^{\dagger}&=[\hat H'_{\mathrm{NM+SC}}(-{\bm k})]^*,\label{eq:s13}\\
\hat{C}\hat H'_{\mathrm{NM+SC}}({\bm k})\hat{C}^{\dagger}&=-[\hat H'_{\mathrm{NM+SC}}(-{\bm k})]^*,\label{eq:s14}
\end{align}%
with $\hat H'_{\mathrm{NM+SC}}({\bm k})=\varepsilon({\bm k})\hat{\sigma}_{0}\hat{\tau}_{3}+\hat H'_{\mathrm{SC}}({\bm k})$, the conventional TRS $\hat{\Theta}=i\hat{\sigma}_{2}\hat{\tau}_{0}$, and PHS $\hat{C}=\hat{\sigma}_{0}\hat{\tau}_{1}$.
We note that the conventional TRS $\hat{\Theta}$ and PHS $\hat{C}$ are also defined in the main text.
Here, we introduce a chiral operator $\hat \Gamma\equiv -i \hat C \hat \Theta=\hat \sigma_2\hat \tau_1$. 
Using the Eqs.\ (\ref{eq:s13}) and (\ref{eq:s14}), we obtain the following chiral symmetry:
\begin{align}
\hat\Gamma\hat H'_{\mathrm{NM+SC}}({\bm k})\hat{\Gamma}^{\dagger}&=-\hat H'_{\mathrm{NM+SC}}({\bm k}),
\end{align}%
where $\hat\Gamma$ anticommutes $H'_{\mathrm{NM+SC}}({\bm k})$. 
Since $H'_{\mathrm{NM+SC}}({\bm k})$ is the Hamiltonian obtained from the local gauge transformation, we go back to the original gauge:
\begin{align}
\hat U_{\phi_{\bm k}}^\dagger
\hat\Gamma
\hat U_{\phi_{\bm k}}
\hat H_{\mathrm{NM+SC}}({\bm k})
\hat U_{\phi_{\bm k}}^{\dagger}
\hat{\Gamma}^{\dagger}
\hat U_{\phi_{\bm k}}
&=-\hat H_{\mathrm{NM+SC}}({\bm k}).\label{gamma_k}
\end{align}%
Then, we obtain the chiral operator for the original gauge: {
\begin{align}
    \hat\Gamma_{\bm k}=\hat U_{\phi_{\bm k}}^{\dagger}\hat{\Gamma}\hat U_{\phi_{\bm k}}.
\end{align}
}%
On the other hand, this anticommutation relation in Eq.\ (\ref{gamma_k}) is no longer satisfied when we consider 
$\hat H_{\mathrm{AM}}^\alpha({\bm k})$ due to the broken TRS:
\begin{align}
\hat{\Theta}\hat H^\alpha_{\mathrm{AM}}({\bm k})\hat{\Theta}^{\dagger}&=-[\hat H^\alpha_{\mathrm{AM}}(-{\bm k})]^*,\label{eq:s15}\\
\hat{C}\hat H^\alpha_{\mathrm{AM}}({\bm k})\hat{C}^{\dagger}&=-[\hat H^\alpha_{\mathrm{AM}}(-{\bm k})]^*.
\end{align}%
Nevertheless, $\hat H_{\mathrm{AM}}^\alpha({\bm k})$ has the magnetic mirror symmetry (MMS), which is the product of pTRS and the mirror symmetry. 
The corresponding mirror plane depends on the symmetry of AMs. 
In the main text, since we discuss the chiral operator $\hat \Gamma_{\bm k}$ by focusing on the total Hamiltonian $\hat{H}^{\alpha}_\mathrm{BdG}(\bm{k})$, we introduce the pseudo-magnetic mirror symmetry (pMMS): $\hat{\Theta}^{'}_{\bm k}=\hat{U}_{\phi_{\bm k}}^\dagger\hat{\Theta} \hat{U}_{\phi_{\bm k}}$.
While here, we demonstrate the chiral operator $\hat \Gamma_{\bm k}$ for each term in $\hat{H}^{\alpha}_\mathrm{BdG}(\bm{k})$. 
Thus, we note that we use MMS in the SM, not pMMS. 
In the case of the $d_{xy}$-wave AM, $\hat H_{\mathrm{AM}}^\alpha({\bm k})$ has the following MMS:
\begin{align}
    \hat H_{\mathrm{AM}}^\alpha(k_x, k_y, k_z)&=-\hat H_{\mathrm{AM}}^\alpha(-k_x, k_y, k_z)\nonumber\\
    &=-\hat H_{\mathrm{AM}}^\alpha(k_x, -k_y, k_z),\label{eq:s17}
\end{align}%
where the first and the second line denote the mirror symmetry on the $yz$- and $zx$-plane, respectively.
On the other hand,
for the $d_{x^2-y^2}$-wave AM, $\hat H_{\mathrm{AM}}^\alpha({\bm k})$ has diagonal magnetic mirror planes,
\begin{align}
    \hat H_{\mathrm{AM}}^\alpha(k_x, k_y, k_z)&=-\hat H_{\mathrm{AM}}^\alpha(k_y, k_x, k_z)\nonumber\\
    &=-\hat H_{\mathrm{AM}}^\alpha(-k_y, -k_x, k_z).\label{eq:s18}
\end{align}%
For the $g_{xy(x^2-y^2)}$-wave AM, both Eqs. (\ref{eq:s17}) and (\ref{eq:s18}) is available. 
As a result, pTRS in Eq.\ (\ref{eq:s15}) becomes
\begin{align}
    \hat\Theta\hat H_{\mathrm{AM}}^\alpha(k_x, k_y, k_z)\hat{\Theta}^{\dagger}&=\hat H_{\mathrm{AM}}^\alpha(k_x, -k_y, -k_z)\nonumber\\
    &=\hat H_{\mathrm{AM}}^\alpha(-k_x, k_y, -k_z),
\end{align}%
for $d_{xy}$ and $g_{xy(x^2-y^2)}$-wave AMs, and 
\begin{align}
    \hat\Theta\hat H_{\mathrm{AM}}^\alpha(k_x, k_y, k_z)\hat{\Theta}^{\dagger}&=\hat H_{\mathrm{AM}}^\alpha(-k_y, -k_x, -k_z)\nonumber\\
    &=\hat H_{\mathrm{AM}}^\alpha(k_y, k_x, -k_z),
\end{align}%
for $d_{x^2-y^2}$- and $g_{xy(x^2-y^2)}$-wave AMs.
Finally, we obtain the chiral symmetry by the combination of local gauge symmetry, MMS, and PHS:
\begin{align}
    \hat \Gamma_{\bm k}\hat H_{\mathrm{AM}}^\alpha(k_x, k_y, k_z)\hat{\Gamma}_{\bm k}^{\dagger}&=\hat H_{\mathrm{AM}}^\alpha(-k_x, k_y, k_z)\nonumber\\
    &=\hat H_{\mathrm{AM}}^\alpha(k_x, -k_y, k_z),\label{eq:s23}
\end{align}%
for $d_{xy}$ and $g_{xy(x^2-y^2)}$-wave AMs, and 
\begin{align}
    \hat \Gamma_{\bm k}\hat H_{\mathrm{AM}}^\alpha(k_x, k_y, k_z)\hat{\Gamma}_{\bm k}^{\dagger}&=\hat H_{\mathrm{AM}}^\alpha(k_y, k_x, k_z)\nonumber\\
    &=\hat H_{\mathrm{AM}}^\alpha(-k_y, -k_x, k_z),\label{eq:s24}
\end{align}%
for $d_{x^2-y^2}$- and $g_{xy(x^2-y^2)}$-wave AMs.
Eqs.\ (\ref{eq:s23}) and (\ref{eq:s24}) say that
$H_{\mathrm{AM}}^\alpha({\bm k})$ anticommutes with $\hat\Gamma_{\bm k}$ on the magnetic mirror plane. 
The chiral operators defined on each magnetic mirror plane are as follows:
\begin{align}
\hat{\Gamma}_{k_x=0}&=\hat\sigma_2\hat\tau_2,\\
\hat{\Gamma}_{k_y=0}&=\hat\sigma_2\hat\tau_1,\\
\hat{\Gamma}_{k_x=k_y}&=\hat\sigma_2(\tau_1+\tau_2)/\sqrt{2},\\
\hat{\Gamma}_{k_x=-k_y}&=\hat\sigma_2(\tau_1-\tau_2)/\sqrt{2}.
\end{align}%
% Please check these expressions. I will do it.
Using these chiral operators, we can calculate the winding number on each magnetic mirror plane. When the winding number is nonzero, the system becomes topologically nontrivial, and the crossed flat bands and/or surface arcs appear. 
Therefore, the analysis carried out in this subsection demonstrates the topological origin of the crossed flat bands and surface arcs discussed in the main text.

\section{S2: Calculation of conductance in a hybrid system formed by a superconducting altermagnet and a normal metal} \label{Sec_S2}

In the main text, we have discussed the conductance in superconducting AM/normal metal junctions.
Here, in this section, we show the details of how the conductance is calculated in superconducting AM junctions.
We also discuss the details of the power law in the momentum-dependent conductance.

\subsection{S2.1: Model of the hybrid junction} \label{Sec_S2_1}

We hereby provide the model and formulation in superconducting hybrid junctions.
Then we assume the semi-infinite superconducting AM and normal metal regimes.
In superconducting AM/normal metal junctions, we consider the Hamiltonian:
\begin{align}
    \hat{\mathcal{H}}=\hat{\mathcal{H}}^{\alpha}_\mathrm{SC+AM}+\hat{\mathcal{H}}_\mathrm{J}+\hat{\mathcal{H}}_\mathrm{NM}.
\end{align}%
In $\hat{\mathcal{H}}^{\alpha}_\mathrm{SC+AM}$, when we consider the junction along the $w=z,x$-direction, the local $\tilde{u}^{\alpha}_\mathrm{sc}(\tilde{\bm{k}})$, nearest-neighbor (NN) $\tilde{t}^{\alpha}_\mathrm{sc1}(\tilde{\bm{k}})$, and second NN hopping terms $\tilde{t}^{\alpha}_\mathrm{sc2}(\tilde{\bm{k}})$ with $\tilde{\bm{k}}=(k_u,k_v)$ for $u=x,y$ and $v=y,z$ are given by
\begin{align}
    \tilde{u}^{\alpha}_\mathrm{sc}(\tilde{\bm{k}})&=
    \begin{pmatrix}
        \hat{u}_{0}(\tilde{\bm{k}})+\hat{u}^{\alpha}_\mathrm{AM}(\tilde{\bm{k}}) & \hat{u}_\mathrm{\Delta}(\tilde{\bm{k}})\\
        \hat{u}^{\dagger}_{\Delta}(\tilde{\bm{k}}) & -\hat{u}^{*}_{0}(-\tilde{\bm{k}})-\hat{u}^{\alpha*}_\mathrm{AM}(-\tilde{\bm{k}})
    \end{pmatrix},
\end{align}%
\begin{align}
    \tilde{t}^{\alpha}_\mathrm{sc1}(\tilde{\bm{k}})&=
    \begin{pmatrix}
        \hat{t}_{0}(\tilde{\bm{k}})+\hat{t}^{\alpha}_\mathrm{AM1}(\tilde{\bm{k}}) & \hat{t}_\mathrm{\Delta}(\tilde{\bm{k}})\\
        \bar{t}_{\Delta}(\tilde{\bm{k}}) & -\hat{t}^{*}_{0}(-\tilde{\bm{k}})-\hat{t}^{\alpha*}_\mathrm{AM1}(-\tilde{\bm{k}})
    \end{pmatrix}.
\end{align}%
\begin{align}
    \tilde{t}^{\alpha}_\mathrm{sc2}(\tilde{\bm{k}})&=
    \begin{pmatrix}
        \hat{t}^{\alpha}_\mathrm{AM2}(\tilde{\bm{k}}) & 0\\
        0 & -\hat{t}^{\alpha*}_\mathrm{AM2}(-\tilde{\bm{k}})
    \end{pmatrix}.
\end{align}%
In the case of the junction along the $z$-direction, we obtain
\begin{align}
    \hat{u}_{0}(k_x,k_y)&=[-\mu-2t\cos{k_x}-2t\cos{k_y}]\hat{\sigma}_{0},\\
    \hat{u}^{\alpha}_\mathrm{AM}(k_x,k_y)&=\hat{H}^{\alpha}_\mathrm{AM}(k_x,k_y),
\end{align}%
\begin{align}
    \hat{t}_{0}&=-t\hat{\sigma}_{0},\\
    \hat{t}^{\alpha}_\mathrm{AM1}(k_x,k_y)&=\hat{t}^{\alpha}_\mathrm{AM2}(k_x,k_y)=0,
\end{align}%
\begin{align}
    \hat{u}_{\Delta}&=0,\\
    \hat{t}_{\Delta}&=i\frac{\Delta}{2}\left[\sin{k_x}-i\sin{k_y}\right]i\hat{\sigma}_{2},\\
    \bar{t}_{\Delta}&=-i\frac{\Delta}{2}\left[\sin{k_x}+i\sin{k_y}\right]i\hat{\sigma}_{2},
\end{align}%
and along the $x$-direction,
\begin{align}
    \hat{u}_{0}(k_y,k_z)&=[-\mu-2t\cos{k_y}-2t\cos{k_z}]\hat{\sigma}_{0},\\
    \hat{u}^{d1}_\mathrm{AM}(k_y,k_z)&=0,\\
    \hat{u}^{d2}_\mathrm{AM}(k_y,k_z)&=-t_{d2}\cos{k_y},\\
    \hat{u}^{g}_\mathrm{AM}(k_y,k_z)&=0,
\end{align}%
\begin{align}
    \hat{t}_{0}&=-t\hat{\sigma}_{0},\\
    \hat{t}^{d1}_\mathrm{AM1}(k_y)&=-i\frac{t_{d1}}{2}\sin{k_y},\\
    \hat{t}^{d2}_\mathrm{AM1}(k_y)&=\frac{t_{d2}}{2},\\
    \hat{t}^{g}_\mathrm{AM1}(k_y)&=2i{t_{g}}\sin{k_y}\cos{k_y},\\
    \hat{t}^{d1}_\mathrm{AM2}(k_y)&=\hat{t}^{d2}_\mathrm{AM2}(k_y)=0,\\
    \hat{t}^{g}_\mathrm{AM2}(k_y)&=-it_{g}\sin{k_y},
\end{align}%
\begin{align}
    \hat{u}_{\Delta}&=i\Delta\sin{k_y}\sin{k_z},\\
    \hat{t}_{\Delta}&=\left[i\frac{\Delta}{2}\sin{k_y}\right]i\hat{\sigma}_{2},\\
    \bar{t}_{\Delta}&=\left[-i\frac{\Delta}{2}\sin{k_y}\right]i\hat{\sigma}_{2}.
\end{align}%
In the normal metal $\hat{H}_\mathrm{NM}$, the local $\tilde{u}_\mathrm{NM}(\tilde{\bm{k}})$ and nonlocal terms $\tilde{t}_\mathrm{NM}(\tilde{\bm{k}})$ are given by
\begin{align}
    \tilde{u}_\mathrm{NM}(\tilde{\bm{k}})&=
    \begin{pmatrix}
        \hat{u}_{0}(\tilde{\bm{k}}) & 0\\
        0 & -\hat{u}^{*}_{0}(-\tilde{\bm{k}})
    \end{pmatrix},\\
    \tilde{t}_\mathrm{NM}(\tilde{\bm{k}})&=
    \begin{pmatrix}
        \hat{t}_{0}(\tilde{\bm{k}}) & 0\\
        0 & -\hat{t}^{*}_{0}(-\tilde{\bm{k}})
    \end{pmatrix}.
\end{align}
The tunneling Hamiltonian is defined as
\begin{align}
    \hat{\mathcal{H}}_\mathrm{J}&=\hat{t}_\mathrm{J}\otimes\hat{\tau}_{3},
\end{align}%
with the Pauli matrices in Nambu space $\hat{\tau}_{i=0,1,2,3}$.
Here, $\hat{t}_\mathrm{J}$ denotes the tunneling Hamiltonian in the electron space:
\begin{align}
    \hat{t}_\mathrm{J}&=-t\hat{\sigma}_{0},
\end{align}%
along the $z$ and $x$-directions.
These equations shown here are used in the next subsection to calculate the conductance.

\subsection{S2.2: Details of calculation of conductance} \label{Sec_S2_2}

In this subsection, we show the details of how we calculate the conductance in superconducting AM junctions. 
The conductance is obtained based on the Lee-Fisher formula~\cite{Lee_Fisher,OhashiPRB2024,fukaya2025PUM,lu2025subgap}:
\begin{align}
    \sigma(eV)&=\frac{1}{2\pi}\int^{\pi}_{-\pi}\int^{\pi}_{-\pi}d\tilde{\bm{k}}\bar\sigma(\tilde{\bm{k}},eV), \label{sigma}
\end{align}%
with
\begin{align}
    \bar\sigma(\tilde{\bm{k}},eV)=\mathrm{Tr'}[\tilde{g}_{a}(\tilde{\bm{k}},eV)+\tilde{g}_{b}(\tilde{\bm{k}},eV)-\tilde{g}_{c}(\tilde{\bm{k}},eV)-\tilde{g}_{d}(\tilde{\bm{k}},eV)]. \label{bar_sigma}
\end{align}%
Eq.\,(\ref{bar_sigma}) is given by Eq.\,(8) of the main text.
Here, we present the explicit expression for Green's function $\tilde{g}_{a,b,c,d}(\tilde{\bm{k}},eV)$. 
For $E=eV$ with the energy $E$ and the bias voltage $V$, $\tilde{g}_{a,b,c,d}(\tilde{\bm{k}},eV)$ are calculated by
\begin{align}
    \tilde{g}_{a}(\tilde{\bm{k}},E)&=\tilde{G}_{11}(\tilde{\bm{k}},E)\hat{V}_{10}\tilde{G}_{00}(\tilde{\bm{k}},E)\hat{V}^{\dagger}_{10},\\
    \tilde{g}_{b}(\tilde{\bm{k}},E)&=\tilde{G}_{00}(\tilde{\bm{k}},E)\hat{V}^{\dagger}_{10}\tilde{G}_{11}(\tilde{\bm{k}},E)\hat{V}_{10},\\
    \tilde{g}_{c}(\tilde{\bm{k}},E)&=\tilde{G}_{01}(\tilde{\bm{k}},E)\hat{V}_{10}\tilde{G}_{01}(\tilde{\bm{k}},E)\hat{V}_{10},\\
    \tilde{g}_{d}(\tilde{\bm{k}},E)&=\tilde{G}_{10}(\tilde{\bm{k}},E)\hat{V}^{\dagger}_{10}\tilde{G}_{10}(\tilde{\bm{k}},E)\hat{V}^{\dagger}_{10}.
\end{align}%
Here, $\hat{V}_{10}$ is defined as $\hat{V}_{10}=-\hat{t}_\mathrm{0}\otimes\hat{\tau}_{0}$ for obtaining the conductance owing to the current operator, and this equation is already provided in the main text.
$\tilde{G}_{ij}(\tilde{\bm{k}},E)$ with $i,j=0,1$ are numerically calculated by the retarded (R) and advanced (A) Green's function $\tilde{G}^\mathrm{R}_{ij}(\tilde{\bm{k}},E)$ and $\tilde{G}^\mathrm{A}_{ij}(\tilde{\bm{k}},E)$:
\begin{align}
    \tilde{G}_{00}(\tilde{\bm{k}},E)&=-\frac{i}{2}[\tilde{G}^\mathrm{A}_{00}(\tilde{\bm{k}},E)-\tilde{G}^\mathrm{R}_{00}(\tilde{\bm{k}},E)],\\
    \tilde{G}_{11}(\tilde{\bm{k}},E)&=-\frac{i}{2}[\tilde{G}^\mathrm{A}_{11}(\tilde{\bm{k}},E)-\tilde{G}^\mathrm{R}_{11}(\tilde{\bm{k}},E)],\\
    \tilde{G}_{01}(\tilde{\bm{k}},E)&=-\frac{i}{2}[\tilde{G}^\mathrm{A}_{01}(\tilde{\bm{k}},E)-\tilde{G}^\mathrm{R}_{01}(\tilde{\bm{k}},E)],\\
    \tilde{G}_{10}(\tilde{\bm{k}},E)&=-\frac{i}{2}[\tilde{G}^\mathrm{A}_{10}(\tilde{\bm{k}},E)-\tilde{G}^\mathrm{R}_{10}(\tilde{\bm{k}},E)].
\end{align}%
Local $\tilde{G}_{i,i}(\tilde{\bm{k}},E)$ and nonlocal Green's functions $\tilde{G}_{i,j}(\tilde{\bm{k}},E)$ are calculated by
\begin{align}
    \tilde{G}_{00}(\tilde{\bm{k}},z)&=[\{\tilde{G}^\mathrm{L}_{0}(\tilde{\bm{k}},z)\}^{-1}-\tilde{t}_\mathrm{NM}\tilde{G}^\mathrm{R}_{1}(\tilde{\bm{k}},z)\tilde{t}^{\dagger}_\mathrm{NM}]^{-1},\\
    \tilde{G}_{11}(\tilde{\bm{k}},z)&=[\{\tilde{G}^\mathrm{R}_{1}(\tilde{\bm{k}},z)\}^{-1}-\tilde{t}^{\dagger}_\mathrm{NM}\tilde{G}^\mathrm{L}_{0}(\tilde{\bm{k}},z)\tilde{t}_\mathrm{NM}]^{-1},\\
    \tilde{G}_{01}(\tilde{\bm{k}},z)&=\tilde{G}^\mathrm{L}_{0}(\tilde{\bm{k}},z)\tilde{t}_\mathrm{NM}\tilde{G}_{11}(\tilde{\bm{k}},z),\\
    \tilde{G}_{10}(\tilde{\bm{k}},z)&=\tilde{G}^\mathrm{R}_{1}(\tilde{\bm{k}},z)\tilde{t}^{\dagger}_\mathrm{NM}\tilde{G}_{00}(\tilde{\bm{k}},z),
\end{align}%
where $z=E\pm i\delta$ with the infinitesimal value $\delta$ are the retarded (R) $z=E+i\delta$ and advanced (A) parts $z=E-i\delta$.
By using the recursive Green's function method~\cite{Umerski97,TakagiPRB2020}, we get the surface Green's functions:
\begin{align}
    \tilde{G}^\mathrm{L}_{0}(\tilde{\bm{k}},z)&=[z-\tilde{u}_\mathrm{NM}(\tilde{\bm{k}})-\tilde{t}^{\dagger}_\mathrm{J}\tilde{G}^\mathrm{L}(\tilde{\bm{k}},z)\tilde{t}_\mathrm{J}]^{-1},\\
    \tilde{G}^\mathrm{R}_{1}(\tilde{\bm{k}},z)&=[z-\tilde{u}_\mathrm{NM}(\tilde{\bm{k}})-\tilde{u}_\mathrm{b}-\tilde{t}_\mathrm{NM}\tilde{G}^\mathrm{R}(\tilde{\bm{k}},z)\tilde{t}^{\dagger}_\mathrm{NM}]^{-1}, \label{GR0}
\end{align}%
with the semi-infinite Green's functions $\tilde{G}^\mathrm{L}(\tilde{\bm{k}},z)$ and $\tilde{G}^\mathrm{R}(\tilde{\bm{k}},z)$, and $\tilde{u}_\mathrm{b}=U_\mathrm{b}\hat{\sigma}_{0}\otimes\hat{\tau}_{3}$ for the strength of the potential barrier $U_\mathrm{b}$.
We obtain the semi-infinite Green's functions for both retarded ($z=E+i\delta$) and advanced ($z=E-i\delta$) parts by using the local $\hat{u}$ and hopping terms $\hat{t}$, see Refs.\ \cite{Umerski97,TakagiPRB2020,san2013multiple}.

For the semi-infinite Green's function, when only NN hopping is included, we choose the local $\hat{u}$ and the hopping terms $\hat{t}$ as
$\hat{u}\rightarrow\tilde{u}_\mathrm{sc}$ and $\hat{t}\rightarrow\tilde{t}_\mathrm{sc1}$ with the NN hopping $\tilde{t}_\mathrm{sc1}$.
In the case of including the second NN hopping term $\tilde{t}_\mathrm{sc2}$ like the $g_{xy(x^2-y^2)}$-wave AM along the $x$-direction, we replace the local $\hat{u}$ and the hopping term $\hat{t}$ as~\cite{Umerski97}
\begin{align}
    \hat{u}\rightarrow
    \begin{pmatrix}
        \tilde{u}_\mathrm{sc} & \tilde{t}_\mathrm{sc1}\\
        \tilde{t}^{\dagger}_\mathrm{sc1} & \tilde{u}_\mathrm{sc}
    \end{pmatrix},\label{usc2}
\end{align}%
\begin{align}
        \hat{t}\rightarrow
    \begin{pmatrix}
        \tilde{t}_\mathrm{sc2} & 0\\
        \tilde{t}^{\dagger}_\mathrm{sc1} & \tilde{t}_\mathrm{sc2}
    \end{pmatrix}.\label{tsc2}
\end{align}%
To connect $\hat{V}_{10}$, if the second NN hopping term is included in the formulation, e.g., $g_{xy(x^2-y^2)}$-wave AM along the $x$-direction, we pick up the subspace of the semi-infinite Green's function $\tilde{G}^\mathrm{L}_{0}=\tilde{G}_{bb}$, which is defined as:
\begin{align}
    \tilde{G}=\left(
    \begin{array}{c|c}
        \tilde{G}_{aa} & \tilde{G}_{ab}  \\\hline
        \tilde{G}_{ba} & \tilde{G}_{bb}
    \end{array}\right),
\end{align}%
in the basis $[\hat{\psi}^{\dagger}_{a},\hat{\psi}^{\dagger}_{b}]$ with $\hat{\psi}^{\dagger}_{l=a,b}=[c^{\dagger}_{l\uparrow},c^{\dagger}_{l\downarrow},c_{l\uparrow},c_{l\downarrow}]$.
We have calculated the conductance in superconducting AM junctions of the main text by using this formulation.

\subsection{S2.3: Power law behavior in the momentum dependent conductance} \label{Sec_S2_3}

\begin{figure}[t!]
    \centering
    \includegraphics[width=12.5cm]{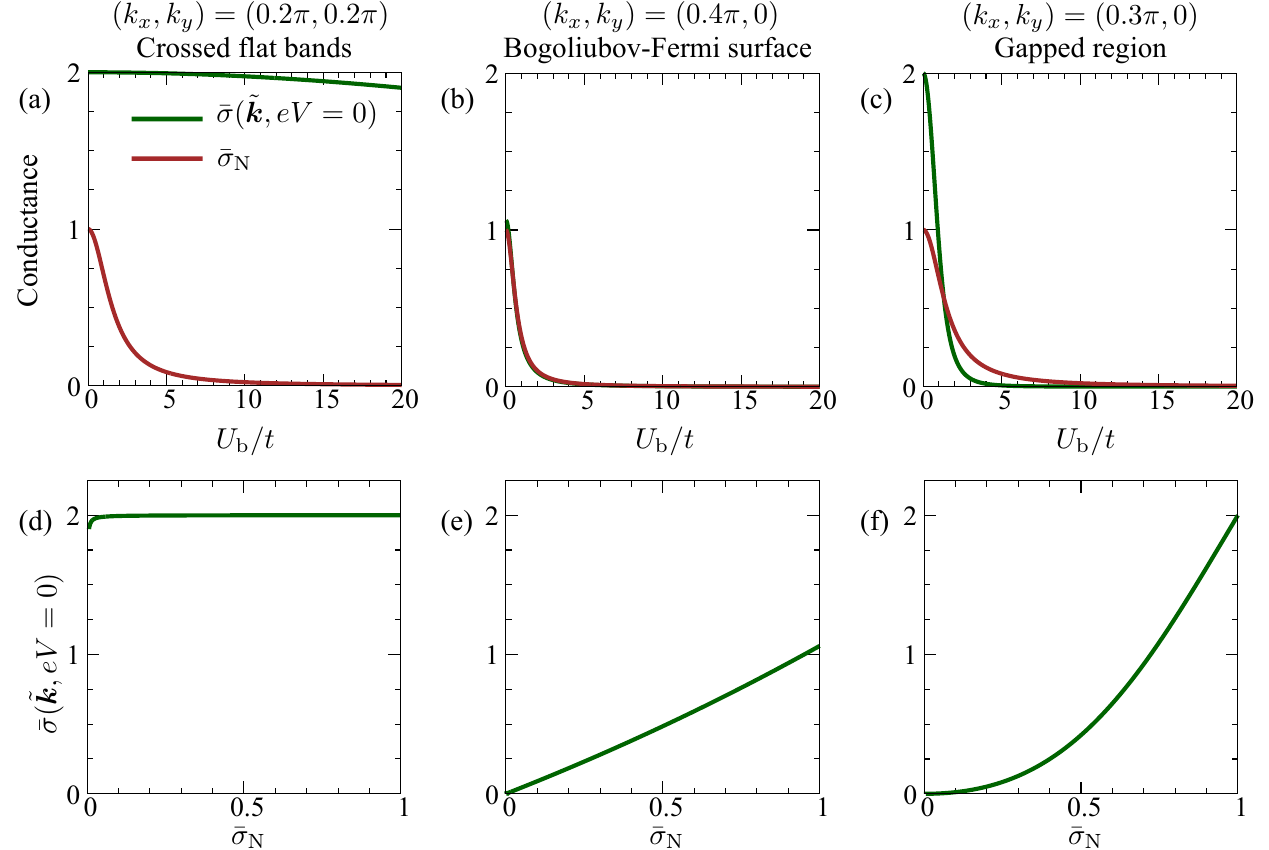}
    \caption{For the $d_{x^2-y^2}$-wave AM, (a-c) zero-bias conductance $\bar{\sigma}(\tilde{\bm{k}},eV=0)$ and zero-bias conductance in the normal state $\bar{\sigma}_\mathrm{N}$ along the $z$-direction as a function of the potential barrier $U_\mathrm{b}$ at $(k_x,k_y)$.
    (d-f) $\bar{\sigma}(\tilde{\bm{k}},eV=0)$ along the $z$-direction as a function of $\bar{\sigma}_\mathrm{N}$ at $(k_x,k_y)$.
    We choose the momentum as (a,d) $(k_x,k_y)=(0.2\pi,0.2\pi)$ (crossed flat band), (b,e) $(0.4\pi,0)$ [Bogoliubov-Fermi surface (BFS)], and (c,f) $(0.3\pi,0)$ (gapped region). 
    We select the parameters as $\mu=-4.5t$, $\Delta=0.01t$, $t_{d2}=\Delta$, and $\delta=0.0001\Delta$.}
    \label{fig_sm:PL_z}
\end{figure}%
\begin{figure}[t!]
    \centering
    \includegraphics[width=12.5cm]{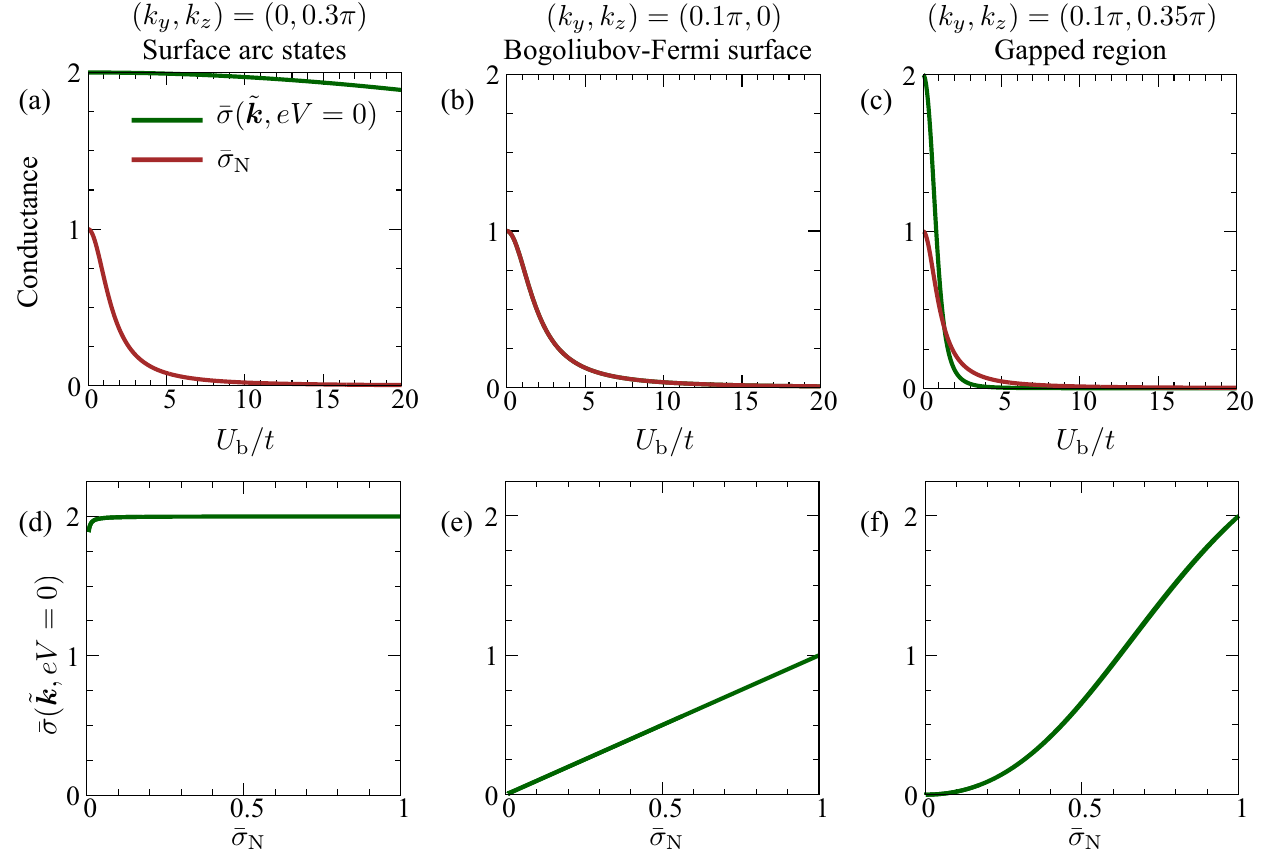}
    \caption{For the $d_{xy}$-wave AM, (a-c) zero-bias conductance $\bar{\sigma}(\tilde{\bm{k}},eV=0)$ and zero-bias conductance in the normal state $\bar{\sigma}_\mathrm{N}$ along the $x$-direction as a function of the potential barrier $U_\mathrm{b}$ at $(k_y,k_z)$.
    (d-f) $\bar{\sigma}(\tilde{\bm{k}},eV=0)$ along the $x$-direction as a function of $\bar{\sigma}_\mathrm{N}$ at $(k_y,k_z)$.
    We choose the momentum as (a,d) $(k_y,k_z)=(0,0.3\pi)$ (surface arc state), (b,e) $(0.1\pi,0)$ (BFS), and (c,f) $(0.1\pi,0.35\pi)$ (gapped region). 
    We select the parameters as $\mu=-4.5t$, $\Delta=0.01t$, $t_{d1}=\Delta$, and $\delta=0.0001\Delta$.}
    \label{fig_sm:PL_x}
\end{figure}%
In general, the three tunneling processes shown in the main text do not appear simultaneously.
In this subsection, we show that these three fundamental tunneling processes coexist, depending on the momentum parallel to the interface of the junction.
For this purpose, we demonstrate the zero-bias conductance $\bar\sigma(\tilde{\bm{k}},eV=0)$ and the zero-bias normal state conductance $\bar\sigma_\mathrm{N}(\tilde{\bm{k}},eV=0)\equiv\bar\sigma_\mathrm{N}$ at $\tilde{\bm{k}}$ along the $z$ and $x$-directions. 
For the $d_{x^2-y^2}$-wave AM, we plot $\bar\sigma(\tilde{\bm{k}},eV=0)$ and $\bar\sigma_\mathrm{N}$ along the $z$-direction as a function of the potential barrier $U_\mathrm{b}$ in Figs.~\ref{fig_sm:PL_z} (a,b,c) and $\bar\sigma(\tilde{\bm{k}},eV=0)$ along the $z$-direction as a function of $\bar\sigma_\mathrm{N}(\tilde{\bm{k}},eV=0)$ in Figs.~\ref{fig_sm:PL_z} (d,e,f), at $t_{d2}=\Delta$.
Then we choose the momentum as (a,d) $(k_x,k_y)=(0.2\pi,0.2\pi)$ (crossed flat band), (b,e) $(0.4\pi,0)$ (BFS), and (c,f) $(0.3\pi,0)$ (gapped region). 
In Fig.~\ref{fig_sm:PL_z} (a), on the crossed flat bands, $\bar{\sigma}$ is almost 2, and $\bar{\sigma}_\mathrm{N}$ is suppressed for $U_\mathrm{b}$.
In Fig.~\ref{fig_sm:PL_z} (b), on the BFS, we obtain $\bar{\sigma}(\tilde{\bm{k}},eV=0)\sim\bar{\sigma}_\mathrm{N}$.
On the gapped region shown in Fig.~\ref{fig_sm:PL_z} (c), both $\bar{\sigma}(\tilde{\bm{k}},eV=0)$ and $\bar{\sigma}_\mathrm{N}$ strongly suppressed.
When we see the power law of the zero-bias conductance for each momentum, we obtain $\bar{\sigma}(\tilde{\bm{k}},eV=0)\sim2\bar{\sigma}^{0}_\mathrm{N}$ on the crossed flat band [Fig.~\ref{fig_sm:PL_z} (d)], $\bar{\sigma}(\tilde{\bm{k}},eV=0)\sim\bar{\sigma}^{1}_\mathrm{N}$ on the BFS [Fig.~\ref{fig_sm:PL_z} (e)], $\bar{\sigma}(\tilde{\bm{k}},eV=0)\sim\bar{\sigma}^{2}_\mathrm{N}$ [Fig.~\ref{fig_sm:PL_z} (f)].
We note that the power law in the gapped region deviates from $\bar{\sigma}(\tilde{\bm{k}},eV=0)\sim\bar{\sigma}^{2}_\mathrm{N}$ owing to the smearing factor $\delta$ in the lattice Green's function method.
Along the $x$-direction, for the $d_{xy}$-wave AM, we also show $\bar\sigma(\tilde{\bm{k}},eV=0)$ and $\bar\sigma_\mathrm{N}$ as a function of $U_\mathrm{b}$ in Figs.~\ref{fig_sm:PL_x} (a,b,c) and $\bar\sigma(\tilde{\bm{k}},eV=0)$ as a function of $\bar\sigma_\mathrm{N}$ in Figs.~\ref{fig_sm:PL_x} (d,e,f), at $t_{d1}=\Delta$.
As well as along the $z$-direction, on the surface arc state at $(k_y,k_z)=(0,0.3\pi)$, $\bar{\sigma}(\tilde{\bm{k}},eV=0)$ becomes almost 2, and $\bar{\sigma}_\mathrm{N}$ is suppressed by $U_\mathrm{b}$ [Fig.~\ref{fig_sm:PL_x} (a)], on the BFS at $(k_y,k_z)=(0.1\pi,0)$, we obtain $\bar{\sigma}(\tilde{\bm{k}},eV=0)\sim\bar{\sigma}_\mathrm{N}$ [Fig.~\ref{fig_sm:PL_x} (b)], and on the gapped region at $(k_y,k_z)=(0.1\pi,0.35\pi)$, both $\bar{\sigma}(\tilde{\bm{k}},eV=0)$ and $\bar{\sigma}_\mathrm{N}$ strongly suppressed shown in [Fig.~\ref{fig_sm:PL_x} (c)].
Then, we also obtain the power law as $\bar{\sigma}(\tilde{\bm{k}},eV=0)\sim2\bar{\sigma}^{0}_\mathrm{N}$ on the surface arc state [Fig.~\ref{fig_sm:PL_x} (d)], $\bar{\sigma}(\tilde{\bm{k}},eV=0)\sim\bar{\sigma}^{1}_\mathrm{N}$ on the BFS [Fig.~\ref{fig_sm:PL_x} (e)], $\bar{\sigma}(\tilde{\bm{k}},eV=0)\sim\bar{\sigma}^{2}_\mathrm{N}$ [Fig.~\ref{fig_sm:PL_x} (f)].
It is noted that the deviation in the gapped region originates from $\delta$ in the Green's function.
Thus, three distinct types of fundamental power laws $2\bar{\sigma}^{0}_\mathrm{N}$ (zero-energy surface Andreev bound states: perfect resonance of Andreev reflection), $\bar{\sigma}^{1}_\mathrm{N}$ (BFS), and $\bar{\sigma}^{2}_\mathrm{N}$ (gapped region: Andreev reflection without resonance) coexist in the zero-bias momentum dependent conductance in superconducting AM junctions. 
We have also commented on this discussion in the main text.

\subsection{S2.4: Configurations for regions of surface states and Bogoliubov-Fermi surface in the total conductance} \label{Sec_S2_4}

\begin{figure}[t!]
    \centering
    \includegraphics[width=13cm]{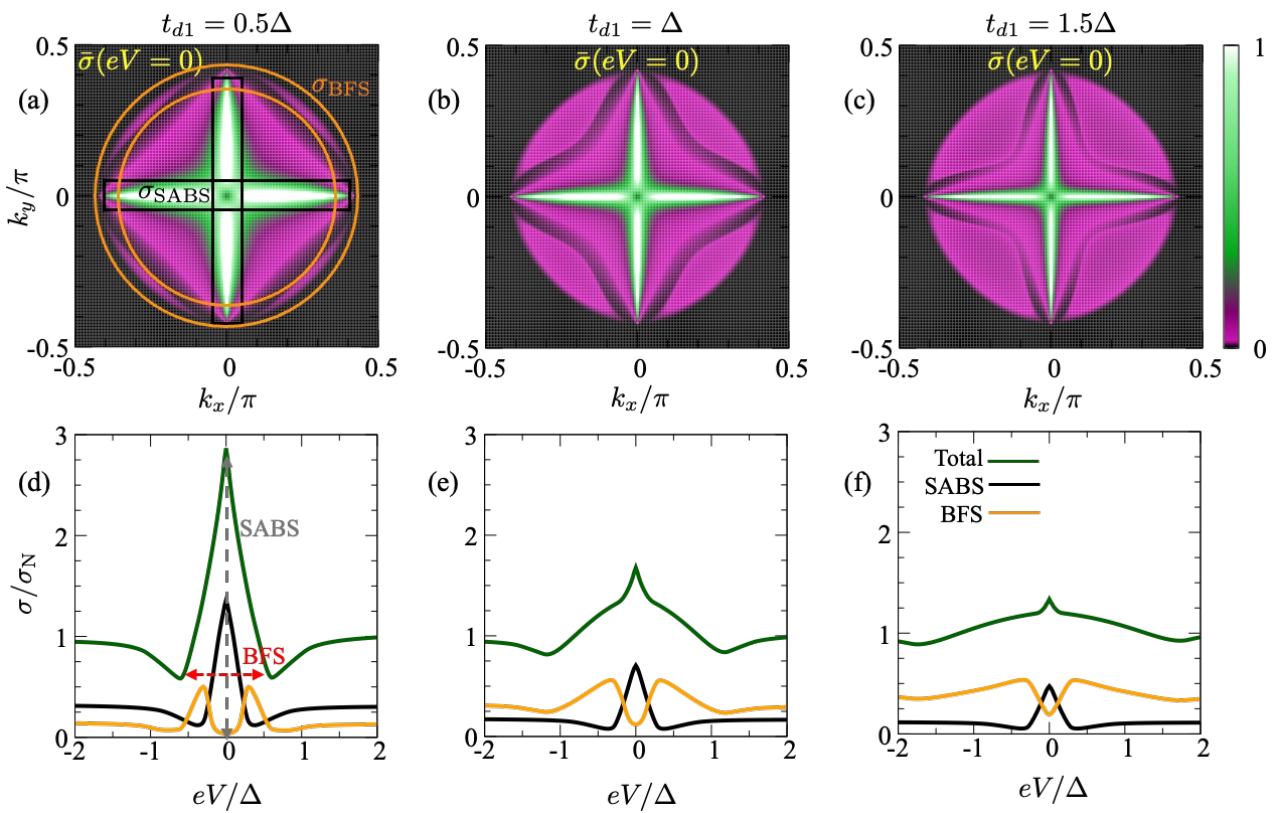}
    \caption{(a,b,c) Zero-bias conductance along the $z$-direction [001] as a function of $k_{x,y}$ for a superconducting AM with the $d_{xy}$-wave.
    (d,e,f) Normalized conductance along the $z$-direction [001] as a function of $eV$ for the configurations of total (green),  crossed flat bands $\sigma_\mathrm{SABS}$ (black), and the Bogoliubov-Fermi surface (orange) $\sigma_\mathrm{BFS}$ with the $d_{xy}$-wave altermagnet.
    In panel (a), we show the schematic image of the three separated regions in the calculations.
    In panel (d), the labels SABS and BFS on the vertical and horizontal dashed arrows indicate that the height and width of the integrated conductance are determined by surface Andreev bound states and Bogoliubov-Fermi surfaces, respectively.
    $\sigma_\mathrm{Gapped}(eV)$ is seen by $\sigma_\mathrm{Gapped}(eV)=\sigma(eV)-\sigma_\mathrm{SABS}(eV)-\sigma_\mathrm{BFS}(eV)$.
    We select the altermagnetic strength as (a)(d) $t_{d1}=0.5\Delta$, (b)(e) $t_{d1}=\Delta$, and (c)(f) $t_{d1}=1.5\Delta$. 
    Parameters: $\mu=-4.5t$, $\Delta=0.01t$,  $U_\mathrm{b}=5t$, $\delta=0.01\Delta$. }
    \label{reply_2}
\end{figure}
\begin{figure}[t!]
    \centering
    \includegraphics[width=13cm]{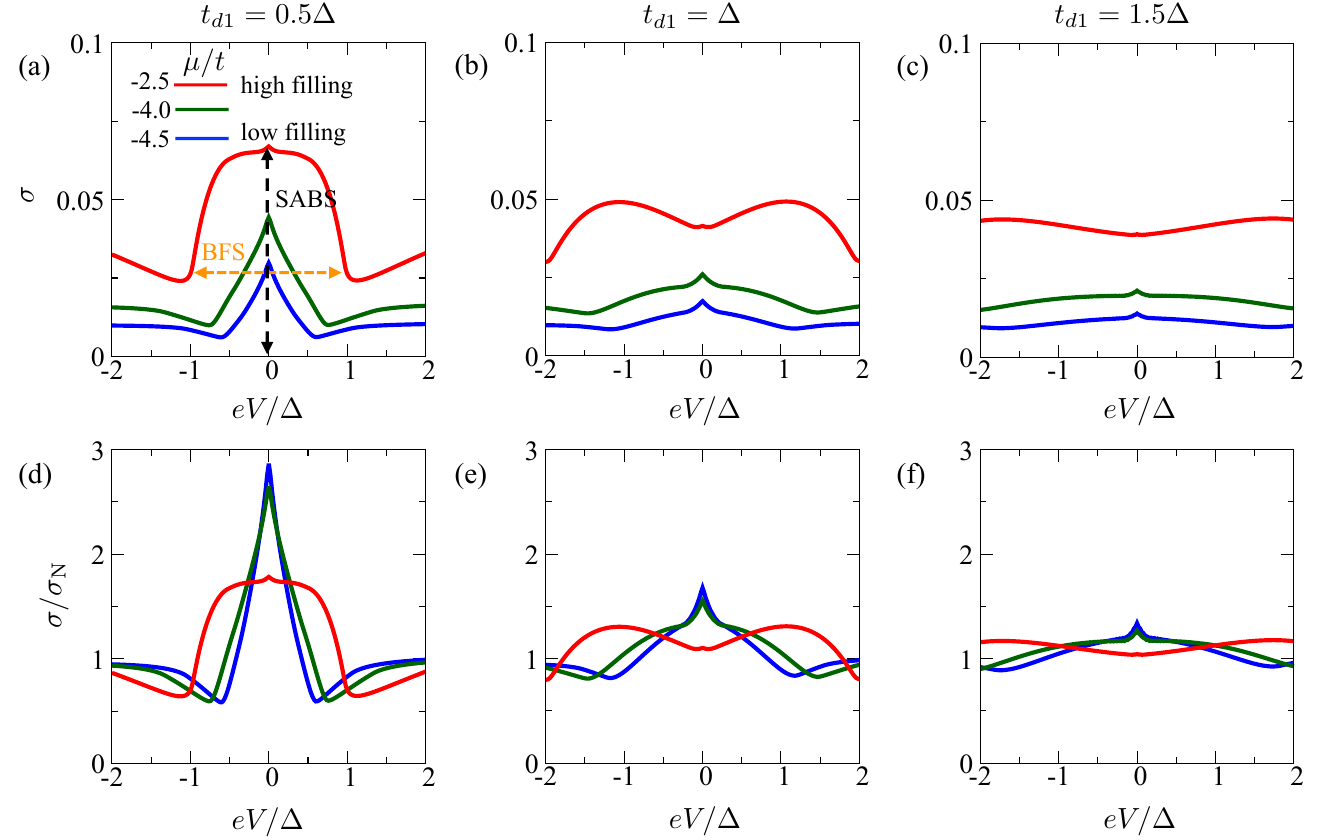}
    \caption{(a,b,c) Total conductance $\sigma(eV)$ along the $z$-direction [001] as a function of $k_{x,y}$ for a superconducting AM with the $d_{xy}$-wave.
    Here, the labels SABS and BFS on the vertical and horizontal dashed arrows indicate that the height and width of the integrated conductance are determined by surface Andreev bound states and Bogoliubov-Fermi surfaces, respectively.
    (d,e,f) Normalized conductance $\sigma(eV)/\sigma_\mathrm{N}$ along the $z$-direction [001] as a function of $eV$ with the $d_{xy}$-wave altermagnet.
    $\sigma_\mathrm{N}$ denotes the normal state conductance at zero bias voltage.
    Blue, green, and red curves correspond to the chemical potential at $\mu=-4.5t$, $-4.0t$, and $-2.5t$, respectively.
    Increasing $\mu$, the distance of the altermagnetic nodal lines becomes larger.
    We select the altermagnetic strength as (a)(d) $t_{d1}=0.5\Delta$, (b)(e) $t_{d1}=\Delta$, and (c)(f) $t_{d1}=1.5\Delta$. 
    Parameters: $\Delta=0.01t$,  $U_\mathrm{b}=5t$, $\delta=0.01\Delta$.}
    \label{reply_2_1}
\end{figure}

\begin{figure}[t!]
    \centering
    \includegraphics[width=13cm]{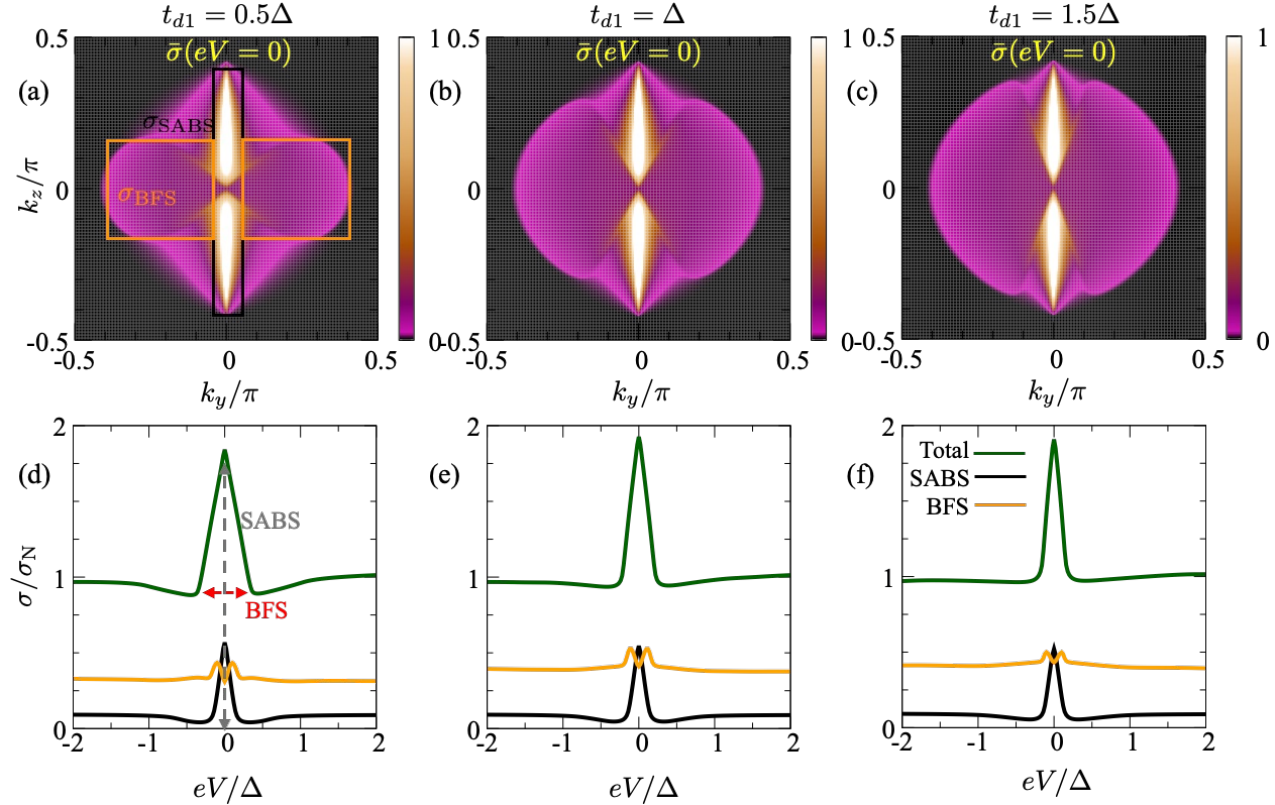}
    \caption{(a,b,c) Zero-bias conductance along the $x$-direction [100] as a function of $k_{y,z}$ for a superconducting AM with the $d_{xy}$-wave.
    (d,e,f) Normalized conductance along the $x$-direction [100] as a function of $eV$ for the configurations of total (green),  surface arc states $\sigma_\mathrm{SABS}$ (black), and the Bogoliubov-Fermi surface (orange) with the $d_{xy}$-wave altermagnet.
    In panel (a), we show the schematic image of the three separated regions in the calculations.
    In panel (d), the labels SABS and BFS on the vertical and horizontal dashed arrows indicate that the height and width of the integrated conductance are determined by surface Andreev bound states and Bogoliubov-Fermi surfaces, respectively.
    $\sigma_\mathrm{Gapped}(eV)$ is seen by $\sigma_\mathrm{Gapped}(eV)=\sigma(eV)-\sigma_\mathrm{SABS}(eV)-\sigma_\mathrm{BFS}(eV)$.
    We select the altermagnetic strength as (a)(d) $t_{d1}=0.5\Delta$, (b)(e) $t_{d1}=\Delta$, and (c)(f) $t_{d1}=1.5\Delta$. 
    Parameters: $\mu=-4.5t$, $\Delta=0.01t$,  $U_\mathrm{b}=5t$, $\delta=0.01\Delta$. }
    \label{reply_3}
\end{figure}%

In this subsection, we demonstrate which region of the momentum-dependent conductance contributes to the total conductance. 
It is useful to decompose it into the contributions coming from the three distinct regimes we found, namely,
\begin{align}
    \sigma(eV)=\sigma_\mathrm{SABS}(eV)+\sigma_\mathrm{BFS}(eV)+\sigma_\mathrm{Gapped}(eV).
\end{align}%
Here, $\sigma_\mathrm{SABS}(eV)$, $\sigma_\mathrm{BFS}(eV)$, and $\sigma_\mathrm{Gapped}(eV)$ correspond to the contributions to the integrated conductance due to:  1) surface Andreev bound states (SABSs): crossed flat bands and surface arc states $\sigma_\mathrm{SABS}(eV)$, 2) the Bogoliubov-Fermi surface (BFS) $\sigma_\mathrm{BFS}(eV)$, and 3) the gapped regions $\sigma_\mathrm{Gapped}(eV)$; see Figs.~\ref{reply_2}(a-c) for a description of how these contributions determine the total integrated conductance along [001] for a $d_{xy}$-wave altermagnet. 
In 1), the SABSs lead to the sharp zero-bias peak in the integrated conductance~\cite{TK95}. 
The profile of the zero bias peak in $\sigma_\mathrm{SABS}(eV)$ is determined by the magnitude of the bulk gap, the distance between superconducting gap nodes in the surface Brillouin zone, and also by the length of the crossed flat bands. 
The latter two effects, the distance between superconducting gap nodes and length of the crossed flat band, can be directly identified by exploiting their dependence on the chemical potential $\mu$: by increasing $\mu$, the distance between superconducting gap nodes and the length of the crossed flat bands increase, which enhances the hight of the zero bias conductance in the total integrated conductance $\sigma(eV)$, which is mainly determined by $\sigma_\mathrm{SABS}(eV)$. 
We demonstrate this enhancement in the total integrated conductance in Figs.~\ref{reply_2_1} (a-c), where we show $\sigma(eV)$ as a function of $eV$ for distinct $\mu$; the enhancement in the total conductance is determined by the contribution from SABSs, as shown in Fig.~\ref{reply_2}(d). 
The height of the zero bias conductance is determined by the SABSs, while the width by the BFS, of importance for identifying the effect of altermagnetism and measuring its strength.  
It is worth noting that larger values of altermagnetism compared to the pair potential tend to reduce the overall integrated conductance [Figs.~\ref{reply_2_1}(a-c)], but still leave clear signals of altermagnetism: for instance, the width of the subgap conductance determines the strength of altermagnetism [Figs.~\ref{reply_2_1}(d-f)], while the shape enables the identification of whether the contribution is coming from SABSs or BFSs [Figs.~\ref{reply_2}(d-f)]. See also vertical and horizontal dashed arrows in Fig.~\ref{reply_2_1}(a).

Furthermore, when looking at the normalized conductance [Figs.~\ref{reply_2_1}(d-f)], which is what we presented in the main text, the zero bias conductance for lower chemical potentials develop larger values: this apparent contradiction can be easily resolved by noting that the integrated normal conductance scales with the 2D area of the Fermi surface on the surface, while the superconducting conductance scales just with the length of the Fermi surface on the surface. Nevertheless, all the remaining features of the integrated conductance remain, and the strength of altermagnetism can be easily extracted [Figs.~\ref{reply_2_1}(d-f)]. 
In 2), the density of states of the bulk Bogoliubov bands near the zero energy increases with the increment of the volume of the BFS~\cite{OhashiPRB2024}, and the region of the BFS becomes larger with the increase of the altermagnetic strength; this is of course seen by noting the increase in the separation between dips at positive and negative energies in the total conductance for $|eV|>0$ as the strength of altermagnetism increases [Fig.~\ref{reply_2_1}]; see also Figs.~\ref{reply_2}(d-f). 
In 3), the gapped region contributes to the V-shape conductance, reflecting the superconducting nodal structures in the bulk~\cite{KT2000}; although it is not very straightforward to identify this contribution, one can use the total conductance and the contributions from crossed flat bands and BFSs to determine the conductance due to the gapped region, see our response to Comment 3.1 of Referee B. 
While Fig.~\ref{reply_2} and Fig.~\ref{reply_2_1} address only $d_{xy}$-wave altermagnet, we have verified that similar conclusions can be drawn for  $d_{x^{2}-y^{2}}$-wave and $g$-wave altermagnets; for these other altermagnets, however, the conductance reflects their distinct nature, such that it is also possible to identify the type of altermagnetism.

To further support the above discussion, in Fig.~\ref{reply_2} and Fig.~\ref{reply_3}, we also present the momentum-dependent conductance as well as the total integrated conductance and their contributions from SABSs and BFSs as a function of $eV$ for distinct altermagnetic strengths.
With the increase of the altermagnetic strength $t_{d1}$, the width of the peak near the zero-bias voltage changes [green and orange lines Figs.~\ref{reply_2} (d-f) and Figs.~\ref{reply_3} (d-f)]. 
This is because the signature of the BFS appears as peak width in the integrated conductance. 
On the other hand, the signature of the altermagnetic nodal lines is exhibited at the height of the peak at zero-bias $eV=0$ in the integrated conductance because $\sigma(eV=0)$ becomes suppressed [green and black lines Figs.~\ref{reply_2} (d-f) and Figs.~\ref{reply_3} (d-f)].

\section{S3: Model and conductance of hybrid junctions based on 3D superconductors with 3D altermagnets} \label{Sec_S3}

In this section, we demonstrate the conductance and bulk density of states (DOS) by using the model of 3D SCs with 3D AMs.
Finally, we study the stability of zero-bias conductance in the case of 3D SCs with 3D AMs.

\subsection{S3.1: Model of 3D superconductors with 3D altermagnets} \label{Sec_S3_1}

In the main text, we focused on the case of 2D AMs.
We now proceed to consider 3D AMs suggested in Ref.\ \cite{EzawaPRB}: $d_{z(x+y)}$, $g_{zx(x^2-3y^2)}$, and $g_{yz(3x^2-y^2)}$-wave, coexisting with superconducting orders.  
We follow the same approach discussed in S2.1 and S2.2 of the SM, but make the following changes.
Instead of Eq.\ (2) in the main text and Eqs.\ (\ref{2D_AM1})-(\ref{2D_AM3}), the Hamiltonian of 3D AM is rewritten as~\cite{EzawaPRB}
\begin{align}
    \hat{H}^{\alpha'}_\mathrm{AM}(\bm{k})&=M^{\alpha'}_{\bm{k}}\hat{\sigma}_{3}\hat{\tau}_{3},\\
    M^{d3}_{\bm{k}}&=t_{d3}\sin{k_z}(\sin{k_x}+\sin{k_y}),\\
    M^{g2}_{\bm{k}}&=t_{g2}\sin{k_z}\sin{k_x}(\sin^2{k_x}-3\sin^2{k_y})\\
    M^{g3}_{\bm{k}}&=t_{g3}\sin{k_y}\sin{k_z}(3\sin^2{k_x}-\sin^2{k_y})
\end{align}%
where $t_{\alpha'}$ with $\alpha'=d3,g2,g3$ are the strength of the $d_{z(x+y)}$, $g_{zx(x^2-3y^2)}$, and $g_{yz(3x^2-y^2)}$-wave AMs, respectively.

For the Fourier transformation, we obtain the local and hopping terms:
\begin{align}
    \hat{u}^{\alpha'}_\mathrm{AM}(k_x,k_y)&=0,\\
    \hat{t}^{d3}_\mathrm{AM1}(k_x,k_y)&=\frac{i}{2}t_{d3}(\sin{k_x}+\sin{k_y}),\\
    \hat{t}^{g2}_\mathrm{AM1}(k_x,k_y)&=t_{g2}\sin{k_x}(\sin^2{k_x}-3\sin^{2}{k_y}),\\
    \hat{t}^{g3}_\mathrm{AM1}(k_x,k_y)&=t_{g3}\sin{k_y}(3\sin^2{k_x}-\sin^{2}{k_y}),
\end{align}%
along the $z$-direction, and
\begin{align}
    \hat{u}^{d3}_\mathrm{AM}(k_y,k_z)&=t_{d3}\sin{k_z}\sin{k_y}, \label{u_am_3D_AM1}\\
    \hat{u}^{g3}_\mathrm{AM}(k_y,k_z)&=t_{g3}\sin{k_y}\sin{k_z}\left[\frac{3}{2}-\sin^{2}{k_y}\right],
\end{align}%
\begin{align}
    \hat{t}^{d3}_\mathrm{AM1}(k_y,k_z)&=\frac{i}{2}t_{d3}\sin{k_z},\\
    \hat{t}^{g2}_\mathrm{AM1}(k_y,k_z)&=\frac{i}{2}t_{g2}\left[\frac{3}{4}\sin{k_z}-3\sin^{2}{k_y}\sin{k_z}\right],\\
    \hat{t}^{g3}_\mathrm{AM1}(k_y,k_z)&=0,\\
    \hat{t}^{d3}_\mathrm{AM2}(k_y,k_z)&=\hat{t}^{g2}_\mathrm{AM2}(k_y,k_z)=0,\\
    \hat{t}^{g3}_\mathrm{AM2}(k_y,k_z)&=-\frac{3}{4}t_{g3}\sin{k_y}\sin{k_z},\\
    \hat{t}^{d3}_\mathrm{AM3}(k_y,k_z)&=\hat{t}^{g3}_\mathrm{AM3}(k_y,k_z)=0,\\
    \hat{t}^{g2}_\mathrm{AM3}(k_y,k_z)&=-\frac{i}{8}t_{g2}\sin{k_z},
\end{align}%
along the $x$-direction, with
\begin{align}
    \tilde{t}^{\alpha'}_\mathrm{sc3}(\tilde{\bm{k}})&=
    \begin{pmatrix}
        \hat{t}^{\alpha'}_\mathrm{AM3}(\tilde{\bm{k}}) & 0\\
        0 & -\hat{t}^{\alpha'*}_\mathrm{AM3}(-\tilde{\bm{k}})
    \end{pmatrix}. \label{t_sc3}
\end{align}%
For including the second NN hopping term $\tilde{t}_\mathrm{sc2}$ like the $g_{yz(3x^2-y^2)}$-wave AM, we adopt Eqs.\ (\ref{usc2}) and (\ref{tsc2}) to obtain the semi-infinite Green's function.
When the third NN hopping term $\tilde{t}_\mathrm{sc3}$ is included like for the $g_{zx(x^2-3y^2)}$-wave AM, then we replace:
\begin{align}
    \hat{u}\rightarrow
    \begin{pmatrix}
        \tilde{u}_\mathrm{sc} & \tilde{t}_\mathrm{sc1}&\tilde{t}_\mathrm{sc2}\\
        \tilde{t}^{\dagger}_\mathrm{sc1} & \tilde{u}_\mathrm{sc} & \tilde{t}_\mathrm{sc1}\\
        \tilde{t}^{\dagger}_\mathrm{sc2} & \tilde{t}^{\dagger}_\mathrm{sc1} & \tilde{u}_\mathrm{sc}
    \end{pmatrix}, \label{u_trans_3}
\end{align}%
\begin{align}
        \hat{t}\rightarrow
    \begin{pmatrix}
        \tilde{t}_\mathrm{sc3} & 0 & 0\\
        \tilde{t}_\mathrm{sc2} & \tilde{t}_\mathrm{sc3} & 0\\
        \tilde{t}_\mathrm{sc1} & \tilde{t}_\mathrm{sc2} & \tilde{t}_\mathrm{sc3}
    \end{pmatrix}.
\end{align}%
We also choose the subspace of the semi-infinite Green's function $\tilde{G}^\mathrm{L}_{0}=\tilde{G}_{cc}$ defined as:
\begin{align}
    \tilde{G}=\left(
    \begin{array}{c|c|c}
        \tilde{G}_{aa} & \tilde{G}_{ab} & \tilde{G}_{ac} \\\hline
        \tilde{G}_{ba} & \tilde{G}_{bb} & \tilde{G}_{bc} \\ \hline
        \tilde{G}_{ca} & \tilde{G}_{cb} & \tilde{G}_{cc}
    \end{array}\right), \label{g_33}
\end{align}%
in the basis $[\hat{\psi}^{\dagger}_{a},\hat{\psi}^{\dagger}_{b},\hat{\psi}^{\dagger}_{c}]$ with $\hat{\psi}^{\dagger}_{l=a,b,c}=[c^{\dagger}_{l\uparrow},c^{\dagger}_{l\downarrow},c_{l\uparrow},c_{l\downarrow}]$, to connect $\hat{V}_{10}$.
This method is applicable when calculating the conductance along the $x$-direction for the $g_{zx(x^2-3y^2)}$-wave AM in S3.3 of the SM.
We calculate the conductance in 3D superconducting AM junctions by this formulation.
The results are shown in the next subsections.

\subsection{S3.2: DOS and conductance along the $z$-direction} \label{Sec_S3_2}

We discussed the conductance in 2D AMs in the main text and the BFS in the End Matter (EM).
In this section, we present the additional results of the conductance caused by the 3D superconducting AMs: $d_{z(x+y)}$, $g_{zx(x^2-3y^2)}$, and $g_{yz(3x^2-y^2)}$-wave.
This additional study aims to support the role of 3D superconducting AMs for hosting exotic zero-energy flat band states.
To obtain the conductance in the case of 3D AMs, we use Eqs.\ (\ref{u_am_3D_AM1})-(\ref{t_sc3}), Eqs.\ (\ref{usc2}) and (\ref{tsc2}), and (\ref{u_trans_3})-(\ref{g_33}).
\begin{figure}[t!]
    \centering
    \includegraphics[width=14cm]{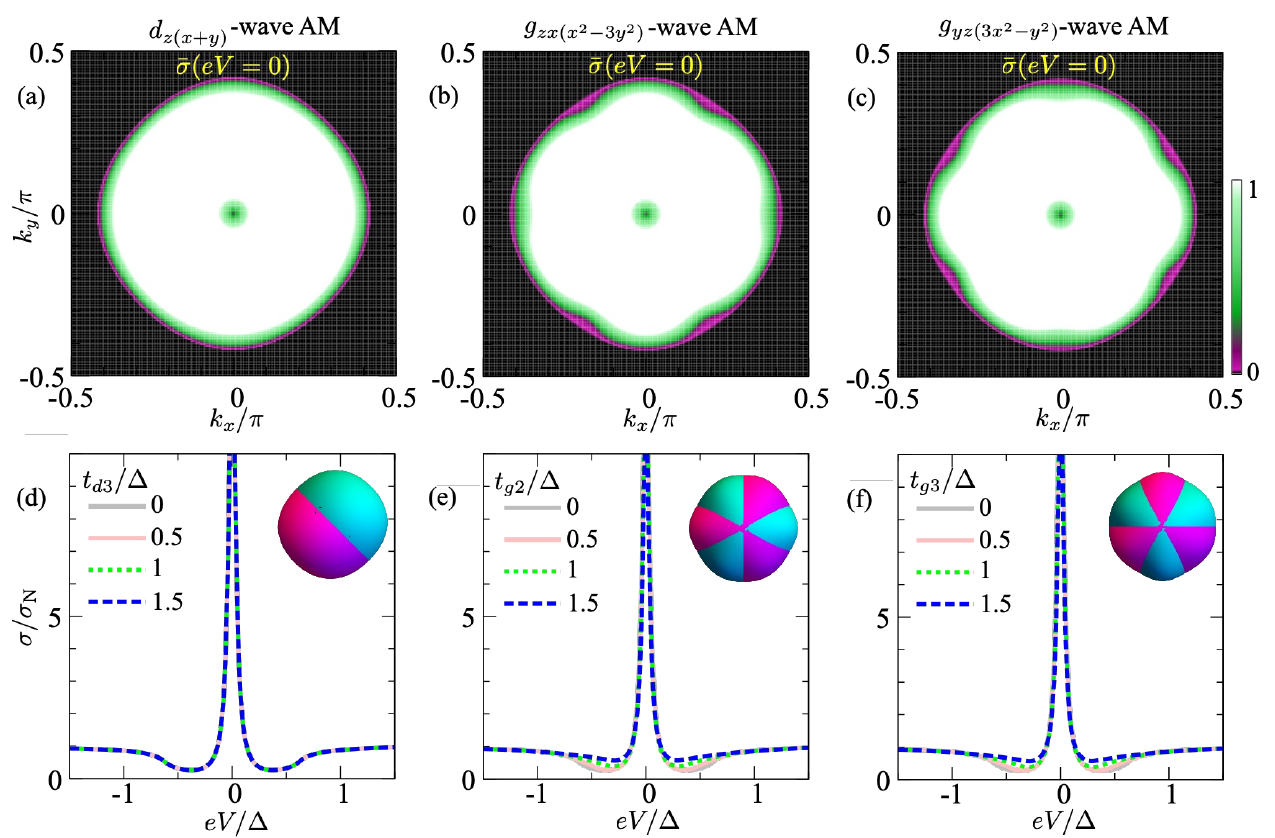}
    \caption{
    (a,b,c) Conductance along the $z$-direction as a function of $(k_x,k_y)$ at $eV=0$ with (a) $d_{z(x+y)}$, (b) $g_{zx(x^2-3y^2)}$, and (c) $g_{yz(3x^2-y^2)}$-wave AM.
    We select the altermagnetic order as (a) $t_{d3}=\Delta$, (b) $t_{g2}=\Delta$, and (c) $t_{g3}=\Delta$.
    (d,e,f) Conductance along the $z$-direction with (d) $d_{z(x+y)}$, (e) $d_{zx(x^2-3y^2)}$, and (f) $g_{yz(3x^2-y^2)}$-wave AM.
    $\sigma_\mathrm{N}$ indicates the conductance in the normal state at $eV=0$.
    Gray, light-red, green, and blue lines indicate the strength of the altermagnetic order as $t_{\alpha'}=0$, $0.5\Delta$, $\Delta$, and $1.5\Delta$ with $\alpha'=d3,g2,g3$.
    In (d,e,f), we show the image of the normal state Fermi surface when we see along the $z$-direction.
    We set parameters as $\mu=-4.5t$, $\Delta=0.01t$, $U_\mathrm{b}=5t$, and $\delta=0.01\Delta$.}
    \label{fig_sm:conductance_z}
\end{figure}%
\begin{figure}[t!]
    \centering
    \includegraphics[width=14.5cm]{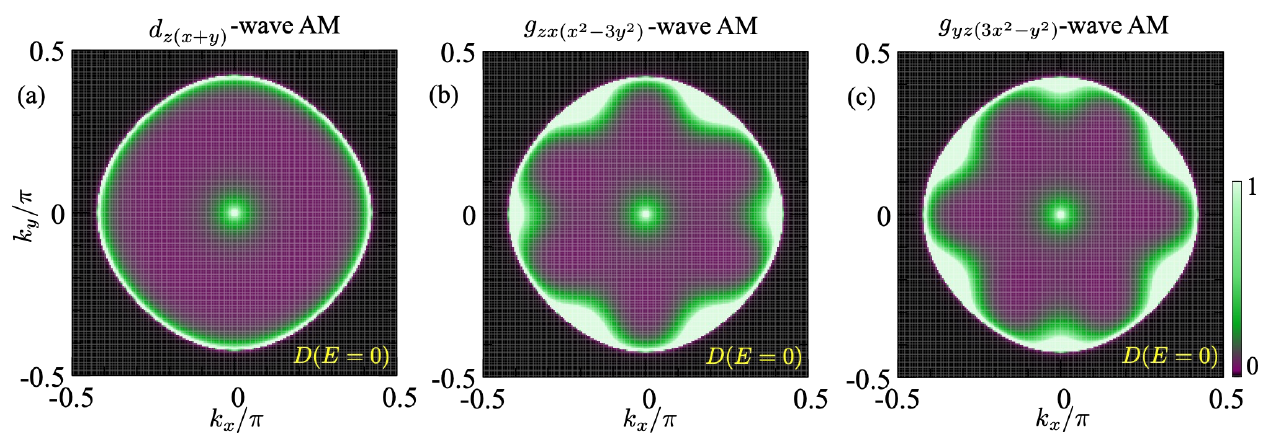}
    \caption{Bulk density of states (DOS) at $E=0$ in 3D chiral $d$-wave SCs integrated for $k_z$ for (a) $d_{z(x+y)}$, (b) $g_{zx(x^2-3y^2)}$, and (c) $g_{yz(3x^2-y^2)}$-wave AM.
    AM is set as $t_{\alpha'}=\Delta$ with $\alpha'=d3,g2,g3$.
    Bright (dark) region indicates the Bogoliubov Fermi surface (crossed flat bands).
    We set parameters as $\mu=-4.5t$, $\Delta=0.01t$, and $\delta=0.01\Delta$.}
    \label{fig_sm:BFS_z}
\end{figure}%
\begin{figure}[t!]
    \centering
    \includegraphics[width=14cm]{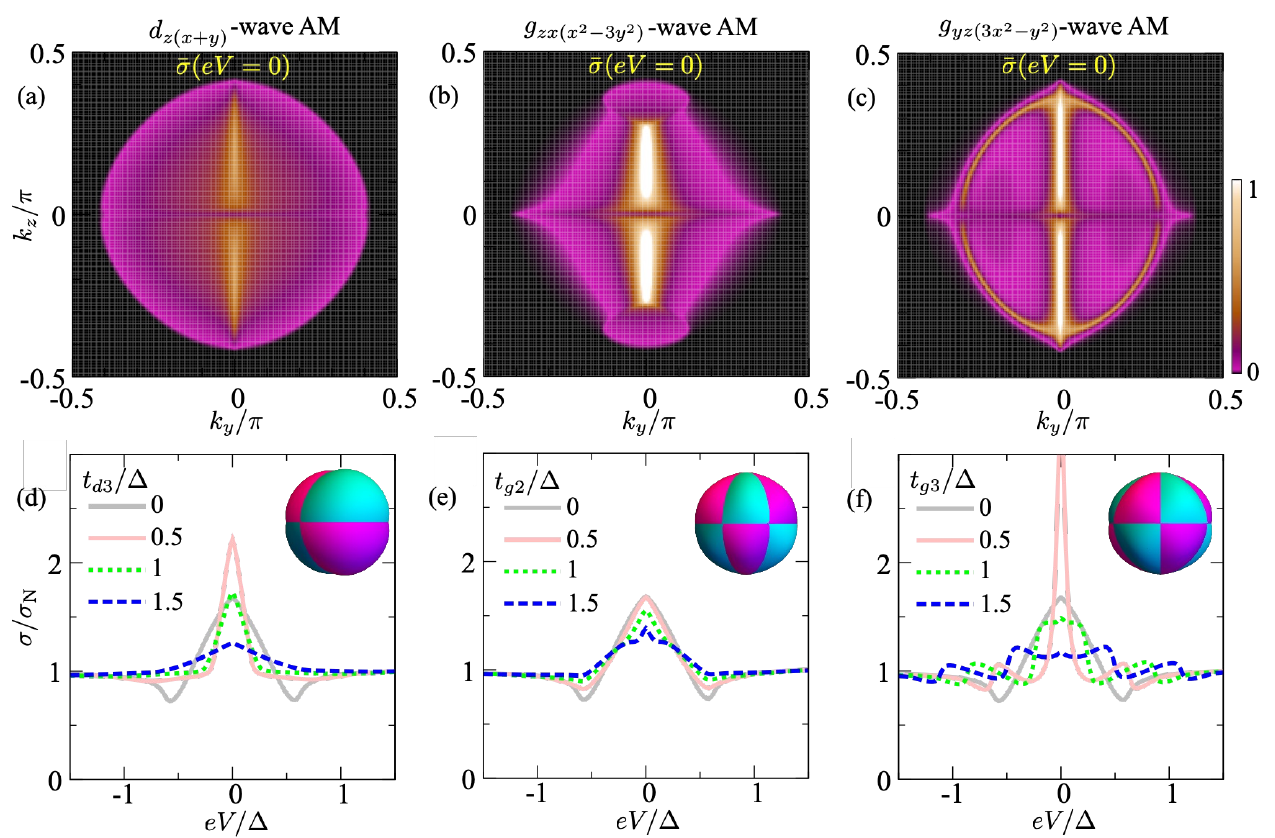}
    \caption{
    (a,b,c) Conductance along the $x$-direction as a function of $(k_y,k_z)$ at $eV=0$ with (a) $d_{z(x+y)}$, (b) $g_{zx(x^2-3y^2)}$, and (c) $g_{yz(3x^2-y^2)}$-wave AM.
    We select the altermagnetic order as (a) $t_{d3}=\Delta$, (b) $t_{g2}=\Delta$, and (c) $t_{g3}=\Delta$.
    (d,e,f) Conductance along the $x$-direction with (d) $d_{z(x+y)}$, (e) $d_{zx(x^2-3y^2)}$, and (f) $g_{yz(3x^2-y^2)}$-wave AM.
    $\sigma_\mathrm{N}$ indicates the conductance in the normal state at $eV=0$.
    Gray, light-red, green, and blue lines indicate the strength of the altermagnetic order as $t_{\alpha'}=0$, $0.5\Delta$, $\Delta$, and $1.5\Delta$ with $\alpha'=d3,g2,g3$.
    In (d,e,f), we show the image of the normal state Fermi surface when we see along the $x$-direction.
    We set parameters as  $\mu=-4.5t$, $\Delta=0.01t$, $U_\mathrm{b}=5t$, and $\delta=0.01\Delta$.}
    \label{fig_sm:conductance_x}
\end{figure}%
\begin{figure}[t!]
    \centering
    \includegraphics[width=14.5cm]{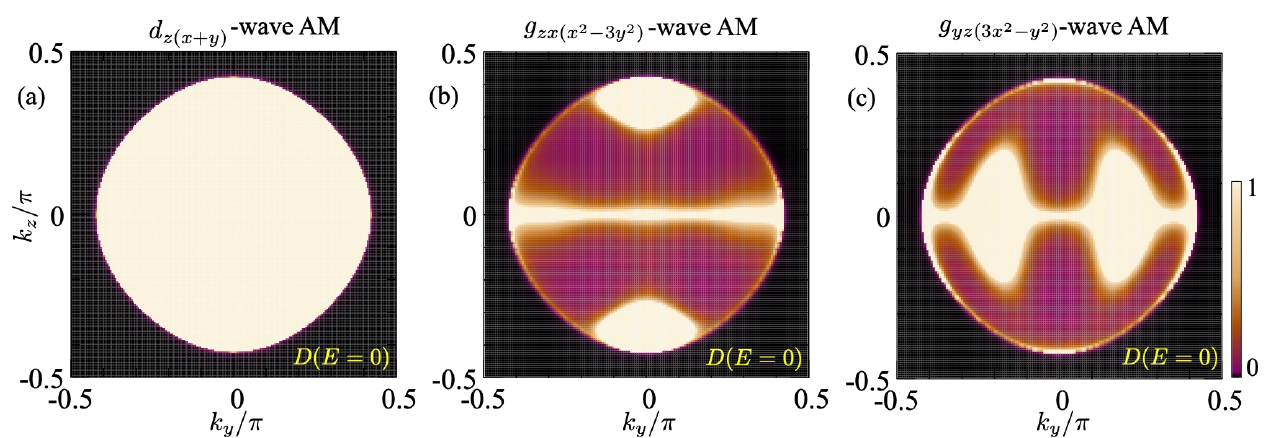}
    \caption{Bulk density of states (DOS) at $E=0$ in 3D chiral $d$-wave SCs integrated for $k_x$ for (a) $d_{z(x+y)}$, (b) $g_{zx(x^2-3y^2)}$, and (c) $g_{yz(3x^2-y^2)}$-wave AM.
    AM is set as $t_{\alpha'}=\Delta$ with $\alpha'=d3,g2,g3$.
    Bright (dark) region indicates the Bogoliubov Fermi surface (surface arc states).
    We set parameters as $\mu=-4.5t$, $\Delta=0.01t$, and $\delta=0.01\Delta$.}
    \label{fig_sm:BFS_x}
\end{figure}%

We plot the momentum-dependent conductance at zero-bias voltage $eV=0$ and the normalized conductance along the $z$-directions for each 3D AM in Fig.~\ref{fig_sm:conductance_z} at $U_\mathrm{b}=5t$.
Then we show the projected Fermi surface with up (magenta) and down spins (cyan) for each transport direction in panels (d,e,f) of Fig.~\ref{fig_sm:conductance_z}.
Along the $z$-direction, for the $d_{z(x+y)}$ AM, because the crossed flat bands do not appear [Fig.~\ref{fig_sm:conductance_z} (a)], the resulting line shape of the conductance does not change within this parameter range [Fig.~\ref{fig_sm:conductance_z} (d)].
While for the $g_{zx(x^2-3y^2)}$ and $g_{yz(3x^2-y^2)}$-wave AM, both BFS and crossed flat bands are obtained [Figs.~\ref{fig_sm:conductance_z} (b,c)].
Then the number of corners of the crossed flat bands corresponds to that of the nodal lines of AM.
These behaviors are the same as those in the case of AMs discussed in the main text.
Nevertheless, the zero-bias conductance peak (ZBCP) is not strongly suppressed with the increase of altermagnetism shown in Figs.~\ref{fig_sm:conductance_z} (e,f). 
Indeed, the crossed flat band is also demonstrated for the $g_{zx(x^2-3y^2)}$ and $g_{yz(3x^2-y^2)}$-wave AMs, not $d_{z(x+y)}$-wave AM.
Next, we discuss the BFS, and then, we calculate the bulk DOS, as described in the case of 2D AM in the EM.
In Fig.~\ref{fig_sm:BFS_z}, we also show how the BFS emerges in the case of 3D AM: $d_{z(x+y)}$, $d_{zx(x^2-3y^2)}$, and $d_{yz(3x^2-y^2)}$-wave, by using the bulk DOS.
As we also mentioned in the EM, the bright (dark) color of the bulk DOS corresponds to the BFS (the crossed flat band and the surface arc state), and the bright point at $(k_x,k_y)=(0,0)$ is the nodal point where the energy gap closes.
For the $d_{z(x+y)}$-wave AM, the structure of DOS does not change [Figs.~\ref{fig_sm:BFS_z} (a)] compared with that without altermagnetism.
It means that the bulk BFS integrated for $k_z$ does not appear along the $z$-direction, and then the corresponding conductance is independent of $t_{d3}$ [Fig.~\ref{fig_sm:conductance_z} (b)].
While the BFS appears for the $g_{zx(x^2-3y^2)}$ and $g_{yz(3x^2-y^2)}$-wave AM by $t_{\alpha'}$ with $\alpha'=g2,g3$ [Figs.~\ref{fig_sm:BFS_z}(b) and Figs.~\ref{fig_sm:BFS_z} (c)].
Then the number of corners corresponds to the nodes of 3D AM.
Hence, 3D AMs lead to the BFS for $g_{zx(x^2-3y^2)}$ and $g_{yz(3x^2-y^2)}$-wave AMs, except for the $d_{z(x+y)}$-wave AM.
We have shown the conductance in superconducting AMs discussed in the main text.

\begin{figure}[t!]
    \centering
    \includegraphics[width=14cm]{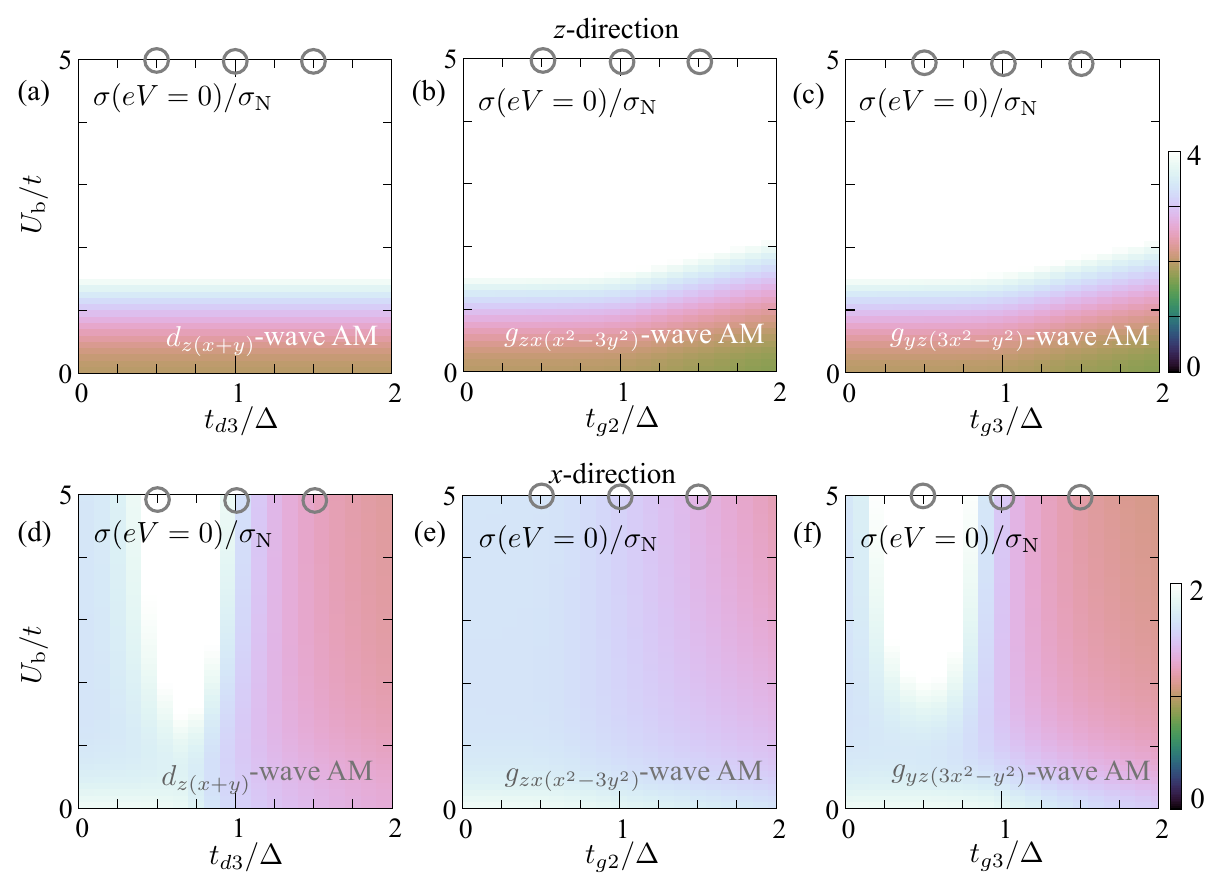}
    \caption{Normalized zero-bias conductance along $z$ (a,b,c) and $x$ (d,e,f) transport directions as functions of $U_\mathrm{b}$ and $t_{\alpha'}$.
    The altermagnetic order: (a,d) $d_{z(x+y)}$, (b,e) $g_{zx(x^2-3y^2)}$, (c,f) $g_{yz(3x^2-y^2)}$-wave.
    $\sigma_\mathrm{N}$ denotes the conductance in the normal state at $eV=0$, and we calculate it for each $t_{\alpha'}$ and $U_\mathrm{b}$.
    Gray circles indicate the parameters shown in Fig.~\ref{fig_sm:conductance_z} and Fig.~\ref{fig_sm:conductance_x}.
    Parameters: $\mu=-4.5t$, $\Delta=0.01t$, and $\delta=0.01\Delta$.
    }
    \label{fig_sm:Ub}
\end{figure}%

\subsection{S3.3: DOS and conductance along the $x$-direction} \label{Sec_S3_3}

Along the $x$-direction, we also show the momentum-dependent conductance at zero-bias voltage $eV=0$ and the normalized conductance along the $x$-directions for each 3D AM in Fig.~\ref{fig_sm:conductance_x} at $U_\mathrm{b}=5t$.
For the $d_{z(x+y)}$-wave AM shown in Figs.~\ref{fig_sm:conductance_x} (a), the intensity of surface arc states becomes weak due to the BFS [Figs.~\ref{fig_sm:BFS_x} (a)], and the conductance at zero-bias voltage is first enhanced and suppressed by $t_{d3}$ in Fig.~\ref{fig_sm:conductance_x} (d).
For the $g_{zx(x^2-3y^2)}$-wave AM, because the point nodes mainly become BFS, see also Figs.~\ref{fig_sm:BFS_x} (b), the region of surface arc states reduces by BFS [Fig.~\ref{fig_sm:conductance_x} (b)].
Indeed, as shown in Fig.~\ref{fig_sm:conductance_x} (e), the zero-bias conductance is suppressed by $t_{d2}$.
While for the $g_{yz(3x^2-y^2)}$-wave AM, spin-split surface arc states appear [Fig.~\ref{fig_sm:conductance_x} (c)], and the resulting zero-bias conductance is strongly suppressed by $t_{d3}$ [Fig.~\ref{fig_sm:conductance_x} (f)].
Indeed, 3D AMs also lead to the modification of the surface states and the resulting conductance, as we worked in the case of AMs shown in the main text.
The energy gap structure becomes gapless at $t_{d3}=\Delta$ [Figs.~\ref{fig_sm:BFS_x} (a)] for the $d_{z(x+y)}$-wave AM. 
While for the $g_{zx(x^2-3y^2)}$-wave AM, at $t_{g2}=\Delta$, not only line node at $k_z=0$ but also point nodes at $k_z=k_\mathrm{F}$ with the Fermi vector $k_\mathrm{F}$ mainly become the BFS [Fig.~\ref{fig_sm:BFS_x} (b)].
As a result, because the region of the surface arc states is small for the BFS, the zero-bias conductance is suppressed by $t_{g2}$ [Fig.~\ref{fig_sm:conductance_x} (e)].
For $g_{yz(3x^2-y^2)}$-wave AM, the line node at $k_z=0$ partly becomes the BFS at $t_{g3}=\Delta$ [Fig.~\ref{fig_sm:BFS_x} (c)]. 
Since surface arc states are spin-split by avoiding the BFS [Fig.~\ref{fig_sm:conductance_x} (e)], the resulting zero-bias conductance is also suppressed, and the line shape of the conductance is also split [Fig.~\ref{fig_sm:conductance_x} (f)].
Indeed, the BFS appears in 3D AMs, as well as AM cases in the main text, and surface arc states modified by the BFS contribute to the line shape of the conductance.
We have also demonstrated the conductance in the case of AMs in the main text.

\subsection{S3.4: Stability of zero-bias conductance in 3D superconducting AM junctions} \label{Sec_S3_4}

In the main text, we discussed the normalized zero-bias conductance for the potential barrier $U_\mathrm{b}$ and the strength of AM $t_{\alpha}$ ($\alpha=d1,d2,g$) space.
We also demonstrate the normalized zero-bias conductance along $z$ [Figs.~\ref{fig_sm:Ub} (a,b,c)] and $x$ [Figs.~\ref{fig_sm:Ub} (d,e,f)] directions for $U_\mathrm{b}$ and the strength of 3D AM $t_{\alpha'}$.
For the $d_{z(x+y)}$-wave AM, because the ZBCP is not suppressed with the increase of $t_{\alpha'}$ along the $z$-direction, the white area does not decrease [Figs.~\ref{fig_sm:Ub} (a)].
Then, crossed flat bands do not appear for the $d_{z(x+y)}$-wave AM.
While for the $g_{zx(x^2-3y^2)}$ and $g_{yz(3x^2-y^2)}$-wave AM, even though the crossed flat bands are realized by AM, the normalized zero-bias conductance is not suppressed [Figs.~\ref{fig_sm:Ub} (b,c)]. 
We note that crossed flat bands emerge even at small $U_\mathrm{b}$ for the $g_{zx(x^2-3y^2)}$ and $g_{yz(3x^2-y^2)}$-wave AM [Figs.~\ref{fig_sm:Ub} (a,b,c)].
Along the $x$-direction, for the $d_{z(x+y)}$ and $g_{yz(3x^2-y^2)}$-wave AM, the normalized zero-bias conductance is enhanced once for $0.25\Delta\le t_{\alpha'}\le\Delta$ with $\alpha'=d3,g3$, and it is suppressed for $t_{\alpha'}>\Delta$ [Figs.~\ref{fig_sm:Ub} (d,f)], see also Figs.~\ref{fig_sm:conductance_x} (d,f).
While for the $g_{zx(x^2-3y^2)}$-wave AM, the normalized zero-bias conductance gradually decreases with the increase of $t_{d2}$ [Figs.~\ref{fig_sm:conductance_x} (e)].
This behavior corresponds to the modification of surface arc states by the BFS.
Therefore, zero-bias conductance can also be modified by 3D AMs, as we have found in the case of AMs in the main text.

\bibliography{biblio_SM}

% \begin{thebibliography}{11}
% \bibitem{LeeFisher} P.\ A.\ Lee and D.\ S.\ Fisher, Anderson localization in two
% dimensions, Phys.\ Rev.\ Lett.\ \textbf{47}, 882 (1981).
% \bibitem{OhashiPRB} R.\ Ohashi, S.\ Kobayashi, S.\ Kanazawa, Y.\ Tanaka, and Y.\ Kawaguchi, Surface density of states and tunneling spectroscopy of a spin-3/2 superconductor with Bogoliubov Fermi surfaces, Phys.\ Rev.\ B \textbf{110}, 104515 (2024).
% \bibitem{Fukaya_arXiv06} Y.\ Fukaya, K.\ Yada, and Y.\ Tanaka, Tunneling conductance in superconducting junctions with p-wave unconventional magnets breaking time-reversal symmetry, arXiv: 2506.13372.
% \bibitem{Bo_arXiv08} B.\ Lu, P.\ Mercebach, P.\ Burset, K.\ Yada, J.\ Cayao, Y.\ Tanaka, and Y.\ Fukaya, Engineering subgap states in superconductors by altermagnetism, arXiv: 2508.03364 (2025).
% \bibitem{Umerski} A.\ Umerski, Closed-form solutions to surface Green’s functions, Phys.\ Rev.\ B \textbf{55}, 5266 (1997).
% \bibitem{Takagi20} D.\ Takagi, S.\ Tamura, and Y.\ Tanaka, Odd-frequency pairing and proximity effect in Kitaev chain systems including a topological critical point, Phys.\ Rev.\ B \textbf{101}, 024509 (2020).
% \bibitem{EzawaPRB} M.\ Ezawa, Third-order and fifth-order nonlinear spin-current generation in $g$-wave and $i$-wave altermagnets and perfectly nonreciprocal spin current in $f$-wave magnets, Phys.\ Rev.\ B \textbf{111}, 125420 (2025).
% % \bibitem{BTK} G. E. Blonder, M. Tinkham, and T. M. Klapwijk, Phys. Rev. B {\bf 25}, 4515 (1982).
% \end{thebibliography}